\documentclass{style}
\usepackage{graphicx}
\usepackage{amssymb}
\usepackage{amsmath}
\usepackage{color} 
\begin{document}

\title{  State-of-the-art of beyond mean field theories with nuclear density functionals}
\author{
  J. Luis Egido\email{J.Luis.Egido@uam.es} \\
  \it  Departamento de F\'{\i}sica Te\'orica, 
Universidad Aut\'onoma de Madrid, 28049 Madrid, Spain}


\date{}

\maketitle

\begin{abstract}
We present an overview of different beyond mean field theories (BMFT) based on the generator coordinate method (GCM) and the recovery of symmetries used in many body nuclear physics with effective forces. In a first step a short reminder of the Hartree-Fock-Bogoliubov  (HFB) theory is given. A general discussion of the shortcomings of any mean field approximation (MFA), stemming either from the lack of the elementary symmetries (like particle number and angular momentum)  or the absence of fluctuations around the mean values, is presented.  \\
              The recovery of the symmetries spontaneously broken in the HFB approach, in particular the angular momentum, is necessary, among others, to describe excited states and transitions.  Particle number projection is also needed to guarantee the right number of protons and neutrons. Furthermore a projection before the variation prevents the pairing collapse in the weak pairing regime.  A whole chapter is devoted to illustrate with examples the convenience of recovering symmetries and the differences between the projection before and after the variation. \\
                       The lack of fluctuations around the average values of the MFA is a big shortcoming inherent to this approach. To build in correlations in BMFT one selects the relevant degrees of freedom of the atomic nucleus. In the low energy part of the spectrum these are the quadrupole, octupole and the pairing vibrations as well as the single particle degrees of freedom.  In the GCM the operators representing these degrees of freedom are used as coordinates to generate, by the constrained (Projected) HFB theory, a collective subspace. The highly correlated GCM wave function is finally written as a linear combination of a projected basis of this space. The variation of the coefficients of the linear combination leads to the Hill-Wheeler equation.\\
                   The flexibility of the GCM Ansatz allows to describe a whole palette of physical situations by conveniently choosing the generator coordinates. We discuss the classical $\beta$ and $\gamma$ 
vibrations by considering the quadrupole operators as coordinates. We present pairing fluctuations by considering the pairing gaps as generator coordinates. The combination of quadrupole and pairing fluctuations mirrors the elementary modes of excitation of the atomic nucleus and provides a realistic  
description of it. Lastly the explicit consideration of the time reversal symmetry breaking (TRSB) in the HFB wave function by the cranking procedure allows the alignment of nucleon pairs opening a new dimension in the BMFT calculations. Abundant calculations with the finite range density dependent Gogny force applied to exotic nuclei illustrate the state-of-the-art of beyond mean field theories with nuclear density functionals. We conclude with a thorough discussion on the potential poles of the theory.

\end{abstract}


\section{Introduction}
\label{Sect:Intro}

The atomic nucleus is a many body system described by a two- and many-body interaction. 
The difficulties caused by the handling of hundred of nucleons and the complexity  of the nuclear
interaction have motivated nuclear models aimed to describe specific aspects of the 
nucleus \cite{BM.88}. Among the most ambitious models the mean field approach with density dependent forces is one of the most  successful. The reason lies  in the microscopic character of the theoretical approach and in the interaction.  With its density dependence the latter effectively represents a many body interaction able to describe global properties of the nucleus.
In the past this approach has been mainly used to describe {\it ground  state} properties such us  binding
energies, mean square radii, deformations, etc. Only recently, when angular momentum (AM) projection and other improvements of the MFA became feasible, the spectroscopy of the nuclei with density functionals started. The different improvements of the MFA gave rise to several theories generically known as Beyond Mean Field Theories  which we will describe in detail below.

In the HFB approach the quasiparticles move independently in the commonly produced mean field potential. To include   additional correlations between the quasiparticles one has to go  beyond the mean field approach.  That means one has to improve the wave function (w.f.) to deal with the residual interactions. This can be done in several ways: \`a la shell model, considering multi-particle multi-hole excitations \cite{RMP_77_427_2005,PPNP_47_319_2001,ref_HARA_SUN}; in the different variations of the Random Phase Approximation \cite{SKy_RPA, GOG_RPA1,GOG_RPA2,PRN.03} and alike; and by  selecting degrees of freedom, to which the energy is specially sensitive, to be used as coordinates to generate correlated wave functions. In the latter class one can consider the generator coordinate method in the Gaussian overlap approximation which provides the Bohr Hamiltonian approach (see \cite{RS.80} for a pedagogical discussion), broadly used recently            
  \cite{kumar,Liebert_Gogny,pole_PQQ,Ring_Rel}, or the GCM by itself. In this article we concentrate on the GCM.  The GCM Ansatz for the nuclear many body system is a linear combination of the wave functions generated by the different coordinates conveniently projected as to conserve the elementary symmetries of angular momentum (AM) and particle number (PN). 

Though the GCM, in principle, can provide the exact solution of the nuclear many body problem, if one blindly chooses many coordinates the CPU time explodes rather soon. In realistic cases it is therefore more practical to concentrate on the kind of physics one wants to describe. In nuclear structure physics the low energy part of the spectrum of even-even nuclei is dominated by collective states representing shape and pairing vibrations. Among these the lowest ones are the quadrupole modes ($\beta$ and $\gamma$) and
the pairing modes.  It  therefore seems that taking these three degrees of freedom a great deal of nuclear states can be properly described\footnote{In several regions of the nuclide chart one can find nuclei where other modes are also important.  For example, the octupole degree of freedom is relevant for some Ra, Th isotopes.}. This approach provides a very good qualitative description of the nuclear phenomena. In particular the experimental  spectra  are well reproduced, though in general somewhat stretched, and the collectivity of the transition probabilities are, occasionally, enhanced. The reason for these deviations has to do with the inability  of the generated wave functions to provide the right mass for the collective motion. Its origin is related to the lack of alignment in the generated wave functions. To correct this situation the cranking procedure has been incorporated to the wave function generation procedure. In particular, the consideration of the cranking frequency as an additional coordinate has recently shown that it cures the mentioned  deficiencies providing a considerable agreement with the experiment and in many cases a quantitative description.

  Of course the increase of the number of coordinates implies a considerable growth in the CPU time of the calculations.  Nevertheless not always a quantitative agreement with the experiment is needed and for this reason it  is important to emphasize one important aspect of the GCM, namely, its physical insight. 
 That means, to analyze a physical situation, for example the triaxiality of a nucleus,  one can pick 
just the operator representing the associated degree of freedom and perform the corresponding GCM 
calculation. A typical case is the shape coexistence. In this case two rather different configurations are needed, which can be generated very easily by the GCM. Since usually one mixes many different configurations in the GCM these calculations are also commonly called configuration mixing (CM) approaches or symmetry conserving configuration mixing (SCCM) methods. The GCM approach is very well suited for the description of nuclei with a rather soft energy surface.
 There has been a parallel and almost simultaneous development of beyond mean field theories with density dependent forces, namely Skyrme, Gogny and relativistic.  The three theories have their pros and cons, but for simple convenience reasons we will illustrate the different approaches with results obtained with the Gogny force. An additional bonus of  these interactions is that they  have been fitted to describe bulk properties of the atomic nuclei along  the nuclide chart. The global character of the interaction provide a high degree of predictability absent in other descriptions like the shell model calculations. In this paper we review the mentioned theories starting with the simplest calculations and increasing  
gradually the degree of complication up to the  the most general situations.  Special emphasis is put on the physical concepts and on clarity  with  examples taken from real calculations. 
We have devoted special care to the discussion of some  tricky issues like how to deal with  the exchange terms of the interaction or the handling  of the density dependent term to avoid the appearance of divergences in the calculations. Special attention has also been given to the relevance of particle number conservation  in beyond mean field calculations.  Two Appendixes have been dedicated to consider all these points.

In Sect.~\ref{Sect:TheApp} we present the theoretical approaches. We start with the plain mean field approach in Subsect.~\ref{Sect:TheMFA}. In Subsect.~\ref{Sect:TheSCMFA} we formulate the symmetry conserving mean field approach.  The next point concerns the theory of the configuration mixing approaches and it is discussed in Subsect.~\ref{Sect:TheCMA}.   In Subsect.~\ref{Sect:TheCMAGC} we discuss the relevant coordinates. The next Sections are devoted to the axially symmetric configuration mixing calculations. After a short introduction in Sect.~\ref{Sect:ASCMC} we discuss the case of the $\beta$ coordinate using as example the Ti isotopes. The next case concerns the pairing degree of freedom which is discussed together with the $\beta$ vibrations in Subsect.~\ref{Sect:ASCMCBPDF}.  The next step in complication is represented by the triaxial calculations which involve three-dimensional angular momentum projection and is introduced in Sect.~\ref{Sect:TC}. The example of the $\beta$ and $\gamma$  coordinates is discussed in  Subsect.~\ref{Sect:TCBGC} for the nucleus $^{24}$Mg. Finally, the case of
$\beta, \gamma$ and the cranking frequency $\omega$ as coordinates is  presented in Subsect.~\ref{Sect:TCBGOC} where the Titanium isotopes and the nucleus $^{42}$Si are discussed.  The conclusions are presented in Sect.~\ref{Sect:CONC}.  We finish with three  Appendices, in A a list of the acronyms used in the text is provided, in B the peculiarities of the GCM formulation for 
density dependent forces is presented  and in C a thorough discussion on the necessity of particle number projection is done.
\section{Theoretical approaches}
\label{Sect:TheApp}
In this Section we describe the different theoretical approaches. We start with the
plain mean field approach and proceed through the forthcoming 
sections increasing the degree of sophistication of the theories.
\subsection{Mean Field Approach}
\label{Sect:TheMFA}
The mean field approximation  is the simplest approach one can imagine to describe a many body
system, namely, its wave function is a product of quasiparticles (or particles).   The most general MFA is the Hartree-Fock-Boboliubov (HFB) based on the use of the most general linear transformation. The reason for the success of the MFA lies in the variational principle used to determine these quasiparticles and in the spontaneous symmetry breaking phenomenon.  
    In the HFB theory \cite{RS.80} the quasiparticle operators are defined by  the general Bogoliubov transformations
\begin{equation}  \label{bogtrans}
\alpha _l =\sum_kU_{kl }^{*}c_k+V_{kl }^{*}c_k^{\dagger},
\end{equation}
with ${c_{k}^{\dagger},c_{k}}$ the particle creation and annihilation operators in the original basis, for example in  the Harmonic Oscillator one. $U$ and $V$ are the Bogoliubov matrices  to be determined by the Ritz variational principle.  Since the Bogoliubov transformation mix creator and annihilator operators the HFB wave function 
$(|\phi \rangle =\prod_k \alpha_k |-\rangle$)  is not an eigenstate of the particle number operator. 
 Furthermore if  the  index $k$, in Eq.(\ref{bogtrans}), is allowed to run indiscriminately over all
states of the basis  all symmetries of the system such as parity, angular momentum etc.  are broken.
The resulting HFB wave function is, obviously, the {\it most general} product wave function that can be obtained within the given configuration space. The price is high, none of the symmetries of the system is conserved.  This is a real
problem since the nuclei have a fixed number of protons and neutrons and the nuclear states are characterized 
by the quantum numbers of  parity and angular momentum. For the moment we will incorporate these
quantum numbers in the MFA in a semiclassical way and later on we will be more rigorous.  For the particle number case and for systems with very large particle number this is not a real problem since for a HFB state the relative fluctuation  $\Delta \hat{N}/N \sim 1/\sqrt{N}$ \cite{RS.80}, with $(\Delta N)^2= \langle \hat{N}^{2}\rangle -N^{2}$. For nuclei or any mesoscopic  system it is a severe problem and one should keep the right number of particles at least on the average in the minimization process, i.e.,
\begin{equation}
 \delta {E^{\prime}}[\phi\{U,V\}]   = 0  \label{min_E1},
\end{equation}
with 
\begin{equation}
{E^{\prime}}= \langle\phi|\hat{H}|\phi \rangle
 - \lambda_N \langle \phi |\hat{N} | \phi \rangle, \label{E_Lagr}
  \end{equation}
 the Lagrange multiplier $\lambda_N$  being determined by the constraint 
 \begin{equation}
 \langle \phi |\hat{N} | \phi \rangle =N.
 \end{equation}
 $\hat{N}$  is the particle number operator and $N$ the number of particle of the system.
  As with the particle number, in the case
of continuous symmetries one can add additional Lagrange multipliers to Eq.~(\ref{E_Lagr}) as to accomplish that all  quantum numbers are at least satisfied on the average. In the case of the angular momentum this is the well known {\em cranking} approximation. In this case the functional to be minimized is
\begin{equation}
{E^{\prime}}= \langle\phi|\hat{H}|\phi \rangle  - \lambda_N \langle \phi |\hat{N}| \phi \rangle - \omega
\langle \phi |\hat{J}_x | \phi \rangle, \label{E_Lagr_ome}
  \end{equation}
 the Lagrange multiplier $\omega$  being determined by the constraint 
 \begin{equation}
 \langle \phi |\hat{J}_x | \phi \rangle =\sqrt{I(I+1)}. \label{cra_cond}
 \end{equation}
The incorporation of a constraint on the angular momentum  in the variational principle implies 
 the time reversal symmetry breaking (TRSB) of the wave function $|\phi\rangle$. The equations above are an example of constrained HFB equations. Similarly one can constrain other operators like deformation parameters or energy gaps to create wave functions {\em \`a la carte}. 

The variational principle  is state dependent, i.e., it determines the self-consistent mean field {\it only for one state}, namely $|\phi\rangle$. As long as we are only interested in the ground state this is alright and we can describe properly many  properties for different nuclei as it was done in the past.  
The HFB theory with density dependent interactions like Skyrme \cite{Bender_RMP_03}, Gogny \cite{GG.83} or relativistic \cite{VAL.05} has been successfully applied in the past to describe many nuclear ground state properties, super-deformed nuclei or high spin states to mention a few examples.
 
 We have mentioned above that for very large particle number the symmetry breaking wave function is a good approach, but what happens in the case of finite systems such as atomic nuclei when the Lagrange multipliers are used~?  Unfortunately the approach does not
perform that well.  For the particle number and in the strong pairing regime it works relatively well.  In the general case, since the pairing phenomenon is not very collective in atomic nuclei -only a few Cooper pairs participate-  the HFB approximation breaks down very soon. The appendix \ref{App:C} is devoted  to the discussion of this point in a detailed way.
\subsection{Symmetry Conserving Mean Field Approaches}
\label{Sect:TheSCMFA}
   Inherent to the HFB Ansatz is the lack of correlations between the quasiparticles and the absence of exact quantum numbers.  As mentioned above 
the variational principle of Eq.~(\ref{min_E1}) determines the coefficients $(U,V)$ of the Bogoliubov transformation, i.e., $|\phi\{U,V\}\rangle$,  and only $|\phi\{U,V\}\rangle$ satisfy on the average the conservation of particle number and/or angular momentum. 
That means,  excited states based on $|\phi\{U,V\}\rangle$, for example,
multi-quasiparticles  states, do not satisfy  the symmetry constraints nor the variational principle \cite{EMR.80}. Nuclear states do have a sharp number of particles and a given angular momentum and parity.
All these facts point to the necessity of recovering the symmetries and introducing correlations in order to describe nuclear states
with good quantum numbers.
The simplest theory in this direction is the so-called symmetry conserving mean field approximation (SCMFA). In this approach the (intrinsic) wave function is still a HFB wave function but the quantum numbers are singled out by means of projectors \cite{RS.80,HHR.82,ER.82_1,ref_HARA_SUN,KARL_1,KARL_2,ETY.99,FD.74}.  Thus, the wave function
\begin{equation}
|\Phi_{M}^{N,I}\rangle =  \sum_K g^{I}_K \hat{P}^I_{MK} | \Phi^{N} \rangle \equiv  \sum_K g^{I}_K \hat{P}^I_{MK} P^N | \phi \rangle,
 \label{Proj_WF}
 \end{equation}
where we have introduced $ | \Phi^{N} \rangle$ with $P^N$  and  $P^I_{MK}$  projectors on the particle number (PNP) and the angular momentum (AMP), respectively,  is an eigenstate of the particle number and the angular momentum operators. By  $P^N$ we mean  $P^NP^Z$,  $P^Z$
is omitted to simplify the notation.  The $g_K$ parameters have to be determined by the variational principle \cite{RS.80}, see below.  
The operator $\hat{P}_{MK}^{I}$ is the angular momentum
projection operator \cite{RS.80} given by
\begin{equation} \label{AMProj}
\hat{P}_{M K}^{I} =
\frac{2I+1}{8{\pi}^{2}}
\int d\Omega 
\mathcal{D}_{M K}^{I *} (\Omega)
\hat{R}(\Omega),
\end{equation}
with $\Omega$ representing the set of the three Euler angles 
$\left(\alpha, \beta, \gamma  \right)$, $\mathcal{D}_{M K}^{I} (\Omega)$ is
the well known Wigner function \cite{Varsh.88} and 
$\hat{R}(\Omega)= 
e^{-i \alpha \hat{J}_{z}} e^{-i \beta \hat{J}_{y}} e^{-i \gamma \hat{J}_{z}}$
is the rotation operator. 
The particle number operator is given by
\begin{equation}{\hat{P}}^N  =  \frac{1}{2\pi}\int_{0}^{2\pi} 
{e}^{i \varphi (\hat{N}-N)} \, d{\varphi},
\label{eq:PN}
 \end{equation}
 the variable $\varphi$ is the canonical conjugated coordinate to $\hat{N}$ in the associated gauge space.
 
The wave function of Eq.~(\ref{Proj_WF}) depends only on the matrices $U$ and $V$ of  the Bogoliubov transformation and on the
coefficients $g_K$.
The proper way to determine them is by the variational principle, i.e., by minimization of the projected energy
\begin{equation}
\delta{E^{N,I}{[U,V,g]}}= \delta \frac{ \langle\Phi^{N,I}_{M}|\hat{H}|\Phi^{N,I}_{M} \rangle}{\langle\Phi^{N,I}_{M} |\Phi^{N,I}_{M}\rangle} =0.\label{V_E_pro}
 \end{equation}
This is known as the {\em variation after projection} (VAP) approach and is the right way the Bogoliubov
matrices should be determined since only states with the right quantum numbers are considered in the variation.  Sometimes the wave function $|\Phi^{N,I}_{M}\rangle$ is determined in the {\em projection after variation} (PAV) approach.  In this case the Bogoliubov matrices are determined by minimization of the unprojected energy, Eq.~(\ref{E_Lagr}), and afterwards the projection takes place. Manifestly the PAV approach is worse than the VAP one. In the case of the PN, for example,  the worst situation occurs in the weak pairing regime where  the HFB wave function collapses to the Hartree-Fock (HF) one and the superfluid  phase is missed, see the discussion of this point in Section (\ref{Sect:ASCMCBDF}). The fact that the variational equation, Eq.~(\ref{V_E_pro}), is highly non-linear and must be solved iteratively \cite{grad}, together with the heavy CPU time consumption of a  three dimensional  angular momentum projection makes the solution of Eq.~(\ref{V_E_pro}) a very difficult task. An affordable calculation, however,  is the one where only the PNP is performed in the VAP (PN-VAP) and the AMP is performed after the variation (AM-PAV). In  this case the situation is simpler because the PN-VAP is relatively easy and in the AM-PAV only the $g_K$ coefficients have to be varied. The resulting equations to determine the $g_K$ coefficients are a particular case of the Hill-Wheeler equation and are given in Eq.~(\ref{HW_K}).
There are several implementations of the PN-VAP approach using either separable forces 
\cite{ER.82_2}, small configuration spaces  \cite{Carlo} or the most recent ones with the Gogny  \cite{AER.01P} and Skyrme functionals \cite{DSN.07,BDL.09}. 

The scope of the SCMFA is clear: one can calculate now global properties with sharp quantum numbers with considerable 
improvements for some observables, for example, one-nucleon separation energies, transition probabilities and so on. The
precedent description, however, is limited to ground states or Yrast states, $I_1$. One can calculate excited states $I_2$ within the SCMFA  using the gradient method to calculate a state orthogonal to  $I_1$ and applying again the variational principle to determine its wave function \cite{EMR.80,KARL_1,KARL_2}.  In principle one can iterate the procedure to calculate the nth  state  $I_n$ but the degree of complication increases considerably after two states.
The T\"ubingen group \cite{KARL_1,KARL_2} has developed a battery of sophisticated approaches with great success.
As mentioned the complexity of these calculations restricts  the application of these approaches to nuclei with a few active shells. Furthermore, collective states, like vibrations,  can be described very well in simpler approaches.

\subsection{Configuration Mixing Approach}
\label{Sect:TheCMA}

 In spite of its many body character and of having the right quantum numbers, the wave function  $|\Phi^{N,I}_{M}\rangle$, behaves in many ways like a product wave function, keeping,  in  some aspects, the properties of the intrinsic wave function $|\phi\rangle$.  Thus, the particle number projected wave function  $|\Phi^{N}\rangle = P^N | \phi \rangle$ has  a many
body character but its quadrupole moment, for example, is very close to the one of $ | \phi \rangle$.  In other words
the projectors restore the corresponding symmetries but leave other properties of the intrinsic wave function unchanged. 

In order to describe correlated ground and excited states not describable by a product wave function (vibrations or shape coexistence for example),  one has to go beyond mean field.
As mentioned in the Introduction a useful way of introducing correlations is to consider operators, to which the energy is specially sensitive, to be used as coordinates to generate correlated wave functions.
The operators commonly used as coordinates are those representing the most relevant nuclear degrees of freedom like the shape operators (multipole operators), pairing, and so on. We will denote these operators by $\hat{A}_i$, $i=1,2,...,M$.

In the GCM the correlated wave function is written as a linear combination of projected mean field wave functions $\phi(\vec{a})$. The latter are provided by the constrained PN-VAP equations, 
\begin{equation}\label{eq:E_TOT}
\begin{split}
\delta{E^{\prime N}[U,V]} &= \delta \left[ \frac{ \langle\Phi^{N}(\vec{a})|\hat{H}|\Phi^{N} (\vec{a})\rangle}{\langle\Phi^{N} (\vec{a})|\Phi^{N} (\vec{a})\rangle}  
- \sum_i\lambda_i \langle \phi(\vec{a}) |\hat{A_i} | \phi (\vec{a}) \rangle \right] \\
&=0, 
\end{split} 
\end{equation}
where we have introduced $\vec{a}\equiv\{a_1, a_2, a_3,...\}$ and $|\Phi^{N} (\vec{a})\rangle= \hat{P}^N | \phi (\vec{a}) \rangle $.  The Lagrange multipliers $\lambda_i$  are determined by the constraints 
 \begin{equation}
 \langle \phi(\vec{a}) |\hat{A}_i | \phi (\vec{a})\rangle =a_i, \;\; \forall i.
 \end{equation}
 Sometimes one does not project on the particle number and in this case   $\phi (\vec{a})$ is determined at the HFB level, see Sect.~\ref{Sect:TCBGC}.

The GCM wave function itself is provided by \cite{Bender_RMP_03}
\begin{equation}
\begin{split}
|\Psi^{N,I}_{M,\sigma} \rangle &=  \; \sum_{\vec{a},K} f^{N,I}_{\sigma}(\vec{a},K)  P^N P^I_{MK} \; |\phi (\vec{a}) \rangle \\
                                           &= \sum_{\vec{a},K}  f^{N,I}_{\sigma}(\vec{a},K) |IMK,N,\vec{a} \rangle \label{eq:GCM_Ansatz}
\end{split}
\end{equation}
where we have introduced $ |IMK,N,\vec{a} \rangle$.  As mentioned in Sect.~\ref{Sect:TheSCMFA}  the AMP is not performed in the variation after projection (AM-VAP) approach but  at the GCM level. Notice that by this Ansatz we are mixing states $|\phi(\vec{a}\rangle$ with different {\em deformations}  $\vec{a}$.

The weights   $f^{N,I}_{\sigma}(\vec{a},K)$ are determined by the variational principle which leads to the 
Hill-Wheeler (HW) equation \cite{Hiwhe}
\begin{equation}
\sum_{\vec{a}',K'} \, \,(\mathcal{H}^{N,I}_{\vec{a}K, \vec{a}'K'} - E^{N,I}_\sigma \mathcal{N}^{N,I}_{\vec{a}K,
\vec{a}'K'}) f^{N,I}_{\sigma}(\vec{a}',K') = 0,
\label{HW}
\end{equation} 
where $\mathcal{H}^{N,I}_{\vec{a}K,\vec{a}'K'}$ and $\mathcal{N}^{N,I}_{\vec{a}K,\vec{a}'K'}$ 
are the Hamiltonian and norm overlaps defined by 
\begin{eqnarray}
\mathcal{H}^{N,I}_{\vec{a}K, \vec{a}'K'} \! & = & \!   \langle IMK,N,\vec{a}  |H | IMK',N,\vec{a}' \rangle \label{hamove} \\
 \mathcal{N}^{N,I}_{\vec{a}K, \vec{a}'K'} \! & = & \!   \langle IMK,N,\vec{a}  | IMK',N,\vec{a}' \rangle \label{normove}.
\end{eqnarray}
We have added the subscript $\sigma $ in $f^{N,I}_\sigma(\vec{a},K)$ and $E^{N,I}_\sigma$
in Eq.~(\ref{HW}), because the diagonalization
of the Hill-Wheeler equation not only provides the ground state $(\sigma=1)$ but also the
 wave functions  $| \Psi^{N,I}_{\sigma} \rangle $ and energies $E^{N,I}_\sigma$  
of the excited states $(\sigma=2,3,...)$. The $I$ dependence in $f^{N,I}_\sigma(\vec{a},K)$ indicates that a different diagonalization
must be done for each $I$ value.
The presence of the norm matrix in Eq.~(\ref{HW}) is due to the linear dependence of the basis states $|IMK,N,\vec{a} \rangle$ of Eq.~(\ref{eq:GCM_Ansatz}) and it is solved by standard techniques \cite{RS.80,BendHFB_SK_08,TE.10}: First, the  norm matrix is diagonalised,
\begin{equation}
\sum_{\vec{a}',K'} \, \,\mathcal{N}^{N,I}_{\vec{a}K, \vec{a}'K'} u^{N,I}_{k}(\vec{a}'K')  = n^{N,I}_k  u^{N,I}_{k}(\vec{a}K),
\label{nat_bas}
\end{equation}
to provide orthogonal states.  States with eigenvalues $n^{N,I}_k$ zero or very close to zero correspond to linearly dependent states and must be eliminated. As a criterium to set a cutoff one choses states such
that $n^{N,I}_k/n^{N,I}_{max} \ge \zeta$.  In this way the orthonormal states, called natural states, are provided by
\begin{equation}
| k^{N,IM}\rangle=\sum_{\vec{a},K} =\frac{u^{N,I}_{k}(\vec{a},K)}{n^{N,I}_k} |IMK,N,\vec{a} \rangle
\label{nat_states}
\end{equation}
The diagonalization of the Hamiltonian in this basis takes the form
\begin{equation}
\sum_{k'} \langle k^{N,I}|\hat{H} |k'^{N,I} \rangle g^{N,I\sigma}_{k'} =E^{N,I}_{\sigma} g^{I\sigma}_k
\end{equation}
and provides the eigenvalues $E^{N,I}_{\sigma}$ of Eq.~(\ref{HW}) and the  eigenvectors $g_{k}^{N,I\sigma}$. The weights $f^{N,I}_{\sigma}(\vec{a},K)$ of Eq.~(\ref{eq:GCM_Ansatz}) are given by
\begin{equation}
f^{N,I}_{\sigma}(\vec{a},K) = \sum_{k} \frac{g_{k}^{N,I\sigma}}{\sqrt{n^{N,I}_k}} u^{N,I}_k(\vec{a},K). \label{f_nat}
\end{equation}

 In addition, the collective w.f.s
\begin{equation}
 p^{N,I\sigma}(\vec{a})=\sum_{K} p^{N,I\sigma}_{K}(\vec{a})=\sum_{k,K} g_{k}^{N,I\sigma} u^{N,I}_k(\vec{a},K)
 \label{coll_wf}
\end{equation} 
 are orthogonal and $|p^{N,I\sigma} (\vec{a})|^2$ can be interpreted as a probability amplitude.  We have also introduced the
 quantity $p^{N,I\sigma}_{K}(\vec{a})$.
 
The presentation above is very general and  is valid for a Hamiltonian formulation. 
 Peculiarities of this formulation associated to neglect the exchange terms of the interaction as well as with density functionals are discussed in Appendix \ref{App:B}.

Finally, the expression for the transition probability  from an initial state ${I_{i}} {{\sigma}_{i}}$ to a final state ${I_{f}} {{\sigma}_{f}}$
 is 
\begin{eqnarray} \label{BE2_GCM_AM-cojo}
\lefteqn{B(E \lambda,{I_{i}} {{\sigma}_{i}} \rightarrow {I_{f}} {{\sigma}_{f}})=} \nonumber \\
&&\frac{e^{2}}{2I_{i}+1} \sum_{M_{i} M_{f} \mu}
\left |
\langle \Psi^{N,I_{f}}_{{M_{f}},{\sigma}_{f}} \mid  
\hat{Q}_{\lambda \mu}
\mid \Psi^{N,I_{i}}_{{M_{i}},{\sigma}_{i}}
\rangle \right | ^{2} \nonumber\\
&& =  \frac{e^{2}}{2I_{i}+1}
\left |  
\sum_{\vec{a}_{i} \vec{a}_{f} }  
\langle I_{f} \sigma_{f} \vec{a}_{f} \mid \mid \hat{Q}_{\lambda} \mid \mid
I_{i} \sigma_{i}\vec{a}_{i} \rangle
\right | ^{2}
\end{eqnarray}
with the reduced matrix elements given by
\begin{eqnarray} \label{RED_QLAMBDA}
\begin{split}
&\langle I_{f}\sigma_{f} \vec{a}_{f} \mid \mid  \hat{Q}_{\lambda} \mid \mid
I_{i} \sigma_{i}\vec{a}_{i} \rangle
 = \frac{(2I_{i}+1)(2I_{f}+1)}{8 {\pi}^{2}}
(-)^{I_{i}- \lambda}   \\
& \times  \sum_{K_{i}K_{f} \nu {\mu}^{'}} (-)^{K_{f}}   
f_{\sigma_{i}}^{*I_{f}}(\vec{a}_{f},{K_{f}})
f_{\sigma_{f}}^{I_{i}}(\vec{a}_{i},{K_{i}})
\left ( \begin{array} {ccc}
        I_{i}  &   \lambda &  I_{f}  \\
        \nu   & {\mu}^{'}   & -K_{f}
        \end{array}  \right)  \\
& \times \int d \Omega \mathcal{D}_{\nu K_{i}}^{I_{i} *}(\Omega)
\langle \phi(\vec{a}_{f}) \mid 
\hat{Q}_{\lambda {\mu}^{'}}
\hat{R}(\Omega)
\mid  \phi(\vec{a}_{i})
\rangle	
\end{split}
\end{eqnarray}
In the same way  the spectroscopic quadrupole moment for the state $(I \ge 2,\sigma)$ is given by
\begin{eqnarray} \label{ESPECT_GCM_AM}
Q^{spec}(I,\sigma) &=&e  \sqrt{\frac{16 \pi}{5}}
\left ( \begin{array} {ccc}
        I & 2 &  I  \\
        I & 0 & -I 
        \end{array}  \right ) \nonumber \\
   &\times&     
\sum_{\vec{a}_{i} \vec{a}_{f} }  \langle I \sigma \vec{a}_{f} \mid \mid \hat{Q}_{2} \mid \mid
I \sigma\vec{a}_{i} \rangle.
  \end{eqnarray}
\subsubsection{The generator coordinates}
\label{Sect:TheCMAGC}

Concerning the operators ${A}_i$  to be used as coordinates in Eq.~(\ref{eq:GCM_Ansatz}),  it seems obvious that the larger the number of coordinates the better the results.  It is also obvious, however, that a  compromise must be found since the dimension of the equations rises exponentially with the number of coordinates. That means, one must choose very carefully the operators ${A}_i$ used as coordinates.
It is well known that the binding energy of an atomic nucleus depends strongly on the shape parameters, it seems then reasonable to start with these operators.  Furthermore, since the lowest collective modes of the nuclei are associated with the quadrupole vibrations, one considers first the quadrupole moments $\hat{Q}_{20}$  and $\hat{Q}_{22}$. They are related to the $(\beta,\gamma)$ deformation parameters by 
\begin{eqnarray}
\beta &=&\frac{1}{3r^{2}_{0}A^{5/3}} \sqrt{20\pi(\langle\hat{Q}_{20}\rangle^{2} + 2 \langle \hat{Q}_{22}\rangle^{2})} \label{be_ga0}\\
 \gamma &=& \arctan\left(\sqrt{2}\frac{\langle \hat{Q}_{22}\rangle}{\langle\hat{Q}_{20}\rangle}\right) \label{be_ga}
\end{eqnarray}
with $r_{0}=1.2$ fm and  $A$ the mass number.  Besides the quadrupole parameters the next relevant
degrees of freedom are the pairing correlations, octupole correlations, etc.

 The simplest approach considers one coordinate, namely the axially symmetric quadrupole operator $\hat{Q}_{20}$ or equivalently  $\beta$, that means,  $\vec{a} \equiv (\beta)$. In this case an additional simplification is provided by the fact that the AMP is one-dimensional. Further axially symmetric calculations consider additional operators like  the octupole deformation $\hat{Q}_{30}$ or $\beta_3$, in this case  $\vec{a} \equiv (\beta,\beta_3)$ \cite{Chinese_Peter}, or the gap parameter.
An interesting case is provided by a two dimensional calculation including triaxial deformation, in this case  $\vec{a} \equiv (\beta,\gamma)$, see for example Ref.~\cite{TE.10}. 

Within these approaches there are some variations at the level of determination of the HFB wave functions:  the simplest one does not project on particle number \cite{RODGUZNPA02}, others  implement an  approximate PNP by means of the Lipkin-Nogami approach  \cite{Li.60,No.64} to generate the wave function $|\phi\rangle$  projecting afterwards \cite{Bender_Kr_06,Niksic_PNAMP_06} and only in  Refs.~\cite{TE.10,Rod_CaTiCr_07} a full VAP of the PN is performed.  Concerning AMP almost all approaches ignore the AMP in the determination of the HFB w.f., AMP is performed in the PAV way. 

In GCM calculations the most often used effective interactions are of Skyrme \cite{Bender_RMP_03}, Gogny \cite {BERGNPA84} or relativistic \cite{Golla96} type.  These calculations, in general, produce rather stretched spectra, see for example Refs.~\cite{TE.10,Rod_CaTiCr_07}.  The main reason for this behavior is the absence of angular momentum dependence in the determination of the HFB w.f.. This is well known since long ago \cite{HHR.82,PRC_29_1056_1984,NPA_435_477_1985,ETY.99,PRC_76_044304_2007}.  We can understand it easily in the following way: In the first order of the Kamlah expansion \cite{Ka.68}, the intrinsic wave function $\phi(\vec{a})$,  corresponding to an AM-VAP approach to angular momentum $I$,  can be obtained   in an approximate way by 
\begin{equation}\label{E_TOT_J}
\begin{split}
\delta{E^{\prime N}{[U,V]}}&= \delta \left[ \frac{ \langle\Phi^{N}(\vec{a})|\hat{H}|\Phi^{N} (\vec{a})\rangle}{\langle\Phi^{N}(\vec{a})|\Phi^{N} (\vec{a})\rangle}  
- \sum_i\lambda_i \langle \phi(\vec{a}) |\hat{A_i} | \phi (\vec{a}) \rangle \right. \\ 
&\left.-  \omega \langle \phi(\vec{a}) |\hat{J}_x | \phi(\vec{a}) \rangle \right] =0, 
\end{split}
  \end{equation}
the parameter  $\omega$ being fixed by the condition
\begin{equation}
 \langle \phi(\vec{a}) |\hat{J}_x | \phi (\vec{a})\rangle = \sqrt{I(I+1)}.
 \end{equation}
 The solution of Eq.~(\ref{E_TOT_J}), however,  is  a time reversal symmetry breaking wave function. The consideration of such w.f. increases considerably the CPU time needed to solve the Hill-Wheeler equation, (\ref{HW}), see Sect.~\ref{Sect:TCBGOC} for more details, and with few exceptions see  \cite{BRE.15,EBR.16}, the cranking term has been generally ignored. That means that the set of wave functions $\phi(\vec{a})$ generated to solve the Hill-Wheeler equations are obtained without this term,  and they satisfy 
\begin{equation}
\langle\phi(\vec{a})|\hat{J}_x|\phi(\vec{a})\rangle =\langle\phi(\vec{a})|\hat{J}_y|\phi(\vec{a})\rangle =\langle\phi(\vec{a})|\hat{J}_z|\phi(\vec{a})\rangle =0.
\end{equation}
 But according to Kamlah the solution obtained in this way represents an approximate AM-VAP  to the case $I=0\;\hbar$.  Therefore, this variational procedure favors the case $I=0\;\hbar$,
the case $I=2 \;\hbar$ is a little less favored, $I=4 \;\hbar$ even less, and so on. The result is a stretched spectrum.  The optimal solution to this problem is well known, one should do AM-VAP instead of AM-PAV in order to get the right moment of inertia. Of course, one can also proceed according to the Kamlah expansion and determine the intrinsic wave function $\phi$ according to 
Eqs.~(\ref{E_TOT_J})  for each $I$-value, or, even better, as proposed by Peierls and Thouless \cite{PT-62}, see also \cite{Eg-83}, to take $\omega$ as an additional coordinate to be included in $\vec{a}$. This project has been recently performed in \cite{EBR.16}, where the coordinates  $(\beta,\gamma,\omega)$ were explicitly considered in the GCM ansatz.

A second reason to obtain an stretched spectrum is to use too few coordinates. According to the variational principle the non-constrained coordinates are determined self-consistently as to minimize  the
HFB (or PN-VAP) ground state energy. But the solution of the Hill-Wheeler equation  provides not only the ground state but also excited states. The excited states, however, cannot take another values of the non-constrained operator than the ones obtained to determine the ground state even if they would like. As a consequence the energy of the excited states rises. For example, if we are constraining only on  $\beta_i$ and $\gamma_i$, the constrained HFB equations determine $\phi (\beta_i,\gamma_i)$ according to Eq.~(\ref{eq:E_TOT}). That means, the values of any other dependence like, for example, the octupole and hexadecupole deformations or the pairing gaps of the wave function $\phi(\beta_i,\gamma_i)$ are self-consistently determined exclusively as to minimize that energy. Consequently the values of the weights of the ansatz Eq.~(\ref{eq:GCM_Ansatz}) are conditioned by this choice. Thus, for example, if the smallest proton gap calculated with the set of wave functions $|\phi (\beta_i,\gamma_i)\rangle$ is 0.75 MeV, none of the states $|\Psi^{N,I}_{\sigma}\rangle$ of Eq.~(\ref{eq:GCM_Ansatz}) will hardly have proton gaps smaller than this one.  This is alright for the ground state ($\sigma=1$) but not necessarily for the excited states ($\sigma > 1$). The excited states may like to have other values for the unconstrained degrees of freedom which would lower their energies but within this framework it is impossible. In general, the higher the excitation energy the larger the difference in the relevant degrees of freedom with the ground state. Consequently we expect a stretched spectrum. The solution to this problem is obviously to include further degrees of freedom as generator coordinates in the ansatz  of Eq.~(\ref{eq:GCM_Ansatz}), see Sect.~(\ref{Sect:TC}).

In the next Sections we report  on state-of-the-art calculations performed in the last years: 
In Sec.~(\ref {Sect:ASCMCBDF}) we present the $\beta$  degree of freedom. In Sect.~(\ref{Sect:ASCMCBPDF}) we report on $\beta$ and pairing fluctuations.
 In Sec.~(\ref {Sect:TCBGC}) we report on fluctuations on the $(\beta,\gamma)$ degrees of freedom and
 lastly in Sec.~(\ref{Sect:TCBGOC}) we discuss $(\beta,\gamma)$ and $\omega$ fluctuations.

\section{Axial Symmetry Configuration Mixing Calculations}
\label{Sect:ASCMC}
  In this section we illustrate the different aspects of the theory presented in the precedent sections. The aspects concerning the 
  particle number and the angular momentum projection out of HFB wave functions are quite general and are clearly independent of other aspects, like which generator coordinates are being considered in the final calculations.  Considering the impact of  the different coordinates on the final calculations, however, depends on the coordinates under consideration. For example,  the effects of  $\beta$ and $\gamma$ are independent because they represent different degrees of freedom.
  As one increases the number of coordinates, however, some effects can be larger or smaller depending on the coordinates under
  consideration. For instance the impact of adding pairing fluctuations  on the final spectrum will be larger or smaller depending whether we add them to a set  $(\beta,\gamma)$ or to  $(\beta,\gamma,\hbar \omega)$.
  
\begin{figure}[tb]
\begin{center}
\includegraphics[angle=0, scale=.4]{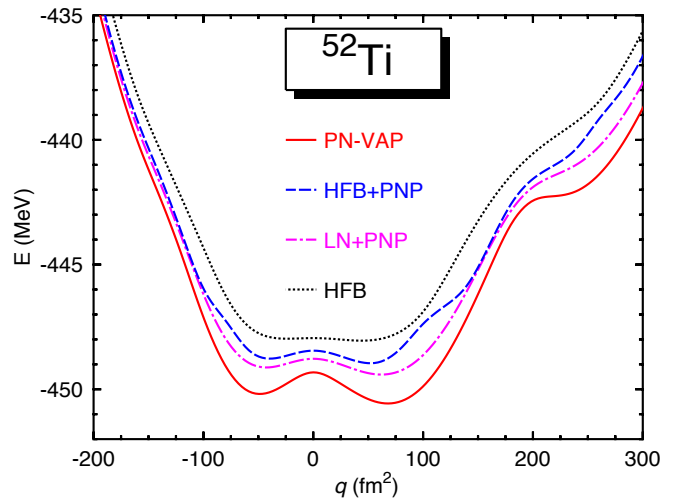}
\end{center}
\caption{(Color Online) Potential energy surfaces as a function of the quadrupole moment in several PNP approaches. The HFB energy is provided as a reference.}
\label{fig:1D_N}
\end{figure}
\begin{figure}[tb]
\begin{center}
\includegraphics[angle=0, scale=.4]{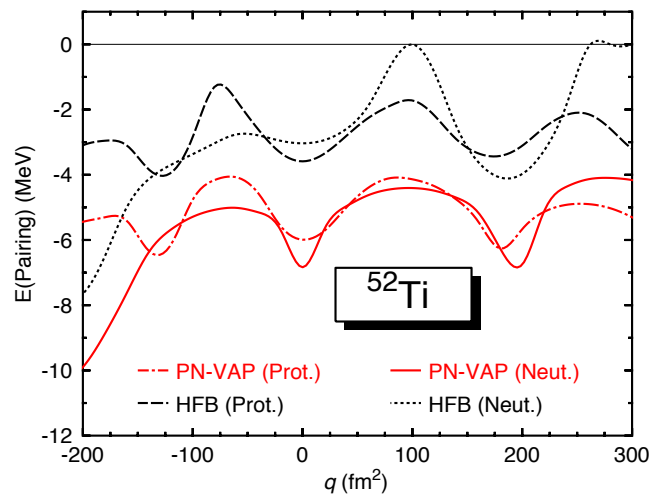}
\end{center}
\caption{(Color Online) Pairing Energies for protons and neutrons in the HFB and the PN-VAP approaches as a function
of the quadrupole moment.}
\label{fig:pai_Ti}
\end{figure}

\subsection{ The $\beta$ degree of freedom}
\label{Sect:ASCMCBDF}

Most of the currently used beyond mean field calculations with effective forces include both symmetry restoration, i.e. particle number and angular momentum projection (PNAMP), and configuration mixing along the {\bf axial} quadrupole deformation \cite{Bender_RMP_03,Rod_CaTiCr_07,Niksic_PNAMP_06}. This approach (axial GCM-PNAMP) has been successfully applied to study many phenomena like, for example, the appearance or degradation of shell closures in neutron rich nuclei \cite{RODGUZNPA02,Rod_Mg_09,Rod_CaTiCr_07,Jung_PLB_08}, shape coexistence in proton rich Kr \cite{Bender_Kr_06} or Pb \cite{Dug_Pb_03,Ray_Pb_04} isotopes or shape transitions in the  $A\sim150$ region \cite{Niksic_PRL_07,Rod_PLB_08}. However, the intrinsic wave functions used there were restricted to have axial symmetry, with $K=0$, because this assumption simplifies considerably the angular momentum projection and decreases significantly the computational burden. This restriction is one of the major drawbacks in the method because it reduces its applicability to systems where the triaxiality does not play an important role.

The simplest calculation consists in taking only one parameter, namely,  $\hat{Q}_{20}$, or $\beta$. In this
case the angular momentum projection is reduced considerably because only $K=0$ components need to be considered and the projection operation $P^I_{MK}$ becomes  $P^{I}_{00}\equiv P^{I}$. In this case the three-dimensional integral implied by a triaxial shape reduces to  a one-dimensional one.

The GCM Ansatz of Eq.~(\ref{eq:GCM_Ansatz}) in our case looks like
\begin{equation}
|\Psi^{N,I}_{\sigma} \rangle = \; \sum_{q} f^{I}_{\sigma}(q)  P^N P^I \; |\phi (q) \rangle
\label{eq:GCM_Ansa_beta}
\end{equation}
with the w.f.s $|\phi (q) \rangle$ being determined by the minimization of the functional
\begin{equation}
{E^{\prime}}^{}= \frac{ \langle\Phi^{}(q)|\hat{H}|\Phi{} (q)\rangle}{\langle\Phi^{}(q)|\Phi^{} (q)\rangle}  - \lambda_{q} \langle \phi |\hat{Q}_{20} | \phi \rangle, \label{E_Lagr_bet}
  \end{equation}
and the Lagrange multiplier $\lambda_{q}$  being determined by the constraint 
 \begin{equation}
 \langle \phi |\hat{Q}_{20} | \phi \rangle =q. 
 \end{equation}
Once we have determined the w.f. $|\phi(q)\rangle$  we can calculate the PNAMP projected energy 
\begin{equation}
 E^{N,I}(q)=  \frac{\langle\Phi^{}(q)|\hat{H}P^IP^N|\Phi^{} (q)\rangle} {\langle\Phi^{}(q)|P^I P^N |\Phi^{} (q)\rangle}.
\label{eq:pnamp}
\end{equation}
In this subsection the wave function $|\Phi\rangle$ can be either $| \Phi \rangle =| \phi\rangle$, in which case we are solving the
plain HFB equations\footnote{ In this case
we have to add to Eq.~(\ref{E_prime})  a term $-\lambda_{N} \hat{N}$ to keep the particle number right on the average, with $\lambda_{N}$ fixed by the constraint $\langle\phi | \hat{N}| \phi\rangle=N $.}, or $| \Phi \rangle \equiv | \Phi^{N} \rangle= P^{N} | \phi\rangle$ in which case we are solving the PNVAP equations.
To illustrate the one-dimensional GCM with the quadrupole moment as a coordinate we have performed calculations
for the Titanium isotopes.  We consider a configuration space of 8 oscillator shells and an interval $-240\;{\rm fm}^{2}\leq q \leq 400$ fm$^2$
with a step size of $\Delta q= 20$ fm$^2$. 

As discussed in Sec.~(\ref{Sect:TheSCMFA}) there are two ways to determine the intrinsic wave function $| \phi \rangle$ in Symmetry Conserving Mean Field Approaches. In the PAV
one minimizes the HFB energy and then performs the projection and in the VAP the projected energy is minimized. Since we are not able to perform AM-VAP, we can have two intrinsic
w.f., the plain HFB and the PN-VAP one. To analyze the impact of using different w.f.s we have performed axially symmetric $\hat{Q}_{20}$-constrained calculations for the nucleus $^{52}$Ti. 
 According to our two intrinsic wave functions we can calculate expectation values in the first case in the approaches HFB+PNP (only PNP), HFB+AMP (only AMP) and HFB+PNAMP (PN and AMP simultaneously) and in the second one in the PN-VAP and PN-VAP+PNAMP in an obvious notation.    The projected energies  $E^{N}(q)=  \langle\Phi^{}(q)|\hat{H}P^N|\Phi^{} (q)\rangle/\langle\Phi^{}(q)| P^N |\Phi^{} (q)\rangle$ or 
  $E^{N,I}(q)=  \langle\Phi^{}(q)|\hat{H}P^IP^N|\Phi^{} (q)\rangle/\langle\Phi^{}(q)|P^I P^N |\Phi^{} (q)\rangle$
 as a function of $q$ supply the potential energy surfaces which
provide useful information on the impact of the projections on the energy and a first impression of the relevance of the 
$\beta$ fluctuations. To study the effect of the PNP we present in  Fig.~\ref{fig:1D_N} the potential energy of the nucleus $^{52}$Ti in several approximations as a function of the quadrupole moment.  The dotted line corresponds to the HFB approximation.  For small deformations we observe a flat behavior which increases steeply for $|q| \geq 80$ fm$^2$, and around 220 fm$^2$ we find a shoulder. The dashed line denoted HFB+PNP, corresponds to the PN projected energy calculated with the w.f. determined in the previous HFB case. As compared with the HFB curve we observe a energy lowering of around 1 MeV, a less flat behavior for small $q$-values and a wrinkle around 100 fm$^2$.   The  continuous line corresponds to the PN-VAP approach. Here the energy gain is about 2 to 2.5 MeV.   There are now two clear minima at $q  \approx  \pm  80$ fm$^2$, the shoulder at large deformation is now more pronounced and there is no wrinkle. 
The observable most sensitive to the PNP is the pairing content of both w.f.'s.
In Fig.~\ref{fig:pai_Ti} the  pairing energies of the intrinsic wave functions in the HFB and PN-VAP approaches are displayed.  We
 find an oscillatory behavior as a function of the deformation. In the HFB approach we observe a collapse of the neutron pairing correlations at the $q$-values corresponding to the prolate minimum and for very large deformations, see below, and a weakening of the proton pairing energy at the oblate side. In the PN-VAP approach we do not observe any collapse and we obtain larger absolute values. As one can see in this figure the origin
of the wrinkle in the HFB energy is due to the pairing collapse of the wave function. Though we  will not discuss the Lipking-Nogami (LN) approach in this work \cite{Li.60,No.64}, we also present the particle number projected LN energy \cite{VER.00} 
in this figure.  As one can observe it is rather close to the PHF+PNP energy but it has the advantage of not presenting a pairing collapse.

\begin{figure}[tb]
\begin{center}
\includegraphics[angle=0, scale=.4]{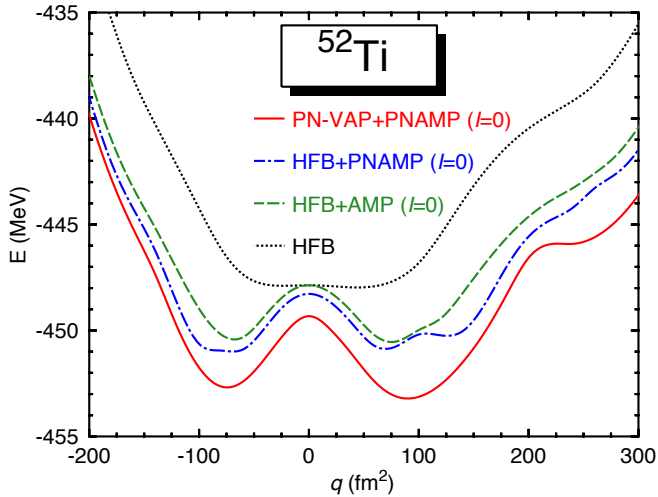}
\end{center}
\caption{(Color Online) Potential energy surfaces as a function of the quadrupole moment in several AMP approaches. The HFB energy is provided as a reference.}
\label{fig:AMP_Ti}
\end{figure}

We now analyze the effect of the AMP on the energy surface. In Fig.~\ref{fig:AMP_Ti}  we present the potential energy curves for the nucleus $^{52}$Ti for $I=0 \hbar$ versus the quadrupole moment.  As a reference we also show the HFB energy.  Compared with the HFB curve,  in the HFB+AMP we observe, in general, an energy lowering of about 3 MeV. Smaller values are found for small deformations and in particular for the 
spherical nucleus at  $q=0$ there is, obviously, no energy correction. The HFB+PNAMP provides an additional energy decrease of 1 MeV and a broadening of the potential energy surface.  It is remarkable that in the calculations with the HFB w.f. we now observe a flattening of the potential around the oblate minimum caused by the weak pairing regime of this w.f. at this deformation.
 In the PN-VAP+PNAMP approach the  nucleus $^{52}$Ti presents well developed coexistent prolate and oblate minima and a prolate  super-deformed shoulder. As compared with the HFB energy one obtains a strong energy lowering, about 5 MeV, and interestingly there are no wrinkles.  As we have seen before in Fig.~\ref {fig:pai_Ti} the pairing energies present an oscillatory behavior as a function of the deformation. The weak pairing regions are associated with a low level density and the strong pairing ones with high level density. Like in the Nilsson model in the constrained HFB equations one goes through sub-shell closures that provide the low level density regions.  In the PN-VAP case  the oscillations are also present but in spite of it the pairing correlations are always large.   
   
\begin{figure}[tb]
\begin{center}
\includegraphics[angle=0, scale=.6]{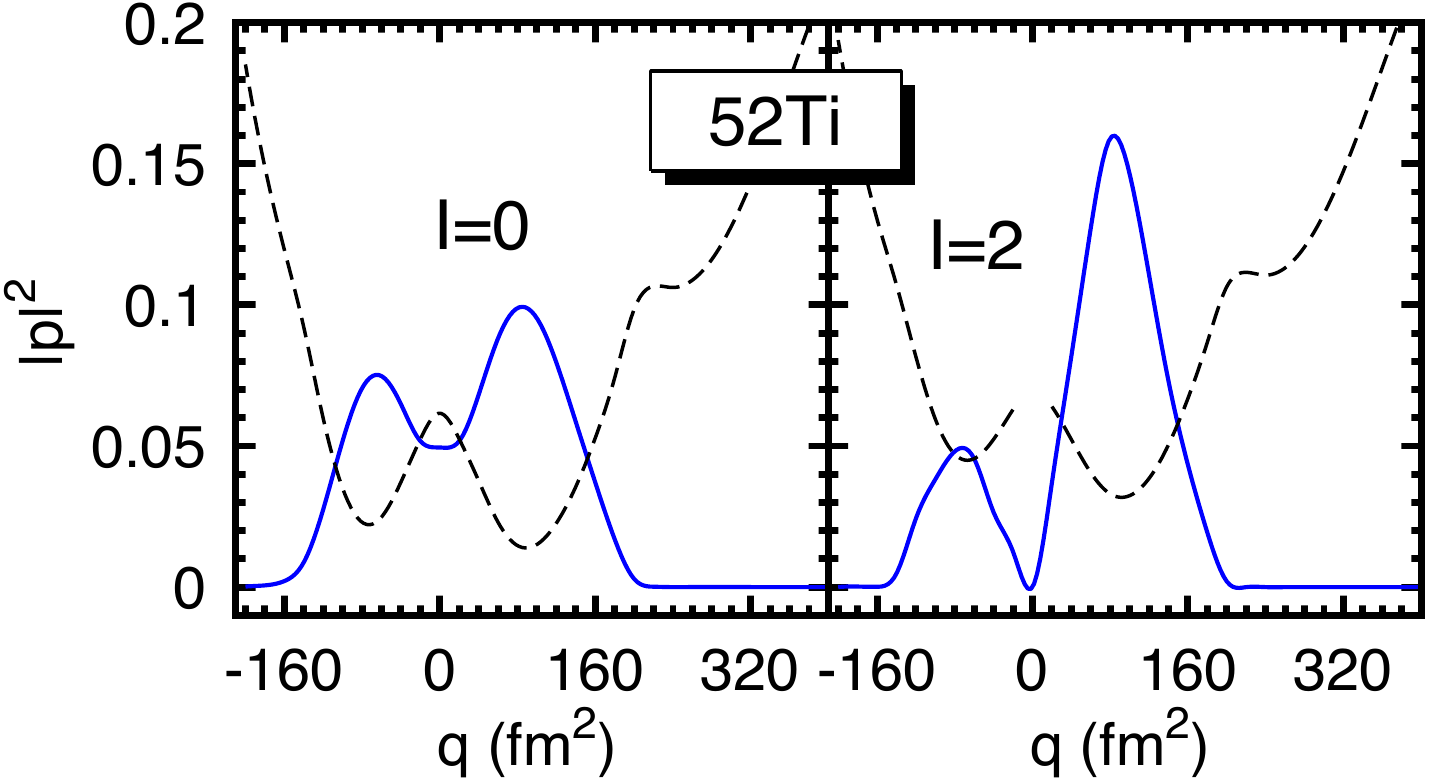}
\end{center}
\caption{(Color Online) Wave functions of the $0^+_1$  and $2^{+}_{1}$ states for $^{52}$Ti, continuous lines.  The corresponding potential energy surfaces, dashed lines, are also depicted.}
\label{fig:wf_Ti_0_2}
\end{figure}

Once we learned about the effect of recovery of symmetries we now turn to the GCM approach. 
The solution of the Hill-Wheeler equation, Eq.~(\ref{HW}), provides the energy eigenvalues $E^{N,I}_{\sigma}$ and the eigenvectors $f^{I}_{\sigma}(q)$ which will allow us to calculate the collective wave functions  of Eq.~(\ref{coll_wf}).  The aspects associated with the convergence of the solution of the Hill-Wheeler equation are discussed in Sect.~\ref{Sect:TCBGC}.
 In Fig.~\ref{fig:wf_Ti_0_2} we display the wave functions of the $0^+_1$ and $2^+_1$ states of $^{52}$Ti in the PN-VAP+PNAMP approach. For clarity we have also included the potential energy surfaces (PES)  of these states. The w.f. of the $0^+_1$  state presents a maximum at the prolate deformation and another oblate one according to the shape of the PES.  The $2^+_1$ state shows a similar distribution but in this case the prolate peak is much
larger. 

\begin{figure}[tb]
\begin{center}
\includegraphics[angle=0, scale=.4]{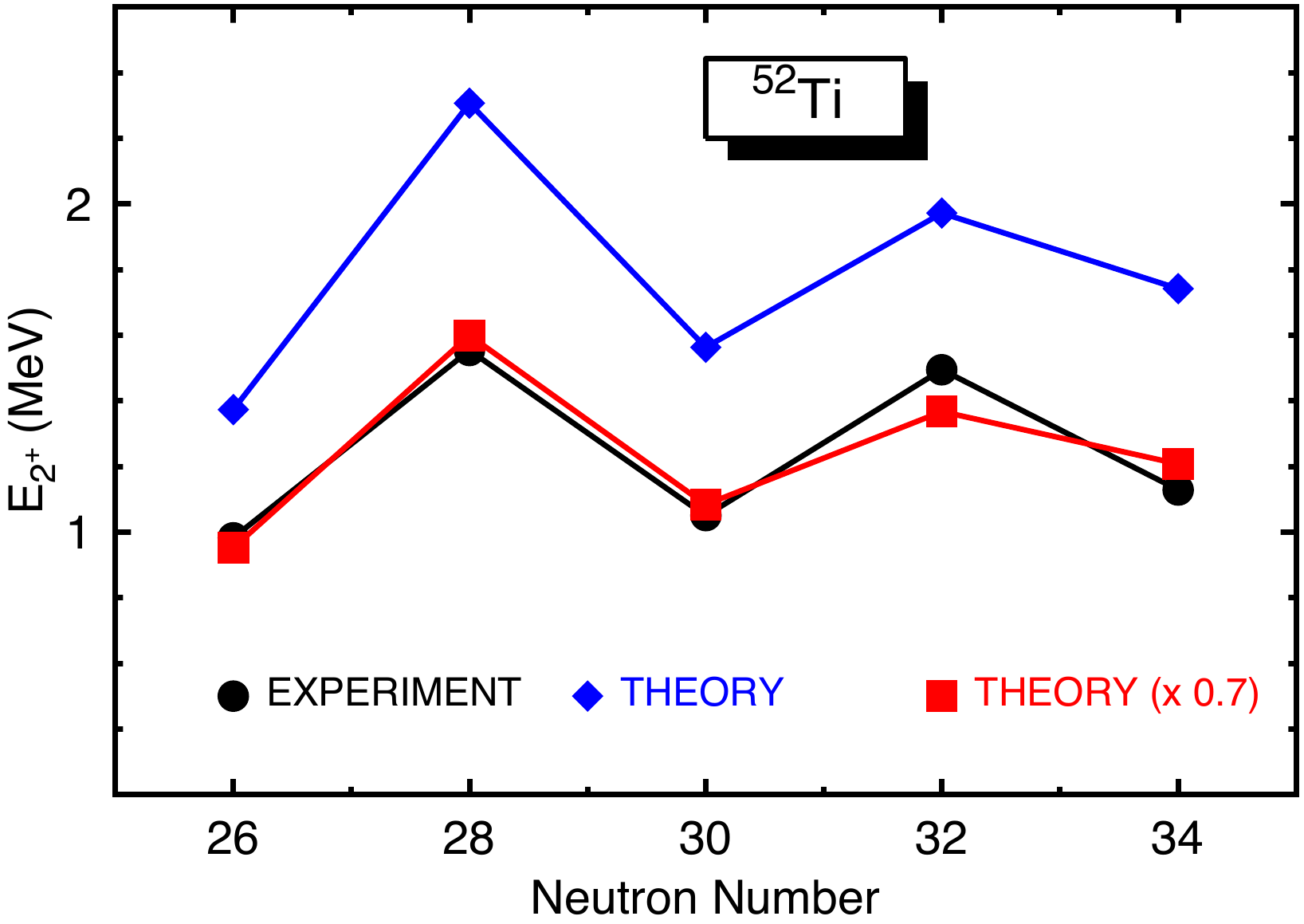}
\includegraphics[angle=0, scale=.4]{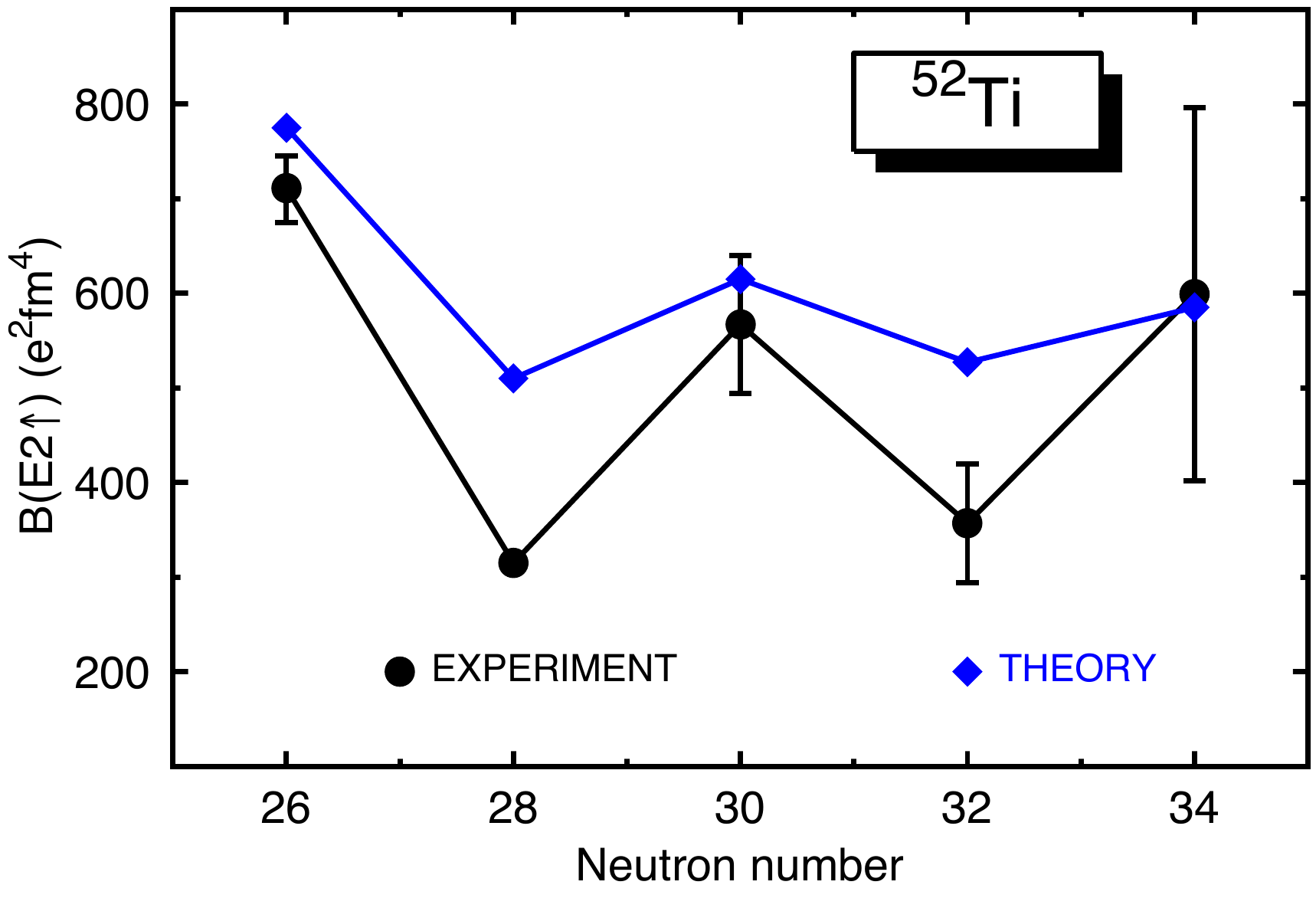}
\end{center}
\caption{(Color Online) Top: Excitation energies of the $2^{+}_{1}$ states for the Titanium isotopes. The experimental results are taken from Refs.~\cite{exp1,exp2,exp3,exp4}. Bottom: $B(E2, 0^{+}_{1} \longrightarrow 2^{+}_{1}$) transition probabilities for the Titanium isotopes. The experimental values are from Ref.~\cite{exp5}.}
\label{fig:E2_AX_TI}
\end{figure}

In the top panel of Fig.~\ref{fig:E2_AX_TI} we display the excitation energies of the $2^{+}_{1}$ states for the Titanium isotopes. A first glance reveals that the theoretical energies behave like the experimental ones but shifted to larger energies.  This illustrates a typical behavior of many 
 GCM calculations \cite{Rod_CaTiCr_07,RE.11, LYV.11,Bender_Kr_06}, namely a stretched spectrum. If
 we multiply the excitation energies by 0.7 we obtain a very good agreement with the experiment. As we will see in Sect.~(\ref{Sect:TCBGOC}) the consideration of the angular frequency as a generator coordinate corrects for this shift.

In the bottom panel of Fig.~\ref{fig:E2_AX_TI} we now plot the $E2$ transition probabilities from the $0^+_1$ to the $2^+_1$ states. 
  Our results present a general trend similar to the experimental data for the isotopic chain. 
  The B(E2) values are correlated with the relative energy of the $2^{+}$ level: the higher the energy is, the smaller is the B(E2). In general, our values are somewhat larger than the experimental ones. Though, it might be a general tendency of the Gogny force (and the Skyrme force, too) to provide larger quadrupole moments than experimentally observed \cite{Rod_CaTiCr_07,Bender_Kr_06}, the consideration of aligned states in the GCM basis states reduces considerably the B(E2) values, see Sect.~\ref{Sect:TCBGOC}. It is interesting to notice that our B(E2) values qualitatively reproduce the experimental zigzag behavior in the Ti isotopes without any need to invoke effective charges.  
\subsection{The $\beta$ and pairing degrees of freedom}
\label{Sect:ASCMCBPDF}
 With increasing number of coordinates the calculations become heavier making the consideration of more coordinates difficult. Only recently in Ref.~\cite{Nuria,Nuria2}   the effect of including fluctuations of the pairing gap in the GCM Ansatz in realistic calculations has been investigated. In this Subsection we mainly extend the study of the previous case of $^{52}$Ti to the case of
 two coordinates, namely the $\beta$ and the pairing degrees of freedom.  In large part this Section is taken from Ref.~\cite{Nuria2}.

Pairing correlations play an important role at the HFB level and in the same way pairing fluctuations are relevant at the SCCM calculations.  For example, if we do not allow pairing fluctuations, the resulting states of the Hill-Wheeler diagonalization cannot have gap values different than those of the basis states. For the ground state this is alright but not for the excited states that would like to have a different one.  However, if fluctuations are allowed each state can take the energetically most convenient value. 
Calculations without fluctuations in general produce stretched spectra. The consideration of pairing fluctuations will furthermore allow the study of pairing vibrations and their coupling with shape vibrations.

There are two collective degrees of freedom associated to pairing. First,  the pairing gap $\Delta$, which measures the amount of  pairing correlations, i.e., the ``deformation'' \cite{Brog_rev} in the associated gauge space. 
Second the angle $\varphi$ which indicates  the orientation of the HFB state in this space. The HFB equations  determines  the w.f. and thereby $\Delta$   while the gauge angle $\varphi$ does not play any role at the mean field level.  The degree of freedom associated  to $\varphi$  has been exploited in the past  \cite{DMP64}:  linear combinations of w.f.s with different orientation in the gauge space  provide a number conserving wave function.  Pairing vibrations, associated with w.f.s with different pairing gaps, around the average gap parameter $\Delta_0$ of the energy minimum, on the other hand, have attracted little attention. As a matter of fact they have been considered only either with very schematic interactions in the framework of the collective Hamiltonian \cite{Bes_coll,Pomo_85}, in  microscopic model calculations \cite{Mafer03,Mafer05}, in reduced configuration spaces \cite{Faess_73} or in earlier BMFT approaches \cite{Meyer,Heenen}.

 With schematic pairing interactions the energy gap $\Delta$ provides a direct measure of pairing correlations in the BCS or HFB approach.  However, to quantify the pairing content of a w.f. with a finite range interaction like the Gogny force, that provides
state dependent gaps,  is not trivial.  A quantity that supplies a measure of the pairing correlations and is easy to handle  is the  mean square deviation of the particle number operator  $(\Delta\hat{N})^2$. This quantity is zero in the absence of pairing correlations and is large for strongly correlated systems. Furthermore, since for a schematic pairing interaction $\langle(\Delta \hat{N})^2\rangle= 4 \sum_{k>0} u_k^2 v_k^2= \Delta^2\sum_{k>0} \frac{1}{E_k^2}$, with $E_k$
the quasiparticle energy,   $ \Delta \propto \langle(\Delta \hat{N})^2\rangle^{1/2}$ and  $(\Delta \hat{N})^2$ provides an indication of the pairing content of the wave function. In the following we will
denote  $\delta = \langle\phi|(\Delta \hat{N})^2|\phi\rangle^{1/2}$ and use it as coordinate to generate wave functions with different pairing correlations  (see the discussion of Fig.~\ref{fig:pair_ener_q} in Appendix C.

In principle the calculations should be 3D with coordinates $(q, \delta_Z, \delta_N)$ with separate constraints for neutrons and protons:
\begin{equation}
\langle \phi|(\Delta\hat{N})^2|\phi \rangle^{1/2} = \delta_N, \;\;\;\; \;\;\;\;      \langle \phi|(\Delta\hat{Z})^2|\phi \rangle^{1/2} = \delta_Z.
\end{equation}
Unfortunately with three constraints  the problem becomes computationally very demanding.  What we have done is to substitute
the above constraints by a single one on $\delta$, the Lagrange multiplier $\delta$ being defined by:
\begin{equation}
\langle \phi|(\Delta\hat{N})^2|\phi \rangle^{1/2}  +   \langle \phi|(\Delta\hat{Z})^2|\phi \rangle^{1/2} = \delta,
\end{equation}
This approximation has been checked in Ref.~\cite{Nuria2} and was found to be rather good. 

\begin{figure}[tb]
\begin{center}
\includegraphics[angle=0, width=\columnwidth]{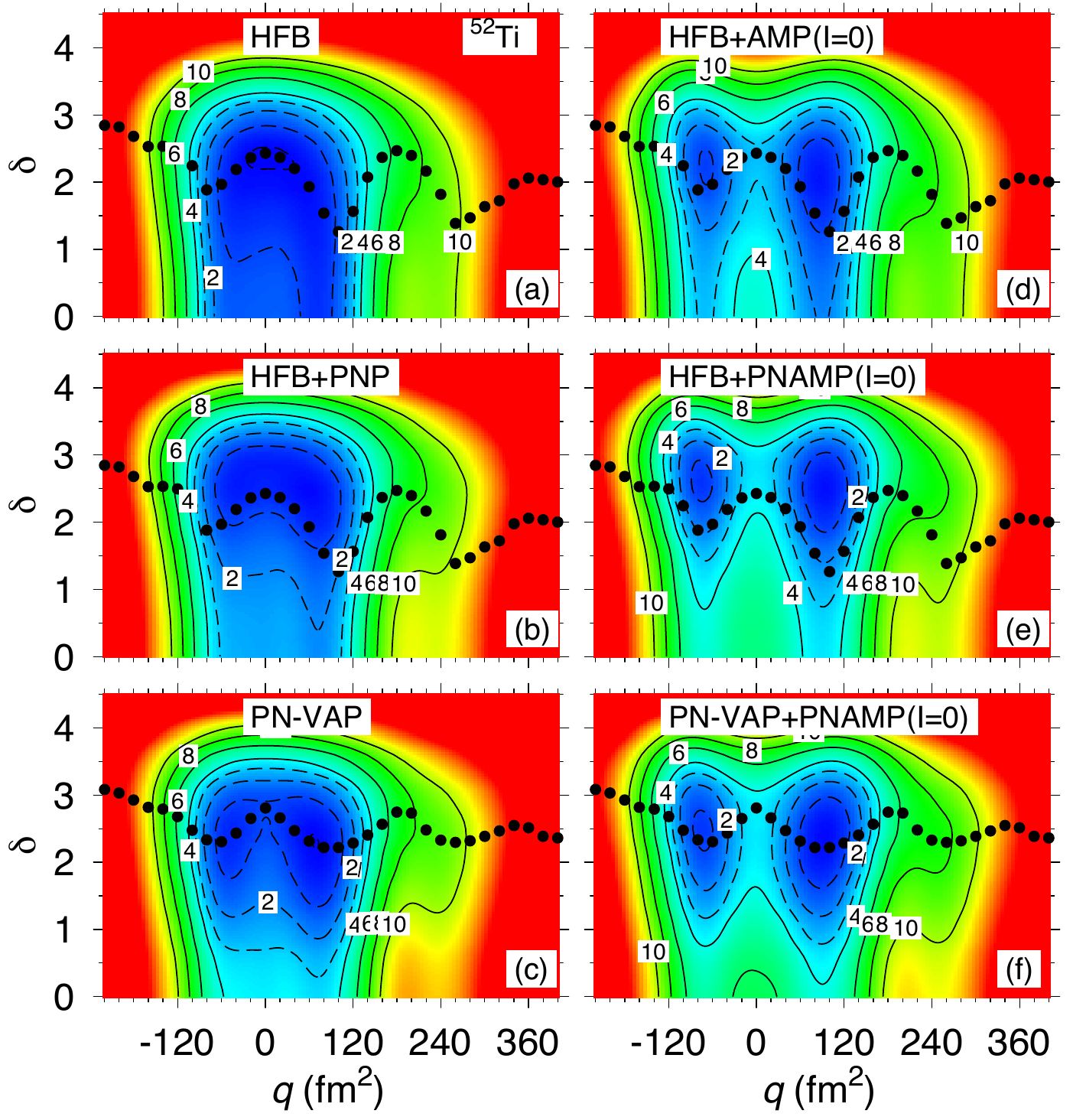}
\end{center}
\caption{(Color online) Potential energy contour plots for $^{52}$Ti in the $(\delta,q)$ plane in different approaches. As dashed lines, equipotential lines from 0 to 3 MeV in step of 1 MeV. As continuous lines, contours from 4 to 10 MeV in steps of 2 MeV. In each panel the energy origin has been chosen independently  and the energy minimum  has been set to zero. The bullets in each panel represent the  $\delta$ values of the self-consistent solution (HFB or PN-VAP) extracted from the 1D ($q$-constrained) approach and are displayed as a discussion guide. Since all HFB based approaches do have the same intrinsic w.f. all of them have the same bullet pattern. The same applies to all PN- VAP based approaches. This figure has been taken from Ref.~\cite{Nuria2}.}
\label{fig:pairfluc_Ti52}
\end{figure}

\begin{figure}[tb]
\begin{center}
\includegraphics[angle=0, width=\columnwidth]{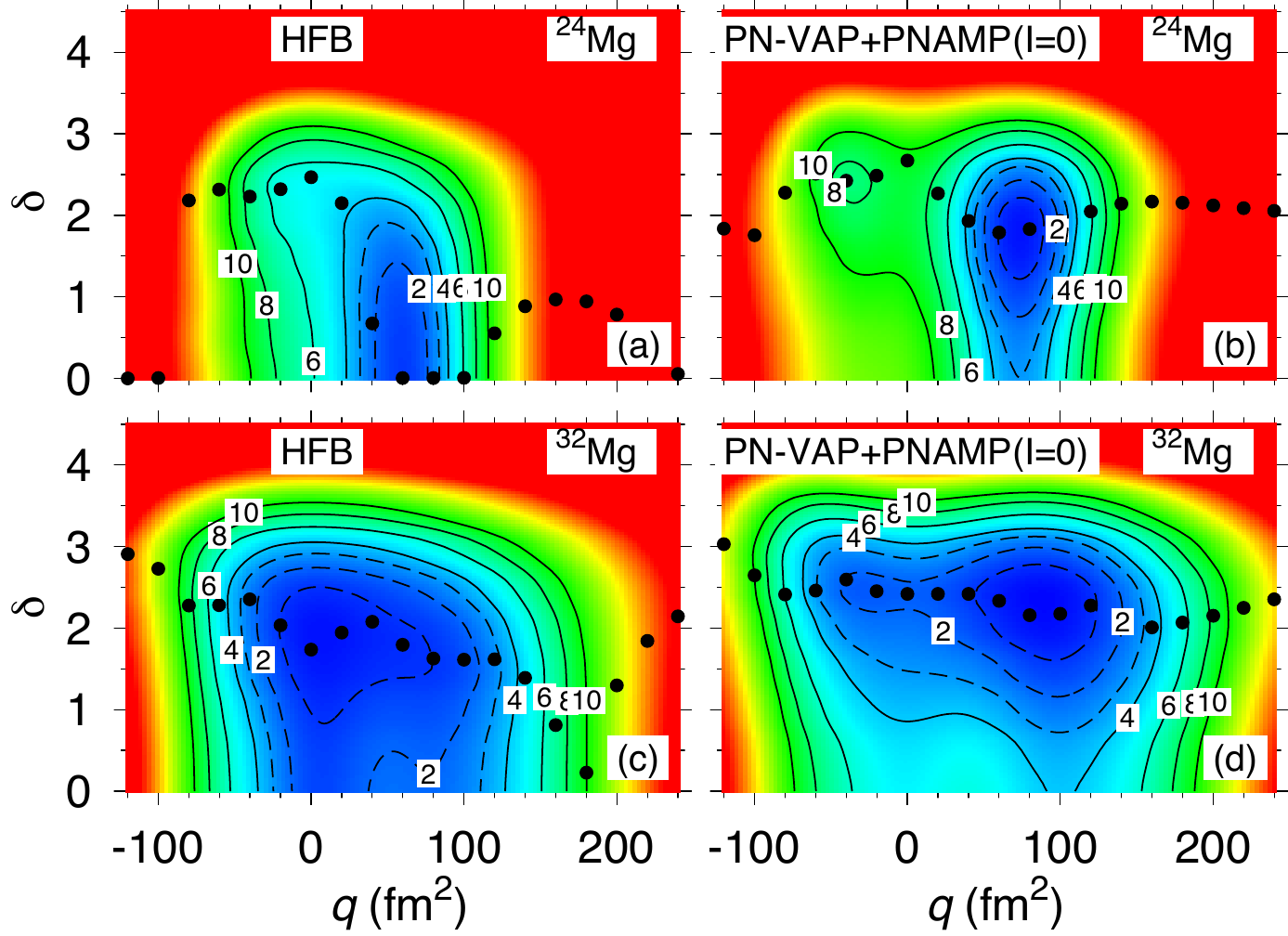}
\end{center}
\caption{Color online. Left panels: Potential energy surfaces for the nuclei  $^{24}$Mg and  $^{32}$Mg in the HFB approach. Right panels: Potential energy surfaces for the nuclei  $^{24}$Mg and  $^{32}$Mg in the PN-VAP+PNAMP  approach for I=0. In both cases the contours follow the same interval as in Fig.~\ref{fig:pairfluc_Ti52}. This figure has been taken from Ref.~\cite{Nuria2}.}
\label{fig:pairfluc_Mg24_32}
\end{figure}
 
In order to implement pairing fluctuations together with axially symmetric quadrupole fluctuations we proceed in the following way:  First, we generate   
intrinsic HFB wave functions $|\phi(q,\delta)\rangle$ with given quadrupole deformation $q$ and  ``pairing deformation'' $\delta$ by solving the variational equation
 \begin{equation}
 \delta {E^{\prime}}[\phi(q,\delta)]   = 0  \label{min_E},
\end{equation}
with 
\begin{equation}
{E^{\prime}}= \frac{\langle\Phi|\hat{H}|\Phi \rangle} {\langle\Phi|\Phi \rangle}   
 - \lambda_q \langle \phi |\hat{Q}_{20} | \phi \rangle- \lambda_\delta \langle \phi|(\Delta\hat{N})^2 +(\Delta\hat{Z})^2|\phi \rangle^{1/2}, \label{E_prime}
  \end{equation}
 and the Lagrange multipliers $\lambda_q$ and $\lambda_\delta$ being determined by the constraints  
 \begin{equation}
 \langle \phi |\hat{Q}_{20} | \phi \rangle =q, \;\;\;\; \;\;\;\;      \langle \phi|(\Delta\hat{N})^2+(\Delta\hat{Z})^2|\phi \rangle^{1/2} = \delta.
 \end{equation}
If in Eq.~(\ref{E_prime}) $|\Phi\rangle\equiv |\phi \rangle$  we are solving the plain HFB equations, as discussed in Sect.~(\ref{Sect:TheMFA}). In this case we have to add an additional Lagrange parameter to keep on the average the right number of particles.  As mentioned before a PNP  and/or  AMP  out of this w.f. would be a PAV. 
However if $|\Phi\rangle\equiv \hat{P}^N|\phi \rangle$, being $\hat{P}^N$ the particle number projector, the determination of 
$|\phi\rangle$ is done in the PN-VAP approach. This method provides a much better description of the pairing correlations in the intrinsic w.f.  \cite{Anguiano_VAP_02} although it is more involved.  Finally, as in the HFB case, an angular momentum projection can be performed afterwards. The variational equations are solved using the conjugate gradient method \cite{grad}.
   Once we have generated the basis states we can proceed with the configuration mixing calculation. 

 We study now the  dependence of the potential energy of these nuclei with respect to the two collective degrees of freedom $(q, \delta)$. In Fig.~\ref{fig:pairfluc_Ti52},  we present contour lines of the potential energy of $^{52}$Ti   as function of the constrained parameters $(q,\delta)$  in different approximations. The bullets represent the $\delta$ values of the self-consistent solutions (HFB or PN-VAP, i.e., without AMP) of the 1D ($q$-constrained) approach of Eq.~(\ref{E_Lagr_bet}). They must  be orthogonal to the equipotential curves in the corresponding approach.  The 1D plots of Figs.~\ref{fig:1D_N},\ref{fig:pai_Ti},\ref{fig:AMP_Ti}
 can be used as a guide in the interpretation of the  2D $(q,\delta)$  plots.

The relationship between the parameter $\delta$ and the pairing energy is rather independent of the $q$-value, see 
Fig.~\ref{fig:pair_ener_q} in  Appendix C. To have a feeling,  for the nucleus $^{52}$Ti and for $q=100$ fm$^2$ in the VAP+PNAMP approach and for $I=0\;\hbar$,  we provide the pairing energy (in parenthesis and in MeV)  corresponding to the preceding $\delta$ values: $0.0 (0.00)$, $0.5 (-0.52)$, $1.0 (-2.11)$, $1.5 (-4.74)$, $2.0 (-8.19)$, $2.5 (-12.53)$, $3.0 (-18.33)$, $3.5 (-26.17)$, $4.0 (-36.71)$, and $4.5 (-50.26)$. We thus see that the $\delta$ range covers a wide  energy interval.

In Fig.~\ref{fig:pairfluc_Ti52} (a)  we display the pure HFB case. Here  we find a region delimited
 from $q=-60$ fm$^{2}$ to $q$=100 fm$^{2}$ in the $X$ axis and from $\delta$=0 to $\delta$=2.5 in the ordinate, where the potential is soft in both directions. That means, for a given value of $q$ (or $\delta$) one does not gain much energy (just around 1 MeV) by increasing the $\delta$ (or $q$) coordinate. However, for the same  $q$ interval but $\delta$ between 2.5 and 4 it takes a considerable amount of energy to increase the pairing correlations of the system. For higher values of $\delta$, the pairing energy gain is huge and the total energy is up to 20 MeV larger. An analogous conclusion is obtained for the region $-140$ fm$^{2}$ $< q< -60$ fm$^{2}$ and 120 fm$^{2}$ $< q<$ 240 fm$^{2}$.  The potential becomes stiff and to deform the nucleus to that values requires a large amount of energy.   This structure is consistent with
 the one dimensional plot shown in the top left panel of Fig.~\ref{fig:1D_N}. 

\begin{figure*}[tb]
\begin{center}
\includegraphics[angle=0,scale=0.75]{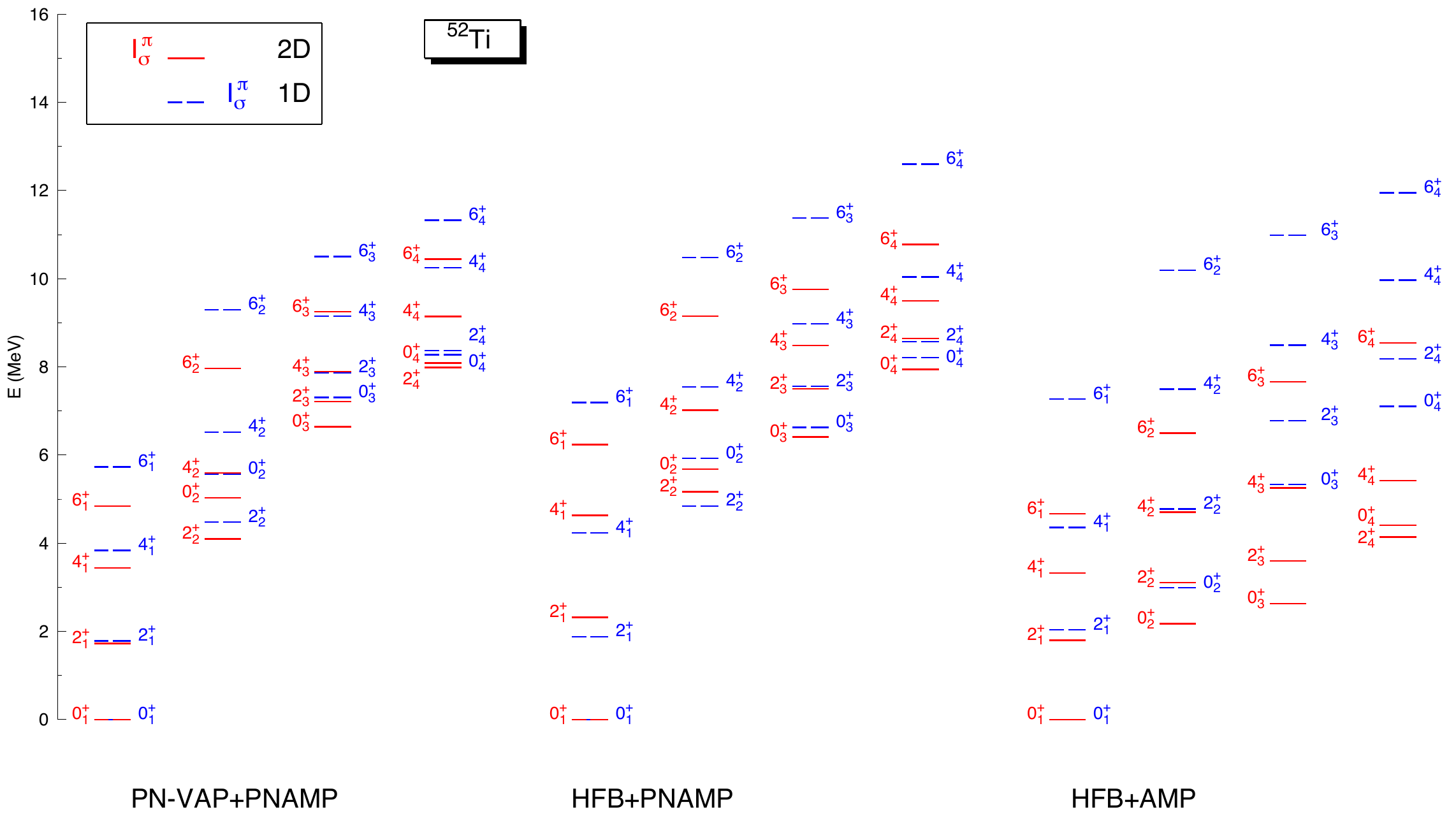}
\end{center}
\caption{(Color Online)  Spectra of $^{52}$Ti in the PN-VAP+PNAMP (left), HFB+PNAMP (middle) and HFB+AMP (right) approaches.  The four lowest states for spin $0^{+}, 2^{+}, 4^{+}$ and $6^{+}$ are represented in the 1D (dashed lines) and 2D(continuous lines). This figure has been taken from 
Ref.~\cite{Nuria2}.}
\label{fig:spect_Ti}
\end{figure*}

Next, in panel (b) we show the effect of particle number projection after the variation, i.e., one takes the HFB wave functions used to generate panel (a) and calculates the PNP energy. One obtains again rather flat minima  but  displaced to $\delta$=2.5. The energy lowering  of the absolute minimum is 1.37 MeV. 
In panel (c) we also represent the effect of the PNP but in this case, the projection is performed before the variation, therefore in that approximation we obtain the energy  with PN-VAP intrinsic wave functions. This plot looks like the previous one and the two trends mentioned  before are present here too: the equipotencial are shifted towards large values of $\delta$ and the minimum is deeper, being now even lower,  1.17 MeV below the PAV absolute minimum. One now observes two minima, one prolate at $(q=60$ fm$^{2}$, $\delta$=2.5) and one oblate at $(q=-40$ fm$^{2}$, $\delta=2.5)$. The PN-VAP approach is the proper way to perform the variation because one minimizes the energy calculated with the right number of particles.  One has to have in mind that, even though the PNP brings the energy minimum of the HFB solution closer to the VAP one,  there are other observables whose values  do not coincide with the self consistent ones provided by the VAP approach.

Now, the angular momentum projection ($I=0\; \hbar$) is performed for the approaches of the left panels  and presented in the corresponding right panels.
We start with the HFB+AMP case, panel (d). In this case since no PNP is performed and since the constraint on the particle number is done at the HFB level,  nothing guarantees that   $\hat{P}^I|\phi(q,\delta)\rangle$  does have the right values for the number of protons and neutrons. In order to correct for this deficiency the usual cranking recipe \cite{RS.80} of minimizing $ \hat{H}^{\prime}=H -\lambda \Delta\hat{N}$ instead of ${H}$ is used, with $\Delta \hat{N} = \hat{N}- \langle \hat{N}\rangle$ This amounts to substitute $ \mathcal{H}$  by $ \mathcal{H}^{\prime}$ in Eq.~(\ref{HW}), see Ref.\cite{Nuria2} for a detailed description.

As seen in Fig.~\ref{fig:AMP_Ti} the AMP increases considerably the depth of the potentials and the $q$-values of the minima.  They move  to larger $q$-values, $-80$ fm$^{2}$ for the oblate minimum and $80$ fm$^{2}$ for the prolate one.   In the HFB+PNAMP, panel (e),  or in the PN-VAP+PNAMP, panel (f),  the effect of the AMP is also to widen the equipotentials and to deepen the minima, the prolate being shifted towards larger value, 100 fm$^{2}$, and the oblate one to 
 $-80$ fm$^{2}$. An interesting point is that in the 2D plot we find that the minima of the energy in the HFB+AMP approach correspond  to  pairing energies of $\delta\approx 2.0$. We find that this is not the case in the PNP approaches where the minima correspond to $\delta\approx 2.5$.  The energy difference corresponding to the different $\delta$ values amounts to a difference in pairing energies of a few MeV \cite{Nuria}. 
 The equipotential surfaces  of panels (e) and (f) look very similar though in detail they are different, c.f.  Fig.~\ref{fig:AMP_Ti}.   The fact that  the minima of the HFB+AMP approach lie in a weak
 pairing region will have important consequences since the masses associated to the dynamics of the system, i.e., the solution of the HW equation, will be much larger than the ones associated to the PN projected approaches, providing a more compressed spectrum.  The energy gain of the absolute minimum in the PN-VAP+PNAMP approach with respect to the HFB (PN-VAP) is 4.53 MeV  (2.71 MeV). 

To further illustrate the role of the pairing fluctuation we display  in Fig.~\ref{fig:pairfluc_Mg24_32} the potential energy surfaces for $^{24}$Mg and $^{32}$Mg just in the HFB and in the PN-VAP+PNAMP approaches and for $I=0 \; \hbar$.  In the first row we find that $^{24}$Mg displays a stiff potential  in the HFB approach. It presents a structure of a deep prolate minimum $(q \approx 80$ fm$^2$) with  $\delta =0$ and a few MeV higher an oblate one ($q \approx -30$ fm$^2$).  We observe that this nucleus is more steep towards larger pairing correlations than the $^{52}$Ti.  In the PN-VAP+PNAMP case the prolate minimum shifts to $q \approx 100 $ fm$^2$ and $\delta \approx 2.0$ and the oblate one to $q \approx -40$ fm$^2$ and $\delta \approx 2.5$, the energy becoming even stiffer around the prolate minimum.
In the second row we display  $^{32}$Mg.  In the HFB approach the energy minimum  has a spherical shape and  $\delta \approx 1.6$.   About 2 MeV higher  there is a prolate shoulder with $q \approx 80$ fm$^2$ and $\delta \approx 1.5$. In the PN-VAP+PNAMP approach, right panel,  we observe two deformed minima, the deepest one at $q \approx 90$ fm$^2$ and $\delta \approx 2.1$ and the secondary oblate one at $q \approx -40$ fm$^2$ and $\delta \approx 2.5$,  about 2 MeV higher. The potential energy surface of the nucleus  $^{32}$Mg is wider and flatter than the one for  $^{24}$Mg. 

We  discuss now the results of the SCCM calculations, for which the Hill-Wheeler equation, Eq.~(\ref{HW}),  has to be solved. Before discussing  the excitation spectra we will comment on the limitations of our approaches. In our description we are considering mainly collective degrees of freedom, namely the quadrupole deformation and the pairing gap. Though we are considering different nuclear shapes and, in principle, single particle degrees of freedom can be expanded as  linear combinations of different configurations, we cannot claim to describe properly genuine single particle states  but only in an approximate way. Collective states, on the other hand, are very well described in our approach. 
The HW equation has to be solved separately for each value of the angular momentum, the diagonalization of this matrix provides the Yrast and the excited states,  $I^{+}_{1}, I^{+}_{2}, I^{+}_{3},...$ for each angular momentum. These energy levels, normalized to the  ground state energy, 
provide the spectrum of the nucleus.  Again, we will study the three cases we are focused on, namely HFB+AMP, HFB+PNAMP and PN-VAP+PNAMP. In order to evaluate the impact of the pairing fluctuations on the different observables we  consider the solutions of the HW equation in 1D, with one coordinate $(q)$,  and in 2D, with two coordinates $(q, \delta)$. We have calculated the four lowest states for each angular momentum.  
We now inspect the excitation spectra, but before making a detailed description let us just mention a very general argument to guide our discussion. 
The comment above on the cranking approximation can also be interpreted in the light of a  {\it quantum} approximation to an  angular momentum VAP method. According to  the  Kamlah expansion \cite{Ka.68}  a VAP of the angular momentum  can be approximated, to first order, in the following way:  the intrinsic HFB wave function, $|\phi \rangle $,  is determined by minimizing the energy $E^\prime= \langle \phi | \hat{H} | \phi \rangle-\omega \langle \phi | \hat{J}_x | \phi \rangle$ with $\omega$ determined by the
 constraint $\langle J_{x}\rangle =  \sqrt{I(I + 1)}$.  The energy is provided by $E^I= \langle \phi | \hat{H} P^I | \phi \rangle/\langle \phi | P^I | \phi \rangle$.  Since for $I=0\;\hbar$,  $\langle \phi | \hat{J}_x | \phi \rangle=0$, the Kamlah prescription does apply in this case  in the three approaches,  but for $I\neq 0\;\hbar$ this is not the case because our w.f. does not break time 
 reversal and thus cannot fulfill the constraint on the angular momentum.  That means that our approaches
 favor the states with $I=0\;\hbar$ because for them an approximate VAP for the angular momentum is performed.  For 
 $I\neq 0\;\hbar$ this is not the case and  we just do plain PAV.  From these arguments and from this perspective it is obvious that the quality of the approach diminishes with growing $I$-values.
 That means,  the relative energy gain will be largest for $I=0\;\hbar$,  and for $I\neq 0\;\hbar$  it will comparatively decrease with increasing $I$. Thus in our current approach we  predict stretched spectra, this will not be the case anymore if we break the time reversal symmetry \cite{BRE.15,EBR.16}. 
 
In Fig.~\ref{fig:spect_Ti} we present the excitation spectrum for $^{52}$Ti  in our three basic  approaches and in the 1D and 2D calculations. The levels are ordered just by the energy. 
In the left hand part we display the most complete  approach, namely the PN-VAP+PNAMP. 
The general trend is that the 1D calculation is more stretched that the 2D one.  This is a clear manifestation of the following fact: Since the 1D and the 2D calculations are self-consistent the ground state energy {\em before} the HW diagonalization, i.e., the minimum of the potential energy surfaces, is the same in both calculations and even after the HW diagonalization they are rather similar.
  This result is a consequence of the fact that the variational principle used to determine the wave functions $|\phi\rangle$ favors ground states.  
In the 1D calculations there is no room for the excited states to change the pairing content of a given w.f., however,  in the 2D calculations the flatness of the pairing degree of freedom opens the possibility of choosing  different pairing energies for a given deformation $q$ allowing thereby  an energy lowering.  We  therefore see that the consideration of additional degrees of freedom partially {\it compensates} the above mentioned problem of approximate VAP for  $I=0\;\hbar$ versus PAV for $I\neq 0\;\hbar$. In reality we are doing a restricted VAP, see Ref.~\cite{PhysRevC.71.044313} for more details.

In the middle of Fig.~\ref{fig:spect_Ti} the HFB+PNAMP spectrum is presented. This spectrum is, in general, more stretched than the PN-VAP+PNAMP one. Another difference is the fact that the ordering of some levels,
in particular the Yrast ones, of the 1D and 2D calculations are inverted as compared with the PN-VAP+PNAMP one.  The reason for this behavior is the lack of self-consistency  (in the sense discussed with PESs)  of this approach.  As we can see in the panel (e) of Fig.~\ref {fig:pairfluc_Ti52} the path of the 1D solution in the $(\delta,q)$ plane, i.e. the bullets line,  goes along lines of smaller pairing correlations than the minima displayed by the 2D contour plots. Consequently, in 1D  the mass parameter associated with the collective motion is  larger than in 2D and the associated  spectrum more compressed in the former than in the latter one. This effect combined with the additional degree of freedom of the 2D discussed above makes that  only the lower levels are inverted. 

Finally in the right part of Fig.~\ref{fig:spect_Ti} the HFB+AMP approach is displayed. First, we observe  very much compressed spectra  as compared with the other approaches.
 It is remarkable  the fact that all states with the same spin are  much closer to each other than in the PNP approaches. One furthermore notices the unusual large lowering of the 2D states as compared with the 1D ones.  These facts seem to indicate, see \cite{Nuria2},  that there is too much mixing in the solution of the HW equation due to spurious contributions stemming from the non-conservation of the particle number symmetry. One also observes that contrary to the inversion of the HFB+PNAMP, the inversion of the 1D and 2D levels does not take place in this case.  This is due to the fact that in this case we are  more self-consistent than in the HFB+PNAMP case.
 Concerning the 2D spectra in the three approaches one can understand the degree
 of compression of the spectra by looking at the right hand panels of 
 Fig.~\ref{fig:pairfluc_Ti52}. We observe that by far the softest surface towards  small pairing correlations is the HFB+AMP,  then, though to a lesser extend,  PN-VAP+PNAMP (in the energetically relevant part, i.e., around the minima)  and finally HFB+PNAMP relatively close to the former one.  Correspondingly we expect the HFB+AMP spectrum to be the most compressed,  followed by  PN-VAP+PNAMP and finally HFB+PNAMP relatively close to the latter one.

\begin{figure}[tb]
\begin{center}
\includegraphics[angle=0, scale=.4]{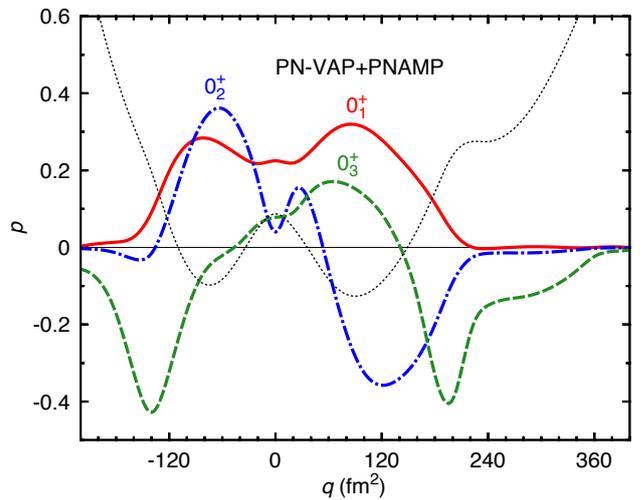}
\end{center}
\caption{Collective wave functions of the three lowest $I=0^+$ states of $^{52}$Ti in the PN-VAP+PNAMP approach in the 1D calculation. The dotted line represents the corresponding PES. }
\label{fig:wf_1D}
\end{figure}

We now discuss the collective wave functions, see Eq.~(\ref{coll_wf}),  solution of the Hill-Wheeler equations in one, 
$p^{I,\sigma}(q)$, and two dimensions,  $p^{I,\sigma}(q,\delta)$,  for the nucleus $^{52}$Ti.
To understand the more interesting 2D case we present in Fig.~\ref{fig:wf_1D} the one-dimensional case in the PN-VAP+PNAMP approach for the three lowest $0^+$ states.  The corresponding potential  energy curve has been plotted also in this figure.  This potential energy curve displays two quasi-coexistent minima, the lowest one prolate and the other one oblate, consequently the w.f.s  (see Fig.~\ref{fig:wf_1D}) of the  $0^{+}_{1}$ and $0^{+}_{2}$ states display a two hump structure with maxima (or maximum and minimum) at these values, the $0^{+}_{2}$ with a node as one would expect for a vibration.  The $0^{+}_{3}$ state, on the other hand, peaks  at large deformations in the prolate and the oblate potential shoulders  and it has a two nodes structure. 
The 1D w.f. in the HFB+PNAMP and in the HFB+AMP, specially the latter one are somewhat different to the PN-VAP+PNAMP, see
Ref.~\cite{Nuria2}.

\begin{figure}[tb]
\begin{center}
\includegraphics[angle=0, width=\columnwidth]{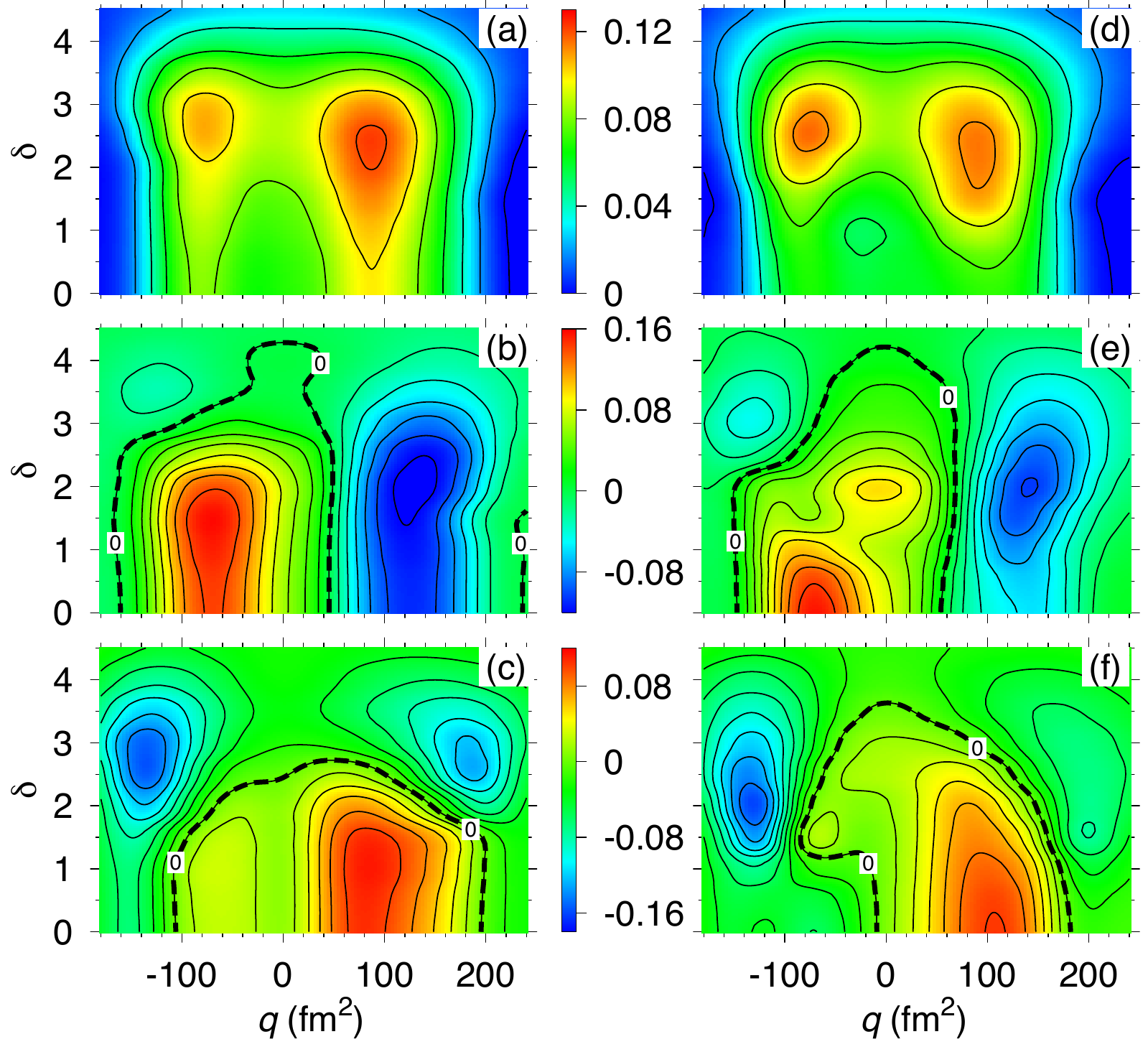}
\end{center}
\caption{ Contour lines of the wave functions of the three lowest $0^+$ states of $^{52}$Ti in the PN-VAP+PNAMP approach (left) and in the HFB+PNAMP (right)  in the 2D calculations.  The contour step size is 0.02. The thick dashed lines correspond to the zeros of the wave function.  To get better resolution the x-axis runs from -180 fm$^2$ up to 240 fm$^2$ at variance with former figures. This figure has been adapted from Ref.~\cite{Nuria2}.}
\label{fig:WF_J0_52Ti}
\end{figure}

Concerning the $(q,\delta)$ calculations,  the potential energy surfaces have been already discussed in Fig.~\ref{fig:pairfluc_Ti52} and the two-dimensional wave functions are presented in Fig.~\ref{fig:WF_J0_52Ti}. We start again with the PN-VAP+PNAMP case. In panel (a) the contour lines of the wave function of the $0^{+}_{1}$ state are shown.
In strong correspondence with the lowest right panel  of  Fig.~\ref{fig:pairfluc_Ti52} it presents a two bump structure, rather soft in the pairing degree of freedom, with a predomination of the prolate side. The bump  maxima are located at $q$-values  close to the 1D case and centered at $\delta$ values close to the self-consistent solution (see bullets in Fig.~\ref{fig:pairfluc_Ti52}). The $0^{+}_{2}$ state, panel (b), displays also   a two bump structure, this time with the maximum on the oblate side and soft in $\delta$. The maxima are located at $\delta$ values smaller than for the $0^{+}_{1}$ state. It presents a nodal line at
$q\approx 50$ fm$^2$ as it corresponds to a $\beta$ vibration in two dimensions. 
  The $0^{+}_{3}$ state, panel (c), presents a three-peak structure, two at large deformations and large pairing correlations and a smaller one around 80 fm$^2$ with smaller pairing correlations. This situation is similar to the 1D case where at similar $q$-values the same peaks are found. The fact that the large deformation peaks do have
  strong pairing correlations is due to the fact that the level density is very high at these
  deformations and that the 2D calculations allow that  a given $q$ value can take different
  pairing content for different collective states.     Looking at  panels (d,e,f) of Fig.~\ref{fig:WF_J0_52Ti} and taking into account the discussion above one can very easily interpret the 2D  wave functions of the HFB+PNAMP approach.
The main difference with the former case is that the beta vibration and the $0^+_3$ state in this case are not as pure as in the PN-VAP+PNAMP case.
As it was the case with the spectrum the  HFB+AMP collective wave functions look more different than the ones of the two former
approximations and will not be discussed here, for more details see Ref.~\cite{Nuria2}.
Interestingly, though the potential energy surfaces in the three cases are rather  similar, see panels (b), (e)  and (f) of Fig.~\ref{fig:pairfluc_Ti52}, the wave functions of the HFB+AMP and the spectrum are rather different from the other two. This has obviously  to do  with the  non-diagonal elements of the Hamilton overlap and the norm overlap, the former through the dynamical corrections and the latter through the linear dependence of the basis states.
 
Though not discussed in this contribution the pairing vibrations play an important role in many nuclear 
processes.  For instance, the consideration of the pairing degree of freedom in the calculation of the  neutrinoless double $\beta$ decay  has resulted in an increase of 10\%-40\% of the magnitude of  the 
corresponding matrix element \cite{Nuria3}.
 We conclude this section  underlining the relevance of the PNP for a proper description of the properties of atomic nuclei.

\section{Triaxial calculations}
 \label{Sect:TC}
In the previous sections we have seen applications of the GCM to axially symmetric problems.
 However, many exciting experimental and theoretical phenomena are closely related to the triaxial degree of freedom, for instance: presence of $\gamma$-bands at low excitation energy and $\gamma$-softness, shape coexistence and shape transitions in transitional regions \cite{Casten_O6_85, Casten_BaE5_00, Regan_ShapeTrans_03, Ita_10Be_02, Clem_Kr_07, Ober_Kr_09, Mo_96}; lowering of fission barriers along the triaxial path \cite{Baran_Fiss_81, Bend_Fiss_98, Warda_Fiss_02}; influence of triaxial deformation in the ground state for the mass models \cite{Naza_MassTriax_05, Moller_MassTriax_06}; triaxiality at high spin \cite{Naza_WS_85, Carl_142Gd_08, Yad_168Hf_08}; 
observation of $K$-bands and isomeric states in Os isotopes \cite{Os_85, Vik_Os_09, Kum_Os_09}; or some other exotic excitation modes such as wobbling motion, chiral bands \cite{Wobb_01, Chiral_04, Chiral_06}. 

From the theoretical point of view some approaches beyond mean field have been proposed to study the triaxial effects. In particular, one of the most widely used is the collective Hamiltonian \cite{RS.80}. It can be derived in the adiabatic approximation to the time-dependent HFB theory  \cite{BV.78}, and in the generator coordinate method with the Gaussian overlap approximation (GOA) \cite{RG.87,BD.90,PR.09}. These two approaches differ in the collective masses and in the zero point energies.  The collective Hamiltonian has been applied with different interactions used to define the collective potential, namely, Pairing-plus-Quadrupole \cite{BKColl_PPQ_68}, Interacting Boson Model \cite{Cast_IBM_84}, Nilsson Woods-Saxon \cite{Naza_WS_85}, Gogny \cite{GGColl_Gogny_83,GDColl_Gogny_Kr_09, GOColl_Gogny_N40_09} or RMF \cite{Niksic_Rel_09},  to describe some of the experimental features listed above. It is however of a limitted scope because it does not allow to include in a simple way  additional degrees of freedom, for example, to deal simultaneously with quadrupole and octupole deformations within a symmetry conserving framework.
On a broader road, a more fundamental approach, free from the approximations of the collective Hamiltonian, using the full GCM and exact microscopic particle number and angular momentum projection has been developed  in the last years.

In the past, exact angular momentum projection with triaxial intrinsic wave functions without GCM has been carried out only for schematic forces and/or reduced configuration spaces. For instance, projection of BCS \cite{RingBCS_PPQ_84} or Cranked Hartree-Fock-Bogoliubov (CHFB) states \cite{ETY.99} with the Pairing-plus-Quadrupole interaction; projection of Cranked Hartree-Fock (CHF) states without pairing with schematic \cite{HeenCHF_SK_84} and full Skyrme interactions \cite{DobaCHF_SK_07} or angular momentum projection before variation with particle number and parity restoration in limited shell model spaces \cite{VAMPIR,MONSTER} have been performed so far.\\
However, the increase of the current computational capabilities has recently allowed  the first implementations of the angular momentum projection of triaxial intrinsic wave functions in the whole $(\beta,\gamma)$ plane with effective forces. In particular, Bender and Heenen reported GCM calculations with particle number and triaxial angular momentum projection (PNAMP) with the Skyrme SLy4 interaction \cite{BendHFB_SK_08}. In this work, the intrinsic wave functions were found by solving the Lipkin-Nogami (LN) equations. On the other hand, Yao \textit{et al.} showed the implementation of the triaxial angular momentum projection \cite{RingAMP_Rel_09} and the extension to the GCM  \cite{RingGCM_Rel_10} for the Relativistic Mean Field (RMF) framework. In these calculations, there is no particle number projection and the mean field states are found by solving RMF+BCS instead of the full HFB or LN equations. These two assumptions could lead to a poor description of important pairing correlations, especially in the weak pairing regime where even spurious phase transitions appear in those cases \cite{Anguiano_VAP_02,Rod_CaTiCr_07}.

  A detailled description of the GCM and the collective Hamiltonian  within the Relativistic approach can be found in Ref.~\cite{NV.11}. An interesting comparison of the full GCM and the collective Hamiltonian has been performed in Ref.~\cite{YH.14}.

\subsection{ The $\beta$ and $\gamma$ coordinates} 
\label{Sect:TCBGC} 
\begin{figure}[tb]
\begin{center}
\includegraphics[angle=0, width=.7\columnwidth]{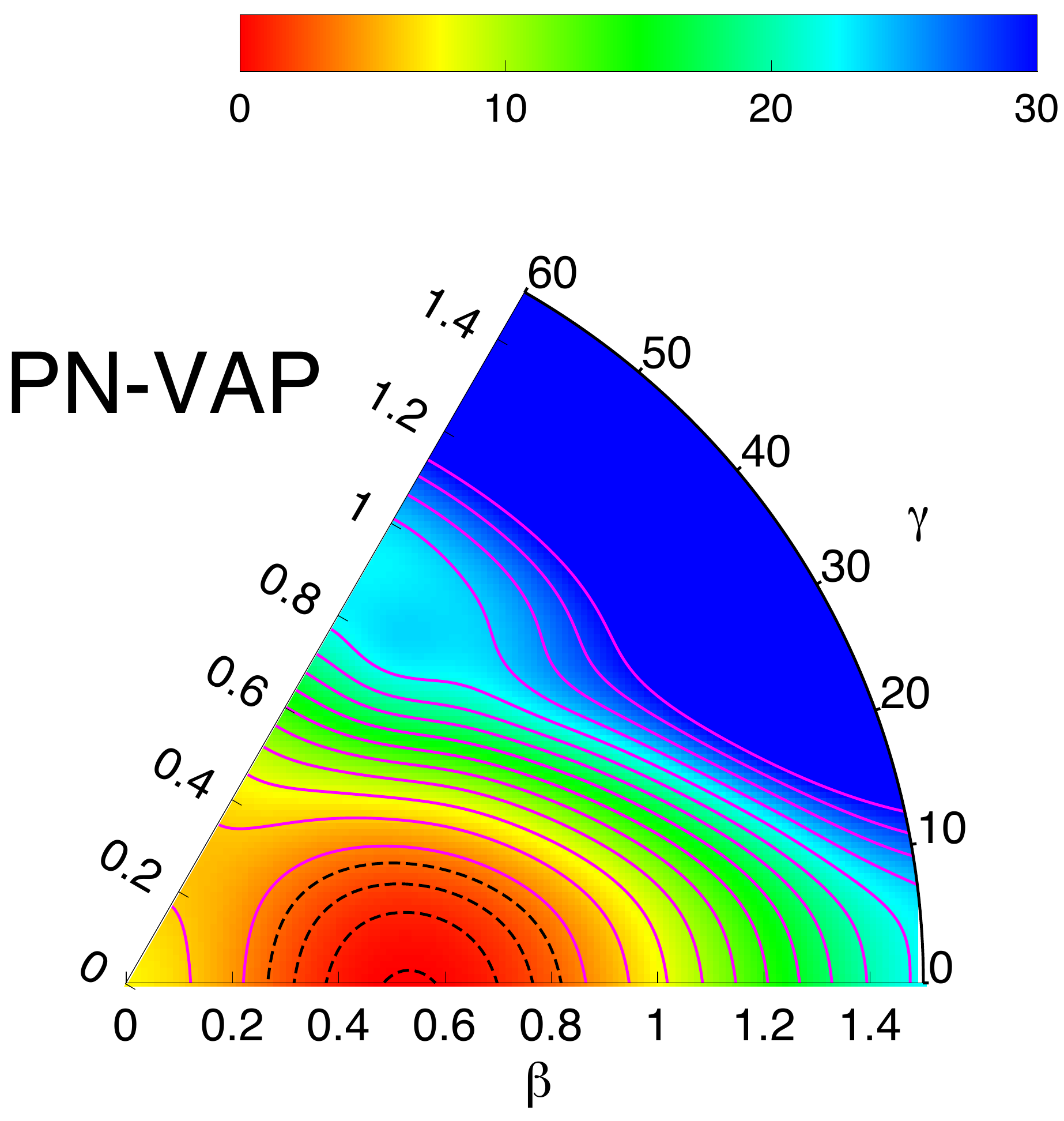}
\end{center}
\caption{PES in the PN-VAP approach for the $^{24}$Mg nucleus. The energy is normalized to the minimum of the PES ($-196.01$ MeV) and the contour lines are divided in 1 MeV (black dashed lines) and 2 MeV steps (continuous magenta lines). This figure has been adapted from Ref.~\cite{TE.10}. }
\label{fig:PN-VAP_24Mg_TRIAX}
\end{figure}

With the coordinates $(\beta,\gamma)$ the GCM Ansatz of Eq.~(\ref{eq:GCM_Ansatz})  looks like
\begin{equation}
|\Psi^{N,I}_{M,\sigma} \rangle =  \; \sum_{\beta,\gamma,K} f^{I}_{\sigma}(\beta,\gamma,K)  P^N P^I_{MK} \; |\phi (\beta,\gamma) \rangle \label{eq:GCM_Ansatz_bet_gam}
\end{equation}

Since we do not break the time reversal symmetry it is sufficient  \cite{RS.80} to consider one sextant of the $(\beta,\gamma)$ plane.
 To discretize the sextant $0^{\circ}\le \gamma \le 60^{\circ}$  we choose a triangular mesh of $N_{points}=99$ in which  we solve the constrained particle number projection before the variation (PN-VAP) equations to determine the HFB wave functions $\phi(\beta,\gamma)$
\begin{eqnarray}
{E^{\prime}}[\phi]= \frac{ \langle\phi^{}|\hat{H}\hat{P}^{N}|\phi{} \rangle}{\langle\phi^{}|\hat{P}^{N}|\phi^{} \rangle} -  \langle \phi |\lambda_{q_{0}}\hat{Q}_{20} + \lambda_{q_{2}}  \hat{Q}_{22} | \phi \rangle, \label{E_Lagr_bet-gam}
  \end{eqnarray}
with  the Lagrange multiplier $\lambda_{q_{0}}$  and $\lambda_{q_{2}}$ being determined by the constraints 
 \begin{equation}
 \langle \phi |\hat{Q}_{20} | \phi \rangle =q_{0}, \;\; \; \langle \phi |\hat{Q}_{22} | \phi \rangle =q_{2}. \label{q0_q2_constr}
 \end{equation}
 The relation between $(\beta,\gamma)$ and $(q_{0},q_{2})$ is provided by Eqs.~(\ref{be_ga0},\ref{be_ga}).
The number of Fomenko \cite{Fo.70} points to perform the integral of the particle number projection is  $N_{Fom}=9$. The intrinsic many body wave functions $|\phi(\beta,\gamma)\rangle$ are expanded in a cartesian harmonic oscillator basis and the number of spherical shells included in this basis is $N_{shells}=7$ with an oscillator length of $b=1.01A^{1/6}$. 

\begin{figure}[tb]
	\begin{center}
		\includegraphics[angle=0, width=\columnwidth]{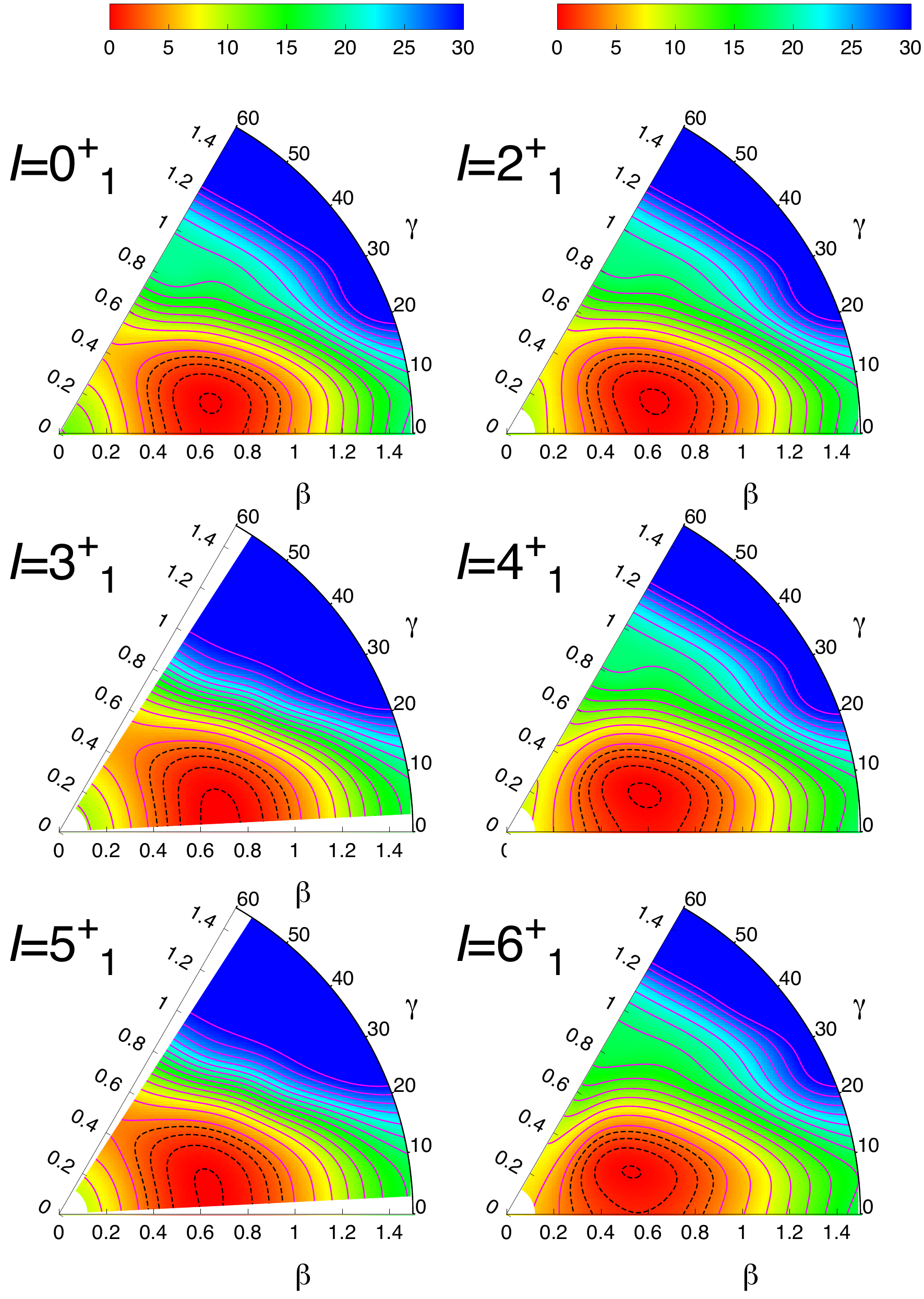}
	\end{center}
	\caption{PNAMP potential energy surfaces including $K$-mixing in the $(\beta,\gamma)$ plane for $I=0-8\; \hbar$ and the lowest eigenvalues in $K$-space. The PES are normalized to the minimum of the surfaces (-200.74, -199.43, -194.04, -196.61, -190.86, -192.27, -186.09, -185.33  MeV for ($I=0,2,3,4,5,6,7,8\;\hbar$) respectively). The contour lines are divided in 1 MeV (black dashed lines) and 2 MeV steps (continuous magenta lines) and states with projected norm less than $10^{-6}$ are removed, see discussion following Eq.~(\ref{nat_bas}). This figure has been adapted from Ref.~\cite{TE.10}.}
	\label{fig:PES_24Mg_TRIAX}
\end{figure}

In this section we present triaxial calculations for the nucleus $^{24}$Mg.  These results are  based, to a large extend, on Ref.~\cite{TE.10}.  

In Fig.~\ref{fig:PN-VAP_24Mg_TRIAX} the PN-VAP energy landscape is plotted showing a single and well defined minimum at $\beta=0.5, \gamma=0^{\circ}$ separated by $\sim 7.7$ MeV from the spherical point and $\sim 6.1$ MeV from the oblate saddle point at $\beta=0.25$. Similar PES are obtained for Skyrme (HFB with particle number projection after variation (PN-PAV) included- \cite{BendHFB_SK_08})  and relativistic (BCS without PNP \cite{RingAMP_Rel_09}) interactions although a softer surface between the spherical point and the minimum is obtained for the Skyrme interaction.

The next step is the simultaneous particle number and angular momentum projection (PNAMP) of the states that conform the PES.
In this case, due to the $g^{I}_K$ dependence\footnote{We use the symbol $g^{I}_{\sigma}((\beta,\gamma)K)$ instead of $f^{I}_{\sigma}(\beta,\gamma,K)$ to indicate that only $K$-components are mixed and not different shapes.} of Eq.~(\ref{Proj_WF}),  in each $(\beta,\gamma)$ point one has to solve a reduced Hill-Wheeler equation, see Eq.~(\ref{HW}), given by
\begin{equation}
\sum_{K'} \, \,(\mathcal{H}_{(\beta,\gamma)K, (\beta,\gamma)K'} - E^{N,I}_\sigma \mathcal{N}_{(\beta,\gamma)K,
(\beta,\gamma)K'}) g^{I}_{\sigma}((\beta,\gamma)K') = 0.
\label{HW_K}
\end{equation} 
Notice that in each $(\beta,\gamma)$ point one can have several eigenvalues $E^{N,I}_\sigma$ labeled by $\sigma$. The Hamiltonian and the norm matrix elements are given by expressions (\ref{hamove}) and (\ref{normove}), respectively.
 The  calculations have been done with the set of integration points in the Euler angles  $(N_{\alpha}=8,N_{\beta}=16,N_{\gamma}=16)$.  In Fig.~\ref{fig:PES_24Mg_TRIAX} we plot the normalized PNAMP energy landscapes in the $(\beta,\gamma)$ plane for the lowest eigenvalue in the $K$-space for each angular momentum $I=0^{+}_{1}-6^{+}_{1}$ (see Eq.~(\ref{HW_K})). In addition, all the points close to the spherical one, and the points close to axiality for odd values of $I$, have been removed for $I\neq0$ because their norm is close to zero. The first noticeable aspect is that the axial minimum of Fig.~\ref{fig:PN-VAP_24Mg_TRIAX} is displaced to triaxial values at larger deformations for all values of the angular momentum, although the barriers between them and the axial prolate saddle points are less than 1 MeV. For $I=0^{+}_{1},2^{+}_{1},3^{+}_{1}$ the minima are located at $(\beta\sim 0.7, \gamma\sim10^{\circ})$ while we observe a softening of the PES with increasing value of the angular momentum and a displacement to larger $\gamma$ and smaller $\beta$ deformation -$(\beta\sim0.65, \gamma\sim19^{\circ})$ for $I=4^{+}_{1},5^{+}_{1}$ and $(\beta\sim0.55, \gamma\sim23^{\circ})$ for $I=6^{+}_{1}$. We  also note that the softening of the PES in the case of odd $I$ values is in the $\gamma$ direction towards the oblate saddle point. The energy difference between the PN-VAP and $I=0^{+}_{1}$ minima is $\sim4.6$ MeV while the gain in energy due to the inclusion of the triaxial degree of freedom, i.e., the difference between the triaxial minimum and the axial saddle point, is $\sim0.7$ MeV. Similar results have been reported with Skyrme and relativistic interactions although these studies of PNAMP-PES only extend to $I=0,2$ and the effect of increasing triaxiality with growing angular momentum were not analyzed.\\

\begin{figure*}[tb]
	\begin{center}
		\includegraphics[angle=0,scale=0.75]{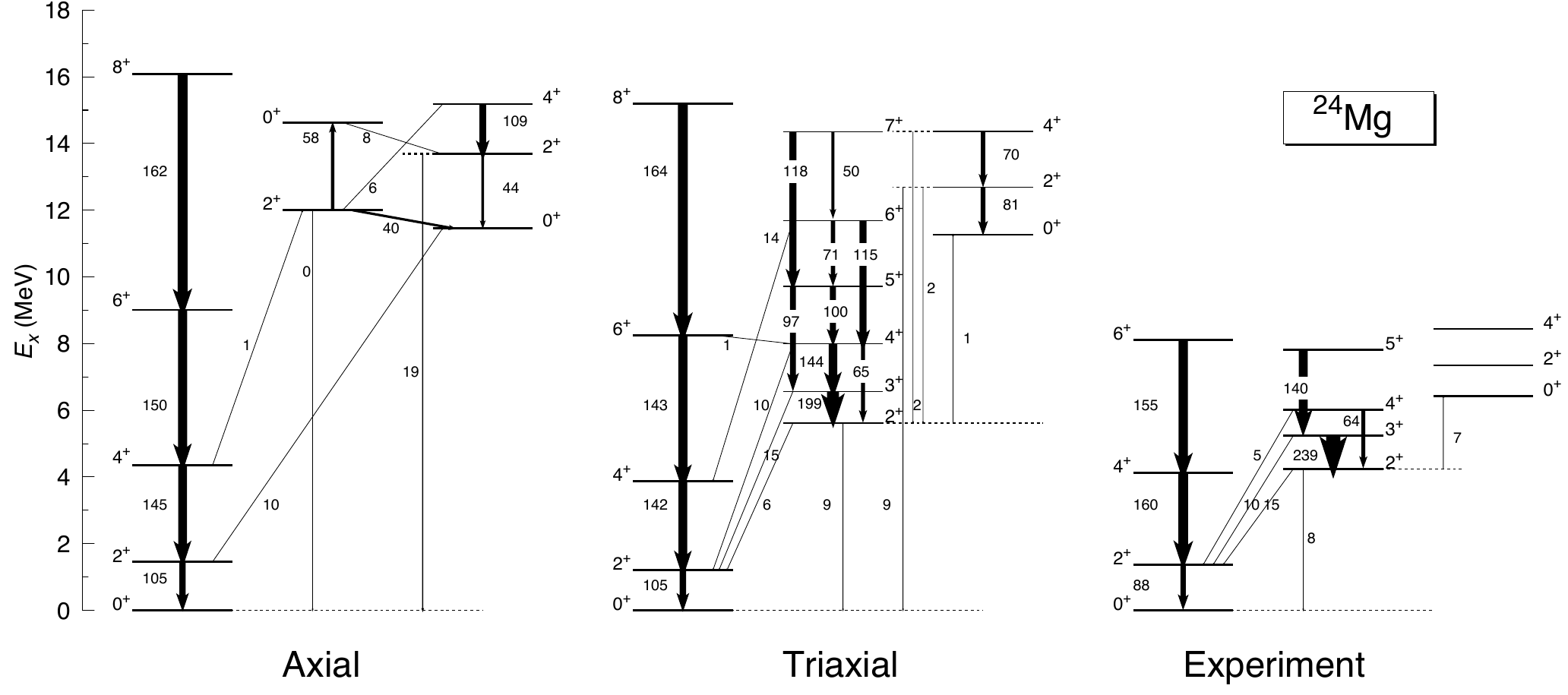}
	\end{center}
	\caption{GCM-PNAMP excitation energies and reduced transition probabilities B(E2) calculated with axial symmetry (left), triaxial (middle) and experimental values (right). The width of the arrows are proportional to value of the corresponding B(E2). The experimental values are taken from \cite{Data24Mg}}
	\label{fig:spect_24Mg}
\end{figure*}

The final step in the calculation to obtain the spectrum is the GCM-PNAMP method, where simultaneous mixing of the different deformations $(\beta,\gamma)$ and $K$ components is performed (see Eq.~(\ref{eq:GCM_Ansatz_bet_gam})). As we mentioned in Sec.~(\ref{Sect:TheSCMFA}), we have to solve the HWG equations separately  for each value of the angular momentum. These generalized eigenvalue problems are solved removing the linear dependence of the states with the definition of the orthonormal natural basis (Eq.~(\ref{nat_states})).  In order to avoid spurious states in this basis, we use the cutoff parameter, $\zeta$ defined below Eq.~(\ref{nat_bas}). The convergence of the PNAMP-GCM method is studied in Ref.~\cite{TE.10}.  The  lowest energies found are represented as a function of the parameter $\zeta$. Here we distinguish a region of large $\zeta$ where the energies are decreasing followed by a range of values where the energies are nearly constant. The appearance of these \textit{plateaus} is the signature of the convergence of the GCM method \cite{BD.90}.  Finally, for small values of $\zeta$ a linear dependence shows up and we obtain senseless values for the energy. The final choice for $\zeta$ is the one in a range for which we observe a large \textit{plateau} for all the levels of interest. This value must be the same for a given angular momentum in order to guarantee the orthogonality of the levels. This analysis has been performed for the different values of the angular momentum, see Ref.~\cite{TE.10}, giving a similar behavior to the previous one. Eventually, we have chosen $\zeta=10^{-3}$ as the final value, similar to the one found in 
Ref.~\cite{RingGCM_Rel_10}.  This procedure can be complemented by  inspection  of the shape of the wave
function as a function of $\zeta$. \\
Once the convergence of the GCM-PNAMP energies has been checked, we plot the definitive spectrum extracted from the triaxial calculations in Fig.~\ref{fig:spect_24Mg} (central part). We classify the different levels in three bands according to the corresponding B(E2) values. The ground state band is formed by a sequence of even values of angular momentum with a level spacing very similar to a rotational band whereas the second one connects states with $I=2,3,4,5$ as it could be expected from a $\gamma$ band. The third band is built with $I$-even states on top of the second $0^{+}_{2}$ state. We observe strong electric quadrupole intraband transitions while the  B(E2)of interband transitions are much smaller. This fact indicates the different underlying structure for each band and the absence of mixing between those states. We can study the nature of these bands decomposing the collective wave functions $|p^{I\sigma}_{K}(\beta,\gamma)|^{2}$ (Eq.~(\ref{coll_wf})) into their $K$ components and summing the contribution of all deformations $(\beta,\gamma)$ for each $K$. The result which is not shown here, see Ref.~\cite{TE.10},  clearly indicates that the first and third are rather pure $K=0$ bands while the second band corresponds mainly to $K=|2|$ states. 

\begin{figure}[tb]
\begin{center}
\includegraphics[angle=0, width=\columnwidth]{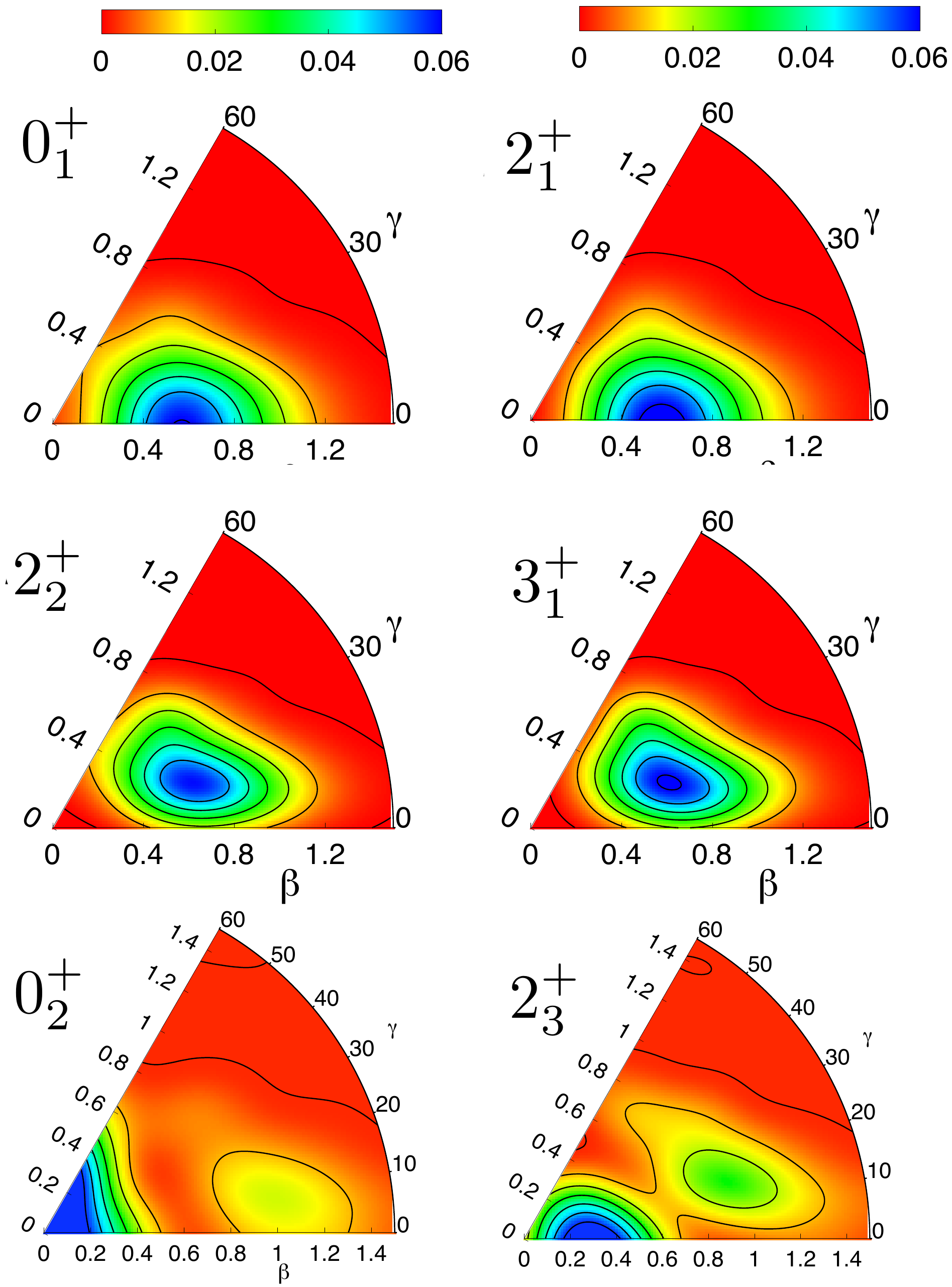}
\end{center}
\caption{GCM-PNAMP collective wave functions $|p^{I\sigma}(\beta,\gamma)|^{2}$ for the two lowest states of the ground state band (top), second (middle) and third (bottom) bands respectively. Contour lines are separated by 0.01 units. This figure has been adapted from Ref.~\cite{TE.10}.}
\label{fig:WF_24Mg_3bands}
\end{figure}
We also plot in Fig.~\ref{fig:WF_24Mg_3bands} the collective wave function, see Eq.~(\ref{coll_wf}),  of each GCM state in the $(\beta,\gamma)$ surface. The most noticeable aspect is that all the states belonging to the same band have a very similar distribution of probability in the plane and the mixing between these states is small, leading to the interband and intraband B(E2) values given in Fig.~\ref{fig:spect_24Mg}. In particular, all the states in the first band have a well defined maximum at $(\beta\sim0.58, \gamma=0^{\circ})$ and the probability drops rather symmetrically in the $\beta$ and $\gamma$ directions. Therefore, although the PNAMP-PES showed triaxial minima (see Fig.~\ref{fig:WF_24Mg_3bands}), the configuration mixing calculations drive the states to axial deformation. This effect has also been reported in Ref.~\cite{RingGCM_Rel_10} with a Relativistic interaction. For the second band, the distribution of probability is concentrated in a region of the plane with $(\beta\in[0.4-1.0], \gamma\in[0^{\circ},35^{\circ}])$ and the maxima are located at $(\beta\sim0.7,\gamma\sim18^{\circ})$. Finally, the states belonging to the third band show a high probability of having spherical $(0^{+}_{2})$ or slightly prolate deformation $(2^{+}_{3},4^{+}_{3},6^{+}_{3})$ 
--$\beta\in[0.0,0.5]$-- combined with a non-negligible mixing of  states with larger deformation in the range of $\beta\in[0.8,1.3],\gamma\in[0^{\circ},30^{\circ}]$. 

In Fig.~\ref{fig:spect_24Mg} we have also compared the triaxial results with axial calculations. In order to understand better the results of this comparison, we  investigate first the relationship between the axial and triaxial collective wave functions. The axial states emerge from the $\gamma=0^{\circ}-180^{\circ}$ path of the $K=0$ component of the corresponding triaxial states. In particular, we can relate the ground state bands in both approaches and also the  $0^{+}_{2},2^{+}_{3},4^{+}_{2}$ states of the axial calculation  with the $0^{+}_{2},2^{+}_{3},4^{+}_{3}$ states of the triaxial one.  For the ground state band, as expected,  the reduced transition probabilities and the energies are similar.

 Nevertheless, the small $K$-mixing for $I\neq0$  is enough to lowers the excitation energies for higher angular momentum. Consequently the first triaxial band is slightly compressed as compared with the axial band. The  axial and triaxial calculations, however, predict  larger differences between the second and third bands. The axial case is unable to describe the $\gamma$-band but also the energies and B(E2) of the third triaxial band with $K=0$  are modified with respect to the corresponding ones in the axial case. This difference is due to both the small $K$-mixing and also to the triaxial configuration around $\beta\sim1.0$ that appears already for $K=0$ 
(see Fig.~\ref{fig:WF_24Mg_3bands}). 
 
The avalaible experimental data for $^{24}$Mg are also displayed in Fig.~\ref{fig:spect_24Mg}. There is a qualitative agreement between theory and experiment both for energies and reduced transition probabilities. The excitation energies for the first band are quantitatively very well described. In addition, it is important to emphasize the quality of the theoretical predictions for the intraband and interband reduced transition probabilities which shows the small mixing between the corresponding bands. Although the triaxial approach improves considerably the axial one,  the band heads of the $\gamma$- and especially the third band are still too high in excitation energy. This is  due to the lack of the correlations associated to the angular momentum restoration before the variation and time-reversal symmetry breaking that are not included in this calculation, see  Ref.~\cite{BRE.15} and next Section. \\

\subsection{The $\beta$, $\gamma$ and $\hbar\omega$ coordinates} 
\label{Sect:TCBGOC}
\begin{figure}[tb]
\begin{center}
\includegraphics[angle=0,scale=0.4]{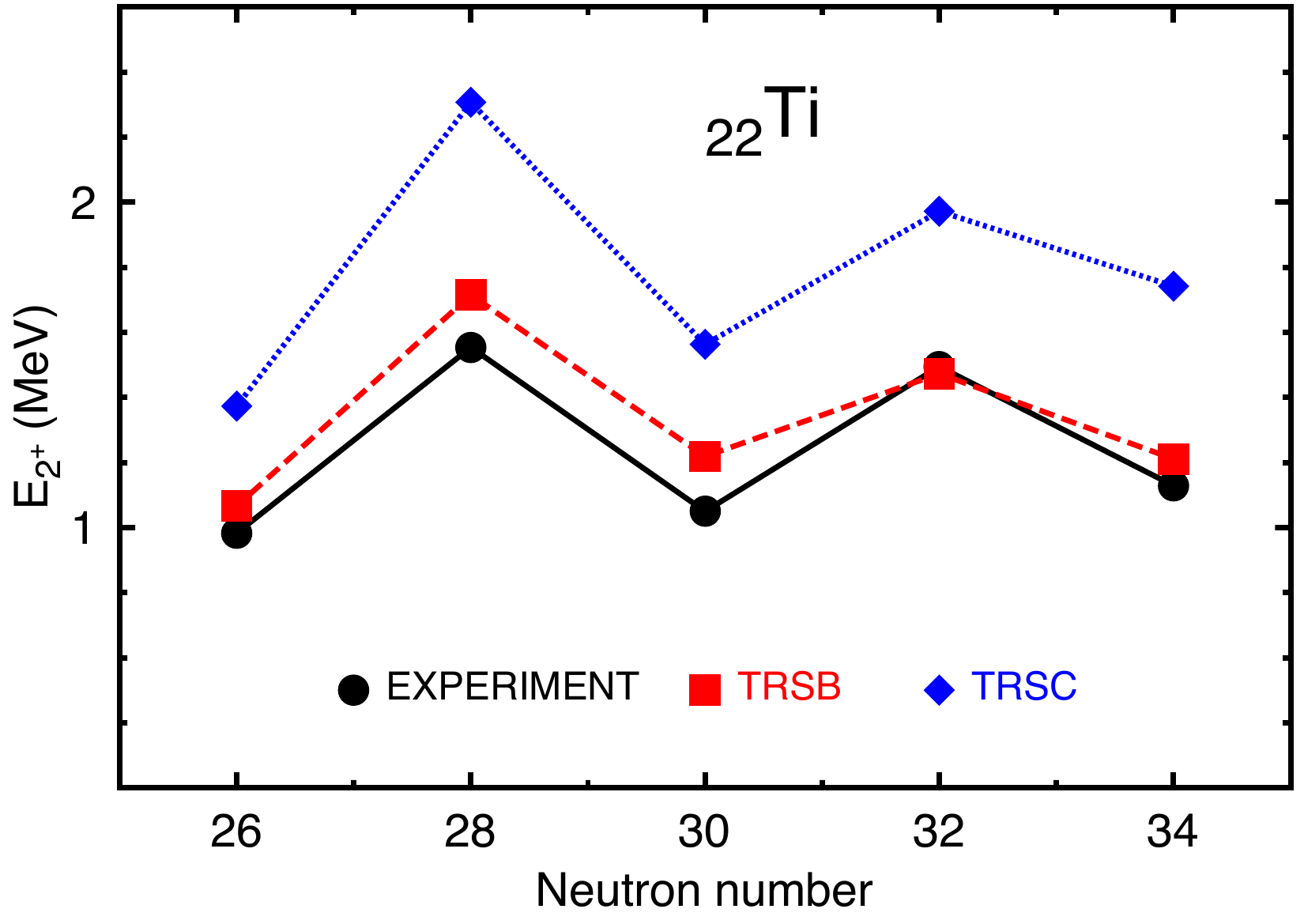}
\includegraphics[angle=0,scale=0.4]{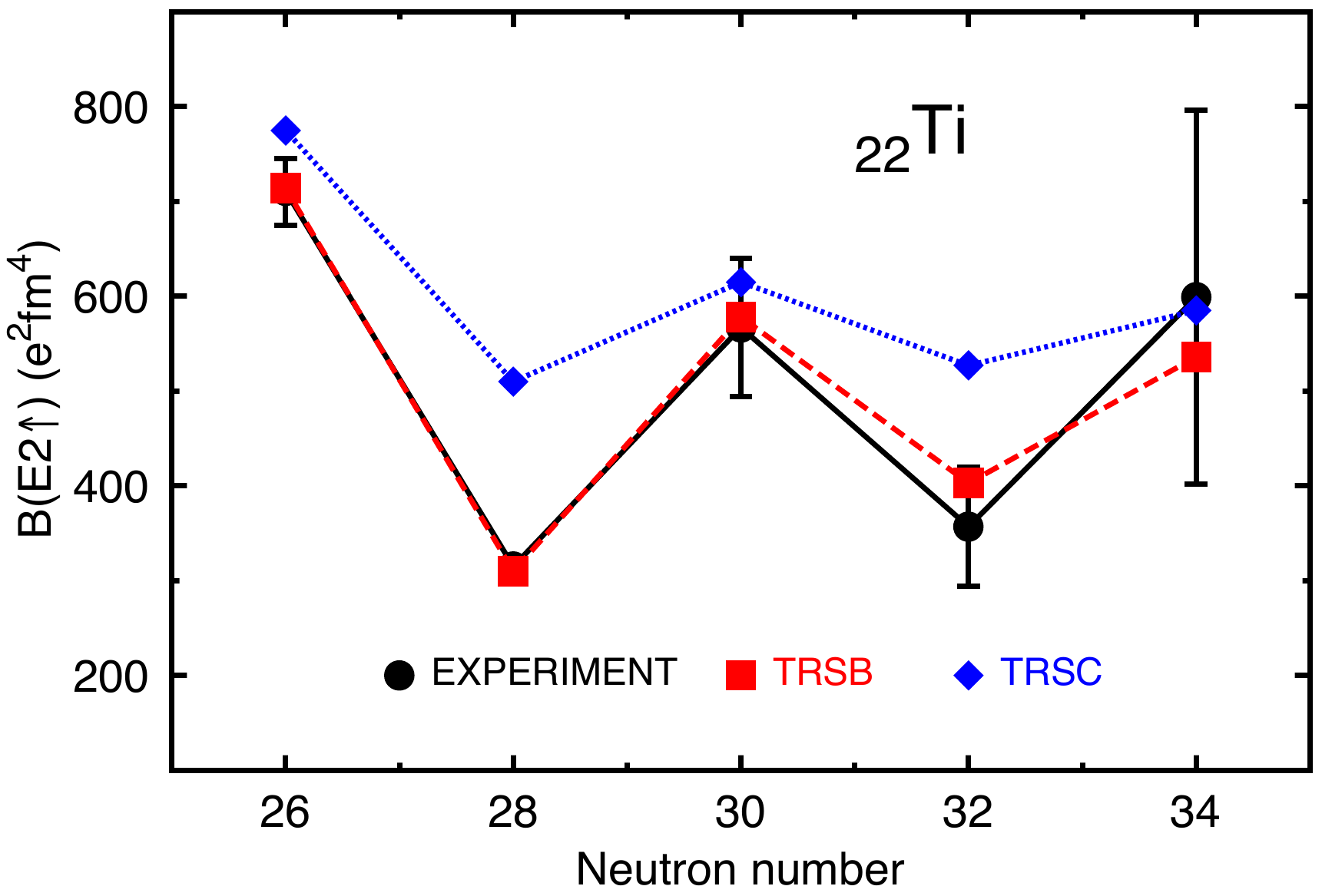}
\end{center}
\caption{(Color Online)  Excitation energies of the $2^{+}_{1}$ states (top) and $B(E2; 0^{+}_1 \longrightarrow 2^{+}_1)$ transition probabilities (bottom) in the Titanium isotopes in two approaches: Time reversal 
symmetry conserving (filled diamonds, blue color) and  time reversal symmetry breaking (filled squares, red color). The experimental values \cite{exp1,exp2,exp3,exp4,exp5} (bullets, black color) are also shown. This figure has been adapted from Ref.~\cite{BE.15}}
\label{fig:Ti_TRSB}
\end{figure}

\begin{figure}[tb]
\begin{center}
\includegraphics[angle=0,scale=0.4]{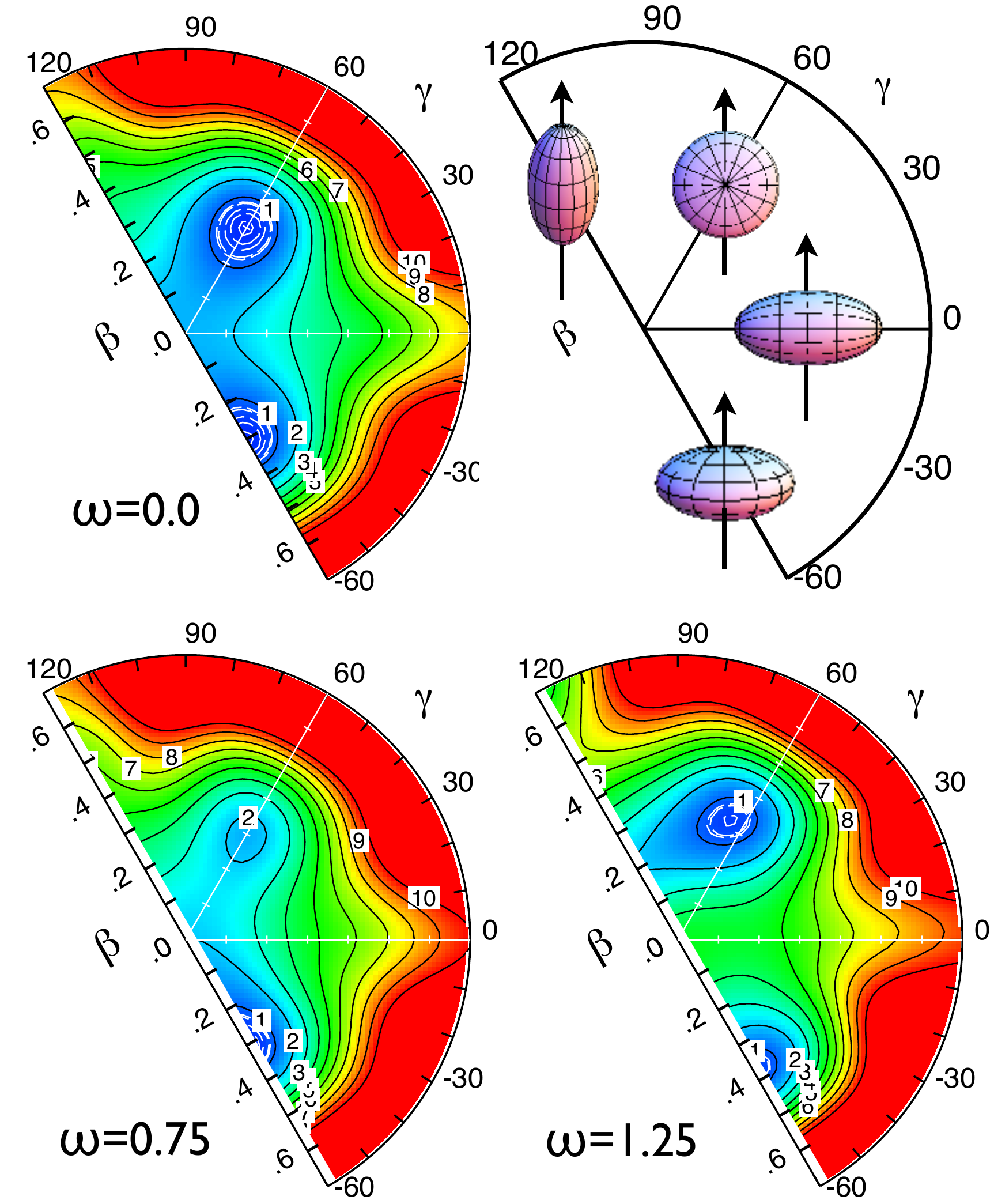}
\end{center}
\caption{(Color Online)  PES of $^{42}$Si in the PNVAP approach for the indicated angular frequency $\hbar \omega$ in MeV in the $(\beta,\gamma)$ plane, $\gamma$ in degrees.  In each case 
the respective  minimum energy has been subtracted. Continuous contour lines  are 1 MeV apart up to a maximum of 10 MeV. To emphasize the minima  white dashed contour lines in steps of 0.2 up to 0.8 MeV have been drawn. The top right panel displays the shapes and orientations in the $(\beta,\gamma)$ plane}
\label{fig:PES_42Si_PNVAP}
\end{figure}

As it has been discussed in previous sections the non-consideration of an angular momentum dependence in the variational principle 
 at the HFB level causes a stretching of the spectrum.  As a matter of fact a phenomenological factor, see for example  \cite{Rod_CaTiCr_07,Jung_PLB_08}, was introduced in some investigations.
 
 In the past, the AM dependence has been implemented  by the cranking technique which entails the time reversal symmetry breaking of the HFB w.f. and single particle alignment. The suitability of this procedure has been shown in the cranked HF \cite{PRC_76_044304_2007} (HFB  \cite{HHR.82,ETY.99}) plus AM projection for Yrast states. 
 In these calculations the  constrained HFB equation, Eq.~(\ref{E_Lagr_ome}) with the cranking condition  
  Eq.~(\ref{cra_cond}),  was solved and
 subsequently the AMP was performed. In general the constraint of Eq.~(\ref{cra_cond}) is used in spite of the fact that, according to its derivation, it is only valid for large, well deformed nuclei with approximate axial symmetry \cite{Ka.68}.  Specially critical is the situation in GCM calculations where one has to consider $(\beta,\gamma)$ deformations of all values.  In this case the condition of Eq.~\ref{cra_cond} does  not apply for most of the points and the best one can do is to avoid the constraint (\ref{cra_cond}) working with fix $\omega$ values.  The optimal solution is the consideration of the angular frequency as a generator coordinate as a generalization of the Peirls-Thouless double projection method \cite{PT-62,Eg-83} . In the case that one considers the  $(\beta,\gamma)$  coordinates together with the $\omega$ one, it  looks like
\begin{equation}
|\Psi^{N,I}_{M,\sigma} \rangle =  \; \sum_{\omega,\beta,\gamma,K} f^{I}_{\sigma}(\omega,\beta,\gamma,K)  P^N P^I_{MK} \; |\phi (\omega,\beta,\gamma) \rangle \label{eq:GCM_Ansatz_bet_gam_ome}
\end{equation}
The HFB wave functions $\phi(\omega,\beta,\gamma)$ are determined by solving the PN-VAP equation 
\begin{eqnarray}
E'[\phi]   =   \frac{ \langle \phi | HP^ZP^N|\phi \rangle } {\langle \phi | P^ZP^N|\phi \rangle }
 -  \langle \phi | \omega {\hat J}_x+ \lambda_{q_0} {\hat Q}_{20}+\lambda_{q_2 }{\hat Q}_{22}|\phi \rangle,  
  \label{E_Lagr_ome-bet-gam}
  \end{eqnarray}
 the Lagrange multipliers  $\lambda_{q_{0}}$  and $\lambda_{q_{2}}$ being determined by the constraints (\ref{q0_q2_constr})
 while the $\omega$  is kept constant during the minimization process. In the $(\beta,\gamma)$
plane the probability amplitude is defined by 
\begin{equation}
|{\cal P}^{I\sigma}(\beta,\gamma)|^2= \sum_\omega |p^{I\sigma} (\beta,\gamma,\omega)|^2,
\label{coll_wf_cr}
\end{equation}
with $p^{I\sigma} (\beta,\gamma,\omega)$  provided by Eq.~(\ref{coll_wf}).

In two recent publications \cite{BRE.15,EBR.16} we have presented the first applications of this theory.  In the first one it was shown that this method describes the excitation energies of the  $2^{+}_1$ and $4^{+}_1$ levels in the $^{24-34}$Mg isotopes very well providing quantitative agreement
 with the experiment. In Ref.~\cite{EBR.16} a complete study of the nucleus $^{44}$S was performed. The calculations provided excitation energies and transition probabilities in very good agreement with the available  experimental data.  An additional comparison with complete spectroscopy results for $^{44}$S, obtained with large scale shell model diagonalization with tuned interactions,  shows the high quality of the calculations provided by the Peirls-Thouless double projection method. 
 In Ref.~\cite{Mafer05}  this method was applied to superconductive grains in the particle number case, again with very good results confirming  the high performance of the double projection method.
 
\begin{figure*}[tb]
\begin{center}
\includegraphics[angle=0,scale=0.65]{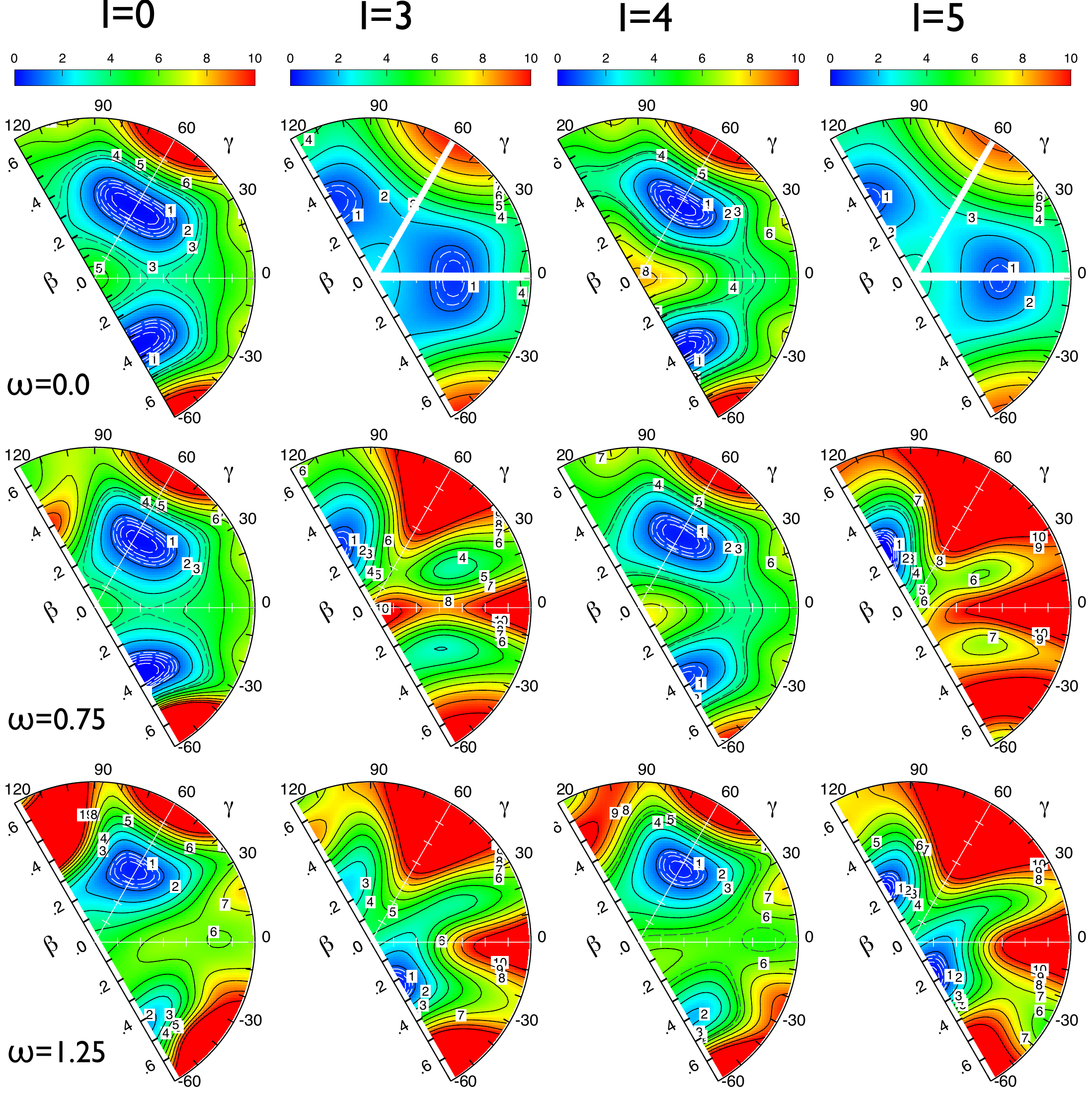}
\end{center}
\caption{(Color Online)  PES of $^{42}$Si in the PNAMP approach for the indicated angular momentum (in $\hbar$) and angular frequency $\hbar \omega$ in MeV in the $(\beta,\gamma)$ plane, $\gamma$ in degrees.  In each case 
the respective  minimum energy has been subtracted. Continuous contour lines  are 1 MeV apart up to a maximum of 10 MeV. To emphasize the minima  white dashed contour lines in steps of 0.2 up to 0.8 MeV have been drawn. Lastly,  to emphasize the prolate saddle point, an extra black dashed contour line between 3 and 4 MeV or 5 and 6 MeV  have been included.}
\label{fig:PES_42Si_3ome}
\end{figure*}

To illustrate how powerful the method is we have performed very simple calculations \cite{BE.15} for the Ti isotopes similar to the ones of Sect.~\ref{Sect:ASCMCBDF}. We just consider one degree of freedom, 
$\beta$,  and the cranking frequency. We do not constraint on  $\gamma$, but the calculations are obviously triaxial. The $\gamma$ values are determined self-consistently by the variational principle.  For a given 
$\beta$ and different $\omega$, in general, we obtain different $\gamma$ values increasing thereby the 
diversity in the mixing.  The configuration space comprises eight oscillator shells. Since we are only interested in the low spin region we consider only two $\hbar \omega$ values, namely $\hbar \omega= 0.0$ MeV and $\hbar \omega= 0.5$ MeV. We use the interval $0 \le \beta \le 0.6$  with a step size of  $0.05$, i.e., 13 points for $\hbar \omega= 0.0$ MeV and 12 points for $\hbar \omega= 0.5$ MeV (the point corresponding to $\beta=0$ is excluded).   That means, we have to solve a Hill-Wheeler equation with 25 points and triaxial angular momentum projection.   In the top panel of Fig.~\ref{fig:Ti_TRSB} we show the
 excitation energies of the $2^{+}_{1}$ states for the Titanium isotopes in two approaches and   the experimental data.  The  simplest
approach is the one of Sect.~\ref{Sect:ASCMCBDF} assuming {\em axial} symmetry, i.e.,  in the calculations only $\hbar \omega =0.0$ MeV and 13 $\beta$ points are considered, these are time reversal symmetry conserving calculations (TRSC). As discussed in Sect.~\ref{Sect:ASCMCBDF}.  These calculations compared with the experiment  provide the right behavior of the energy for the different isotopes but with too large values.  In the second calculation we add the 12 points corresponding to $\hbar \omega= 0.5$ MeV. These are TRSB calculations and a triaxial angular momentum projection must be performed. As we can observe in Fig.~\ref{fig:Ti_TRSB}, the energy lowering is very significant bringing the theoretical results almost in agreement with the experimental ones, i.e., the factor 0.7 introduced in Sect.~\ref{Sect:ASCMCBDF} is not needed anymore. Another aspect of the SCCM calculations which causes some trouble,  is that in general they provide larger collectivity than
experimentally observed.  In the bottom panel of Fig.~\ref{fig:Ti_TRSB} we show the $B(E2; 0^{+}_1 \longrightarrow 2^{+}_1)$ values for the Titanium isotopes in the same two
approximations as before.  The TRSC calculations provide $B(E2)$ values that are too high as compared with the experiment. The TRSB, however, decreases  these values considerably and a very good agreement is obtained

 As an additional application of the method we consider the  $N=28$ isotones because they are very interesting. It  presents many exotic features like shape coexistence, disappearance of old magic numbers, etc. The nucleus  $^{44}$S has been discussed in Ref.~\cite{EBR.16}.  Another very interesting nucleus is  $^{42}$Si, with 14 protons and 28 neutrons.  Long ago a discussion started on whether the weakening of the $N= 28$ shell closure will  cause an enhancement of nuclear collectivity, or whether the shell stability will be restored owing to a possible doubly magic structure.  An early theoretical study of this nucleus from 2002 with the GCM and axial AMP \cite{RGE.02}  predicted a strong oblate deformation for the ground state and a prolate one for the first excited band.  On the experimental side, a study of $^{42}$Si using a two-proton removal reaction with a radioactive $^{44}$S beam \cite{FRI.05}  was interpreted as evidence for a large $Z=14$ sub-shell gap, indicating a nearly spherical shape and a doubly closed-shell structure for $^{42}$Si. Contrary to this result, a disappearance of the $N=28$ spherical shell closure around $^{42}$Si was concluded from other experimental studies performed at GANIL \cite{BG.07} and Riken \cite{Ta.09}.  The earlier theoretical predictions were  reconfirmed by several recent studies with shell-model \cite{NP.09,UO.12} and further mean-field approaches \cite{LY.11}. 
 
 Our  GCM calculation for $^{42}$Si from  2002 \cite{RGE.02} was performed in one dimension ($\beta$) with axial AMP and without PNP. 
 The success of these calculations is a good reason to investigate this nucleus with the state of the art
 of the BMFTs, namely, triaxial calculations, PNVAP approach for the determination of the mean field
 wave functions and breaking of the time reversal symmetry. 
  In the calculation, as before,  the finite range density-dependent Gogny interaction with the D1S parametrization \cite{BERGNPA84}  is used together with a configuration space of eight harmonic oscillator shells, large enough for realistic predictions for $^{42}$Si. Concerning the generator coordinates we take three values of the angular frequency, namely, $\hbar \omega =0.0, 0.75$ and $1.25$ MeV, a discussion on this convergence will be given in Ref.~\cite{RBE.16}. For each $\hbar\omega$ value we take 70 points in the $(\beta,\gamma)$ plane, defined by $0\le \beta \le 0.7$ and $-60^{\circ}\le \gamma \le 120^{\circ}$ -see Fig.~\ref{fig:PES_42Si_PNVAP}. We have to consider this larger $\gamma$ interval instead of the usual $0^{\circ}\le \gamma\le 60^{\circ}$ because, due to the term $-\omega {\hat J}_x$ in Eq.~(\ref{E_Lagr_ome-bet-gam}), the HFB w.f. $|\phi\rangle$ is not time reversal invariant~\cite{BRE.15}.  These extensions increase drastically the computational burden, typically at least by two orders of magnitude.
We notice that rotations close to $\gamma= -60^{\circ}$ and $\gamma= 120^{\circ}$ are non-collective and can excite single particle degrees of freedom.

In Fig.~\ref{fig:PES_42Si_PNVAP} we present the PES for the nucleus $^{42}$Si for angular frequencies  $\hbar \omega =  0.0, 0.75$ and $1.25 \;\hbar$ in the PNVAP approach, i.e., the w.f.s do have a sharp particle number but they are not eigenstates of the angular momentum operator.  These
calculations have been done with fixed $\hbar\omega$ values, that means without the constraint  $\langle \hat{J}_x \rangle = \sqrt{I(I+1)}$ for the angular momentum. The case  $\hbar\omega=0.0$ MeV is a special one for two reasons~: first, because the three sextants are equivalent and second, because since $\langle \hat{J}_x \rangle =\langle \hat{J}_y \rangle =\langle \hat{J}_z \rangle =0$ all $(\beta,\gamma)$ points satisfy the same constraints. For $\hbar\omega \ne 0.0$ MeV these conditions are not satisfied. In particular,
each $(\beta,\gamma)$ point may have different expectation values of the
angular momentum depending on the point and on the $\hbar \omega $ value.
In the left top panel of Fig.~\ref{fig:PES_42Si_PNVAP}, for $\hbar \omega=0$ MeV,  one observes  a  deformed oblate minimum, 1.5 MeV deep with respect to the energy of the spherical shape, with a  deformation parameter of $\beta=0.3$. This nucleus presents a rather soft PES along the $\gamma=60^{\circ}$ axis for small deformations
and very steep for deformations $\beta > 0.45$. Along the prolate axis it does not present any minimum, just a change of curvature at $\beta=0.3$. For small $\beta$ values it is not as soft as along the oblate axis but for larger  $\beta$ values it is softer.  In the triaxial direction  the PES is very steep.
The PES for $\hbar\omega \ne 0.0$  MeV are shown in the lower panels. We first notice that the three sextants are not equivalent anymore. 
For $\hbar =0.75$ MeV we observe that the minimum close to $\gamma= -60^\circ$ is deeper and broader than the one corresponding to  $\gamma= 60^\circ$. For $\hbar =1.25$ MeV the three sextants are significantly different, in particular the softening in the $\gamma$  direction towards  the $\gamma=120^{\circ}$ axis is relevant.

\begin{figure}[tb]
\begin{center}
\includegraphics[angle=0, width=\columnwidth]{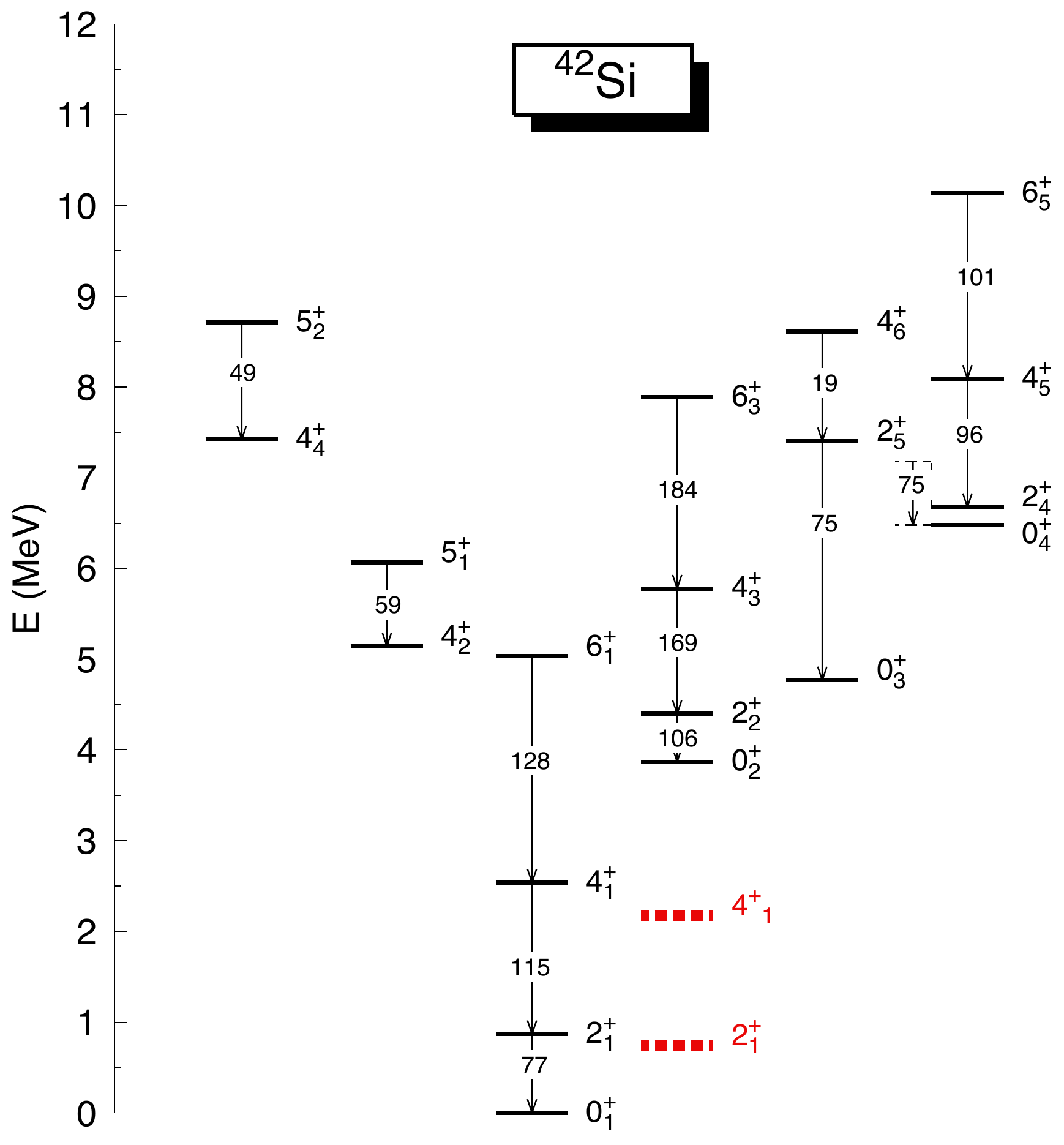}
\end{center}
\caption{(Color Online)  Spectrum of $^{42}$Si. The experimental data, thick dashed lines, are taken from Refs.\cite{BG.07,Ta.09}.}
\label{fig:42Si_spec_th}
\end{figure}

The next step is the simultaneous projection of the particle number and the angular momentum for the possible values of the triads  $(\omega,\beta,\gamma)$. That means, we have to solve an equation similar to Eq.~(\ref{HW_K}), but now considering also the $\omega$ degree of freedom. The results are plotted in Fig.~\ref{fig:PES_42Si_3ome} where we present the PES for the nucleus $^{42}$Si for angular momenta $I = 0, 3, 4, 5 \hbar$  and angular frequencies $\hbar \omega =  0.0, 0.75$ and $1.25 \hbar$.  In this figure we  can learn about the effect of the AMP on the surfaces of Fig.~\ref{fig:PES_42Si_PNVAP}. We discuss first $\hbar \omega =  0.0$ MeV (first row)  for increasing values of the AM. Since for  
$\hbar \omega =  0.0$ MeV the three sextants are equivalent we just concentrate on the $0^\circ\le\gamma\le 60^\circ$ region. Concerning the even $I$-values we find  for $I = 0\;\hbar$  a  5 MeV deep minimum at $\beta=0.35$  (again with respect to the energy of the spherical shape) but softer in the $\gamma$ direction than in the PNVAP case. Along the prolate axis we find a saddle point at $\beta=0.28$ around 3.5 MeV above the oblate minimum. The PES for  $I = 2\;\hbar$ is very similar to the $I = 0\;\hbar$ case and has not been plotted.  For $I = 4\;\hbar$ we find that the PES,  as compared with the one at $I = 0\;\hbar$,  has the oblate minimum about three MeV deeper. It is less soft in the $\gamma$ direction and its prolate saddle point is shifted to larger deformations ($\beta\approx 0.5$).  
The relative energy between the saddle point and the minimum is more or less the same
as  for $I = 0\;\hbar$. 

Concerning the odd $I$-values we first mention that the points close to axiality have been removed because their norm is close to zero. The PES for $I = 3\;\hbar$ is very different compared to the $I = 0\;\hbar$ case. Now the minimum appears close to the prolate axis at $\beta\approx0.35$ and the saddle point close to the oblate one at $\beta\approx0.22$. The PES is softer in the $\gamma$ degree of freedom than in the  $I = 0\;\hbar$ case.  With respect to  $I = 5\;\hbar  $,
we find that the PES is similar to the one for  $I = 3\;\hbar$, the only relevant point is that the minimum gets deeper.

We now turn to the intermediate frequency value $\hbar \omega = 0.75$ MeV, the  corresponding  PESs are plotted in  the panels of the second row of Fig.~\ref{fig:PES_42Si_3ome}.  We start with the even $I$-values. We  notice that now the PES is different in the three sextants.  For $I = 0\;\hbar$  the collective sextant $0^\circ\le\gamma\le 60^\circ$ does not change much as compared with the $\hbar \omega = 0.0$ MeV case and $I=0\hbar$. The only noticeable difference is that the surface is somewhat softer for small $\beta$ values. Larger differences appear for the sextant $60^\circ\le\gamma\le 120^\circ$. Here, close to the $\gamma = 120^\circ$ symmetry axis some single particle states have aligned making  energetically costly to project on $I = 0\;\hbar$. This effect causes a compression of the contour lines around $\gamma = 90^\circ$ not observed at $\hbar \omega = 0.0$ MeV. Concerning the sextant $-60^\circ\le\gamma\le 0^\circ$ we observe a compression of the contour lines at $\beta\approx 0.5$ and $\gamma$ close to $-60^\circ$ probably caused by the same reasons.
For $I=4 \hbar$ and $\hbar \omega = 0.75$ MeV and compared with the same spin and $\hbar \omega = 0.0$ MeV, we find a general softening for smaller $\beta$ values indicating that it is less energetically costly to project to AM $I=4 \hbar$ for collective aligned  (cranked) states than for non aligned ones. We also see that the wedge of  $I=0 \hbar$ and $\hbar \omega = 0.75$ MeV at $\gamma=120^\circ$ as well as the contour line compression in the lower sextant have disappeared as expected.  We now turn to odd $I$-values. For $I=3 \hbar$ and $\hbar \omega = 0.75$ MeV the effect of cranking on the PES is very large as compared with the $\hbar \omega = 0.0$ MeV case. Both PESs are rather different, furthermore the  sextant $60^\circ\le\gamma\le 120^\circ$ is completely different from the other two. Now the minimum is at $\beta\approx 0.3$ close to the axially symmetric $\gamma=120^\circ$ axis. There are also two local minima about 4 MeV above corresponding to triaxial shapes in the other two sextants.
For $I=5 \hbar$ we find a reinforcement of the points commented for $I=3\hbar$.  
We lastly turn to the high angular frequency limit of $\hbar\omega=0.125$ MeV. For $I=0 \hbar$  we find that the single particle alignments make even more energetically costly to project to zero angular momentum. The minimum close to $\gamma=-60^\circ$ has now shifted to higher deformations  and to higher energies as compared with smaller angular frequencies. Also relevant is the presence of a local minimum at $\beta\approx 0.5$ on the $\gamma=0^\circ$ axis, which is the natural evolution of the softening observed for the smaller $\hbar\omega$ values. The $I=4\hbar$ is very similar to the $I=0\hbar$ case.   Concerning the odd $I$-values 
we find in the PES  for $I=3 \hbar$  major changes. First the triaxial minima have disappeared, second the absolute minimum is now close to $\gamma=-60^\circ$ and third,  the minimum close to $\gamma=120^\circ$ is now a secondary minimum 3 MeV above the other one.  For $I=5\hbar$ the only difference with respect to $I=3\hbar$ is that now both minima are more or less at the same energy.

\begin{figure}[tb]
\begin{center}
\includegraphics[angle=0, width=.7\columnwidth]{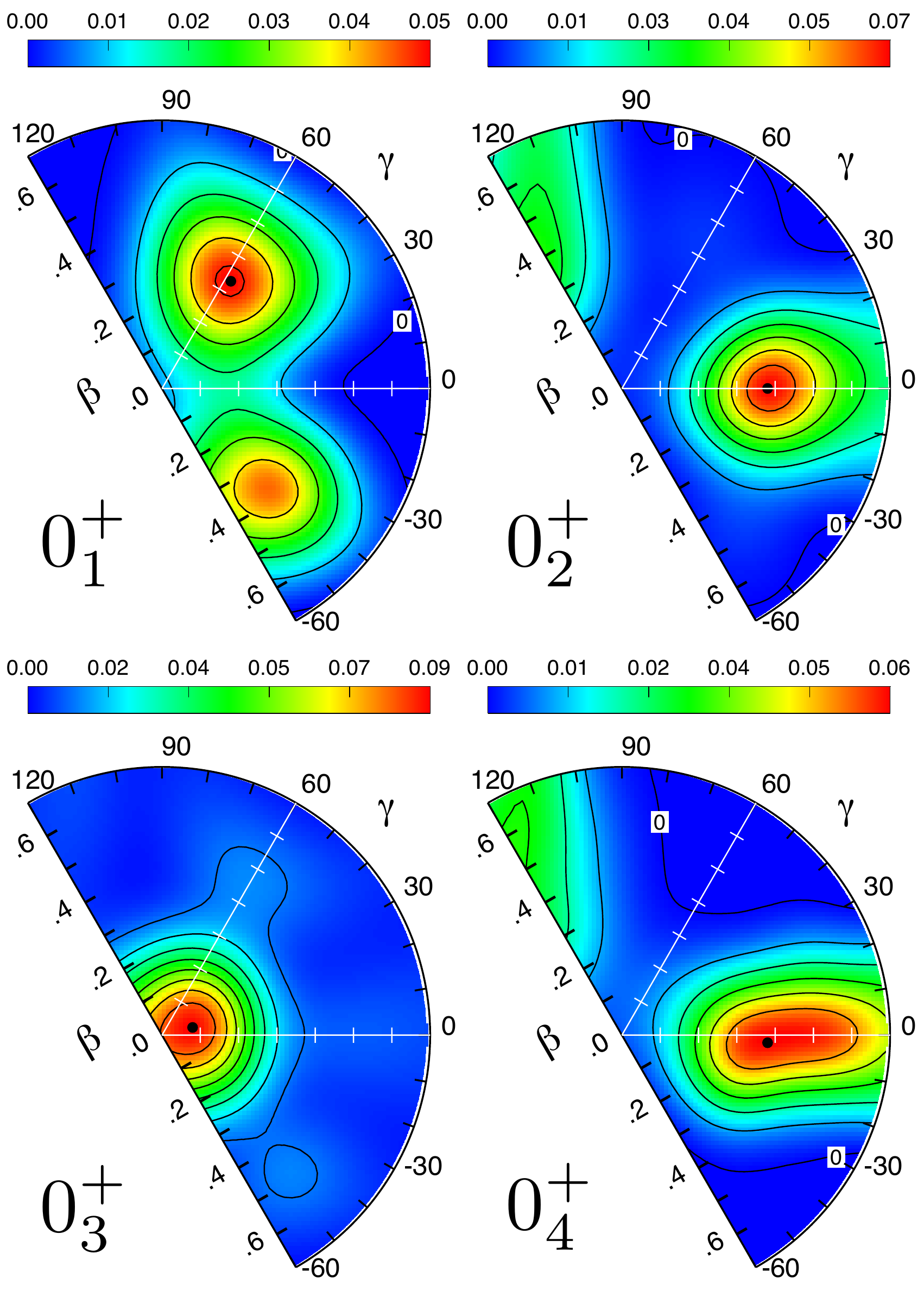}
\end{center}
\caption{(Color Online)  Collective wave functions of the $0^+$ states of the spectrum of the nucleus $^{42}$Si. The scale for each wave function is shown in the corresponding panel. Contour lines are 0.01 units apart.}
\label{fig:42Si_WF_0+}
\end{figure}

The next step is the solution of the Hill Wheeler equation, Eq.~(\ref{HW}),  to obtain the eigenstates and the wave functions of the ground and the excited states. The latter ones will allow the calculation of the transition probabilities, Eq.~(\ref{BE2_GCM_AM-cojo}), and to classify the excited states into bands.
In Fig.~\ref{fig:42Si_spec_th} we plot the spectrum of $^{42}$Si obtained under these premises.
We find four well differentiated $0^+$ bands. Their collective wave functions, see Eq.~(\ref{coll_wf_cr}), are plotted in Fig.~\ref{fig:42Si_WF_0+}.  The ground state band is based on the state $0^+_1$ and shows a clearly rotational spectrum. The wave function of the ground state is plotted in the top-left panel of Fig.~\ref{fig:42Si_WF_0+}. It presents clear maxima 
for the oblate shapes at $\beta\approx 0.35$. It is amazing the wave function decomposition in the cranking frequencies $0.0, 0.75$ and $1.25$  MeV which is $37\%$, $47\%$ and $16\%$, respectively.  A large amount of mixing in spite of being $I=0\hbar$. Looking at the first column of Fig.~\ref{fig:PES_42Si_3ome} it is clear that this wave function corresponds to the  oblate minimum found in this plot. As one can observe in this plot the minimum at $\gamma= -60^\circ$  looses relevance as compared with the one at $\gamma= 60^\circ$ with increasing $\hbar \omega$. This has as a consequence that the concentration of the wave function is larger around $\gamma= 60^\circ$. Also remarkable is the fact that the maximum of the wave function close to $\gamma= -60^\circ$ does not appear exactly on the axis, as it is the case for  $\gamma= 60^\circ$. This has to do with the fact that rotations around symmetry axes are not allowed in quantum mechanics, i.e., only the part of $\hbar\omega=0$ MeV contributes to this
axis. The extension of the w.f. and the broad separation of the contours indicate that this is a collective state.  

The second band is build on the $0^+_2$ state at about 4 MeV excitation energy. It is also a well developed rotational band. Its collective wave function is shown in the top-right panel of Fig.~\ref{fig:42Si_WF_0+}. The wave function decomposition in the cranking frequencies $0.0, 0.75$ and $1.25$  MeV  is $46\%$, $44\%$ and $10\%$, respectively.
 It corresponds to well deformed prolate shapes, the one at $\gamma=0^\circ$
and to a  less extent the one at $\gamma=120^\circ$.  The latter one contributes much less to the w.f. because only the contribution of zero cranking frequency makes sense. Looking at the potential energies of Fig.~\ref{fig:PES_42Si_3ome} this w.f. corresponds to the saddle points found at these plots at the points where the w.f. peaks. The third ''band" is build on the $0^+_3$ state and does not show a typical rotational pattern. A look at Fig.~\ref{fig:42Si_WF_0+} shows that it corresponds to a weakly deformed state, $\beta\approx 0.1$,  whose counterpart is the saddle point close to sphericity in Fig.~\ref{fig:PES_42Si_3ome}. Lastly the fourth band at about 6 MeV excitation energy  is  build on the $0^+_4$ state it corresponds to a prolate shape extending from $\beta=0.3$ up to $\beta=0.6$, see Fig.~\ref{fig:42Si_WF_0+}.
\begin{figure}[tb]
\begin{center}
\includegraphics[angle=0, width=\columnwidth]{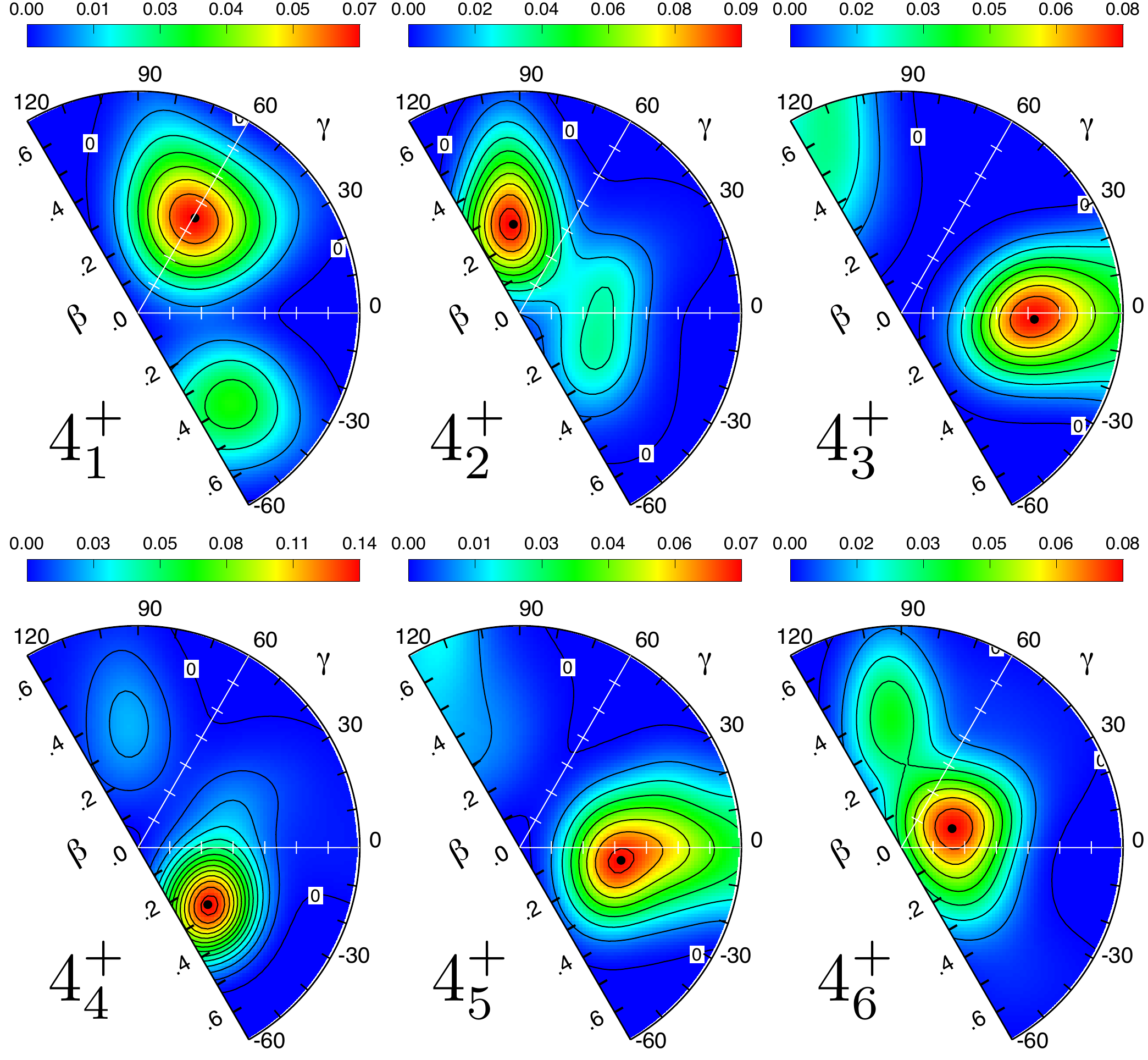}
\end{center}
\caption{(Color Online)  Collective wave functions of the $4^+$ states of the spectrum of the nucleus $^{42}$Si. The scale for each wave function is shown in the corresponding panel. Contour lines are 0.01 units apart.}
\label{fig:42Si_WF_4+}
\end{figure}

Besides these $0^+$ bands we find two additional side bands. We can  clearly assign the $4^+$, $5^+$ sequences while the $2^+, 3^+$ and $6^+$ states are very mixed and therefore difficult to ascribe to one or another. The wave functions of the $4^+_2$ and $4^+_4$  states of these bands are shown in Fig.~\ref{fig:42Si_WF_4+} together with the $4^+$ states of the other bands. We observe that the contours of the $4^+_2$ and $4^+_4$  states are narrower than the ones of the $4^+$ states of the other bands indicating a much less collective character. The wave function of the $4^+_2$ and $4^+_4$  states peak close to the
symmetry axis. Wave functions similar to these were found in $^{44}$S, see Ref.\cite{EBR.16}, and were identified as aligned states obtained by rotations close to symmetry axes.  

Concerning the comparison with the scarce experimental values for the energies, see thick dashed lines in  Fig.~\ref{fig:42Si_spec_th}, we find that our values are slightly too high. The $E2$ transition probabilities along the bands,  see Eq.~(\ref{BE2_GCM_AM-cojo}),  has been also plotted in Fig.~\ref{fig:42Si_spec_th}.
Unfortunately no experimental values are available yet.

\section{Summary and outlook}
\label{Sect:CONC}

In summary, we have presented a detailed description of the different approaches of BMF
theories. Starting with the formulation of the plain mean field we have incorporated the angular momentum and particle number projectors in the theory in the so-called symmetry conserving mean field approach. The correlations beyond mean field have been formulated in the configuration mixing approach
using the generator coordinate technique to produce wave functions corresponding to the different physical situations.  The Hill-Wheeler equation has also been discussed as well as the interpretation of its eigenvalues and eigenvectors. \\
   Several illustrative examples have been presented, starting with the simplest symmetry conserving configuration mixing  in an axially symmetric calculation with just one coordinate to discuss the exotic heavy Titanium isotopes. Later on we have performed a detailed  discussion about the characterization of pairing fluctuations and their influence on the spectrum of $^{52}$Ti.  The general case of triaxial angular momentum projection and $\beta$ and $\gamma$ fluctuations has been studied for the nucleus $^{24}$Mg and compared with the axially symmetric calculations. The incorporation of the $\gamma$
degree of freedom favors the presence of the $\gamma$ band, absent in the axial case, and in general produces a small compression of the spectrum.  Next, the  consideration of the cranking frequency $\hbar \omega$ as a coordinate together with $\beta$ and $\gamma$ brings us to the most general GCM calculations performed so far with effective forces. Two examples are considered: In the Titanium isotopes  the behavior of the excitation energies of the $2^{+}_{1}$ states and their decay  to the ground state is analyzed. We found a lowering of the otherwise too high energies.  Also a large effect on the transition probabilities is obtained. In both cases the effect of considering  $\hbar \omega$ as a coordinate brings the theoretical results to a much better agreement  with the experimental data. As a second example a full calculation was performed for the $N=28$ exotic nucleus $^{42}$Si.  The full spectrum together with the transition probabilities has been obtained. Several bands are found: an oblate ground state band, two prolate bands as well as two aligned bands. \\ 
Finally, in Appendix B a thorough discussion on the potential divergences in density dependent calculations is given and in  Appendix C a very detailed discussion on the need of projection is presented.

Concerning the outlook, it has been shown that the most sophisticated theory using the coordinates $\beta, \gamma$ and $\hbar \omega$ is able to provide high quality spectra and transition probabilities. The main drawback is the large computational time needed to perform the calculations, several weeks for a small nucleus and not too many oscillator shells (six to eight) in a cluster with about 250 cores.  The calculation of larger nuclei with more shells or the calculations of many nuclei seems an arduous task. There are two ways out of this situation, either the use of supercomputers with very large number of CPUs or to reduce considerably the required CPU time. We are convinced that one can  substantially reduce the CPU time in GCM calculations and we are working in that direction.  The success of this work will strongly condition further developments in the future.  

Though not presented in this work, the study of odd nuclei with the coordinates $\beta, \gamma$ and $\hbar \omega$ and the Gogny force has already started and the first results of the calculations will be published soon.  From the computational  and man power requirements
 odd nuclei are much more demanding than even-even ones but also offer the possibility of learning about many aspects of the nuclear many body system.
  
  Though the consideration of the cranking frequency allows the incorporation of single particle degrees of freedom, genuine,  pure two quasiparticle states are not considered in the present state-of-the-art calculations with effective forces. The techniques developed for odd nuclei will help to explore this interesting feature.  All these improvements open many new possibilities that will allow to investigate more and more challenging features of nuclear structure physics.


\begin{ack}
The author gratefully thanks  M. Borrajo, N. L\'opez Vaquero and T. R. Rodr\'{\i}guez for their collaboration in different parts of this article and Fang-Qi Chen for the collaboration in  Appendix B.  Discussions on $^{42}$Si with A. Poves are gratefully acknowledged. This work has been supported from the Spanish Ministerio de  Ciencia e Innovaci\'on under contracts   FPA2011-29854-C04-04 and FPA2014-57196-C5-2-P.

\end{ack}

\appendix
\section{Appendix A:  Acronyms}
\label{App:A}
 As a guide for the reader we put together all  acronyms
  in alphabetical order.
\begin{description}
\item[AM] Angular momentum
\item[AMP]   Angular Momentum Projection
\item[AM-PAV]   Angular Momentum in the Projection After Variation
\item[AM-VAP]   Angular Momentum in the  Variation After Projection
 \item[BCS]    Bardeen Cooper Schrieffer
\item[BMFT] Beyond Mean Field Theory
\item[CM] Configuration Mixing
\item[GCM] Generator Coordinate Method
\item[HF]     Hartree Fock
\item[HFB]    Hartree Fock Bogoliubov
\item[PAV]   Projection After Variation 
\item[PES]  Potential energy surface
\item[PN] Particle number
 \item[PNAMP] Particle number and angular momentum projection
\item[PNP] Particle Number Projection
\item[PN-PAV]  Particle Number in the Projection  After Variation
\item[PN-VAP]  Particle Number in the Variation After Projection
\item[SCCM] Symmetry Conserving Configuration Mixing
\item[VAP]    Variation After Projection
\item[TRSB] Time Reversal Symmetry Breaking
\item[TRSC] Time Reversal Symmetry Conserving
\end{description}
\section{Appendix B: Peculiarities of projected theories and the GCM with the interaction}
\label{App:B}

In Ref.~\cite{AER.01P} it  was shown that in calculations with  a density dependent interaction and in a particle number projected  approach there are two sources for divergences.  The first one is connected  with the omission  of exchange terms. Obviously, these divergences can  be straightened out by including {\em all} missing exchange terms of the interaction. 	The second one has its origin in the density dependent term of the interaction which we shall call $V_{DD}$.  

The ultimate cause for the divergences is that the norm overlap for rotations in the gauge space associated to the particle number operator vanishes under some conditions.  Now, the question is,  if this could also happen in other cases,  for example, with AMP. Though in principle this is possible it seems to be rather unlikely. The reason could be related to the fact that the particle number operator is diagonal in the canonical basis, and in its representation only apper the occupations $\{u_k,v_k\}$ which are the essential ingredients of the norms. This is not the case with the angular momentum operator and as a consequence the potential vanishing of the overlaps only could take place  in very specific situations.

 In this appendix we discuss some aspects of the theory related with the exchange terms of the  interaction and/or a density dependent force, in particular of the Gogny one.  

\begin{figure}[htb]
	\begin{center}
		\includegraphics[angle=0, scale=.4]{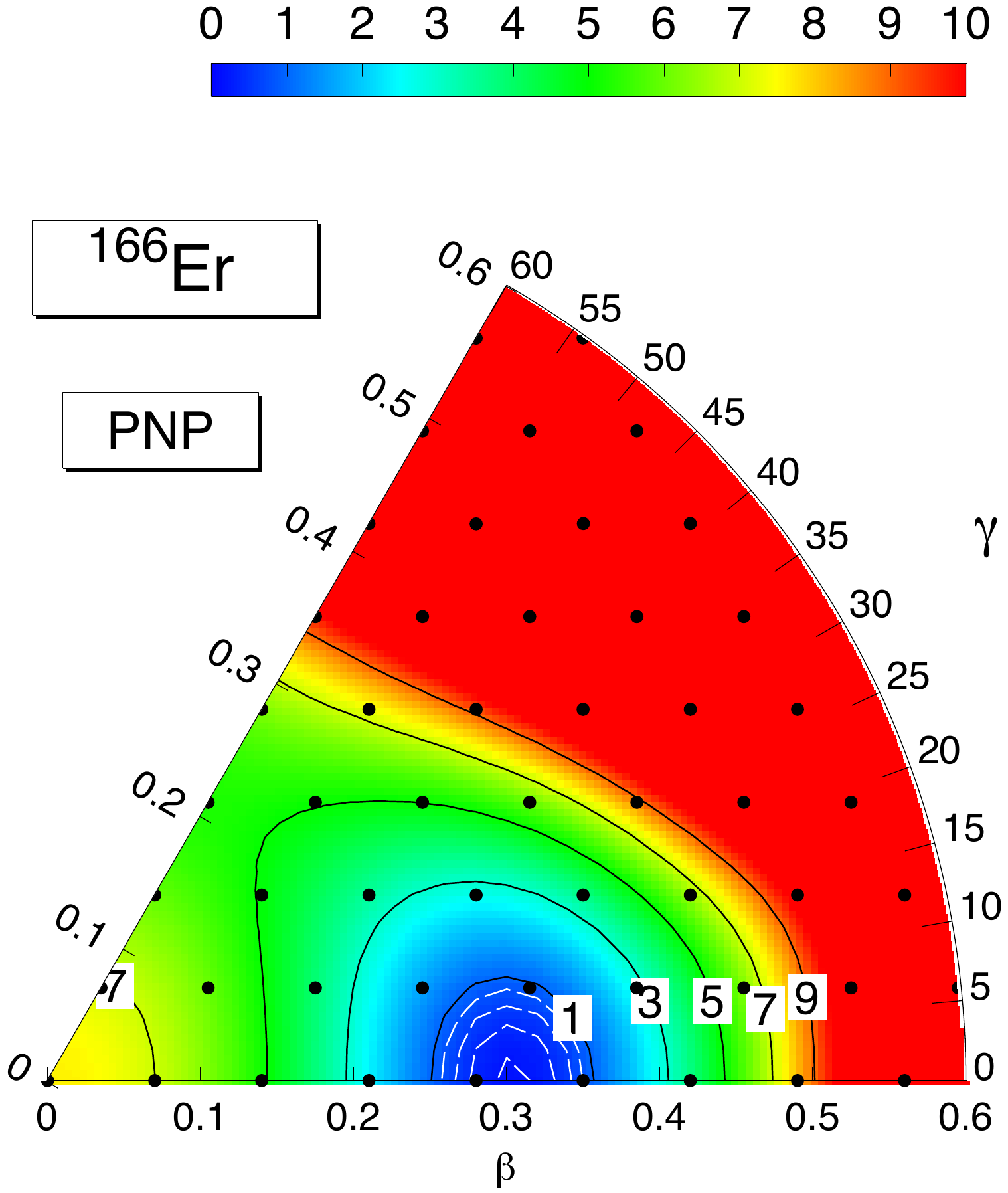}
	\end{center}
	\caption{(Color Online)  Particle number projected PES for the nucleus $^{166}$Er without exchange terms. The bullets represent the mesh points used in the calculations.}
	\label{fig:PNP_mesh}
\end{figure}

\subsection{ The Gogny Force.}

\label{appen:GogF}

In the calculations we used  the Gogny interaction \cite{Go.75} as the effective force.
The main ingredients of this force are the phenomenological density dependent
term which was introduced to simulate the effect of a G--matrix interaction and
the finite range of the force which allows to obtain the Pairing and 
Hartree--Fock fields from the same interaction.  
We use the parametrization D1S, which was fixed by 
Berger {\em et al.}~\cite{Ber91}. The force is given by
\begin{eqnarray}
v_{12} & = & \sum_{i=1}^2 e^{-{(\vec{r}_1 - \vec{r}_2)}^2/\mu_i^2} (W_i +
B_iP_{\sigma} -H_i P_{\tau} -M_i P_{\sigma} P_{\tau} ) + \nonumber \\
& + & W_{LS} (\vec{\sigma}_1 + \vec{\sigma}_2) \vec{k} \times \delta(\vec{r}_1 -
\vec{r}_2) \vec{k} + V_{DD},
\label{eq:vgog}
\end{eqnarray}
and the Coulomb force
\begin{equation}
v_{12}^C = (1+2\tau_{1z})(1+2\tau_{2z}) \frac{e^2} {|\vec{r}_1-\vec{r}_2 | }.
\end{equation}
The density dependent part of the interaction is provided by
\begin{equation}
V_{DD} =  t_3(1+x_0 P_{\sigma}) \delta (\vec{r}_1 -\vec{r}_2) \rho^{1/3} \left (
\frac{1}{2} (\vec{r}_1 + \vec{r}_2 )\right ),
\label{Eq:VDD}
\end{equation}
and the density operator, $\hat{\rho} (\vec{r})$, is given by 
\begin{eqnarray}
\hat{\rho} (\vec{r}) &=&  \sum_{i=1}^{A} \delta (\vec{r} -\vec{r}_i)  =
\sum_{ij} \phi_i^* (\vec{r}) \phi_j (\vec{r}) \langle S_i | S_j \rangle
c_i^{\dagger}c_j \nonumber \\ 
&= &\sum_{ij} f_{ij} (\vec{r}) c_i^{\dagger} c_j \, .
\label{rhodef}
\end{eqnarray}
In the two-body interaction used in the calculations we also include 
the one--body and two--body  center  of mass corrections.
\begin{equation}
\hat{T}  =  \sum_i \frac{{\vec{p}_i}^{\,2}}{2m} \left ( 1 -\frac{1}{A} \right ) -
\frac{1}{Am} \sum_{i>j} \, \vec{p}_i \cdot \vec{p}_j
\end{equation}

\begin{figure*}[tb]
	\begin{center}
		\includegraphics[angle=0, width=\columnwidth]{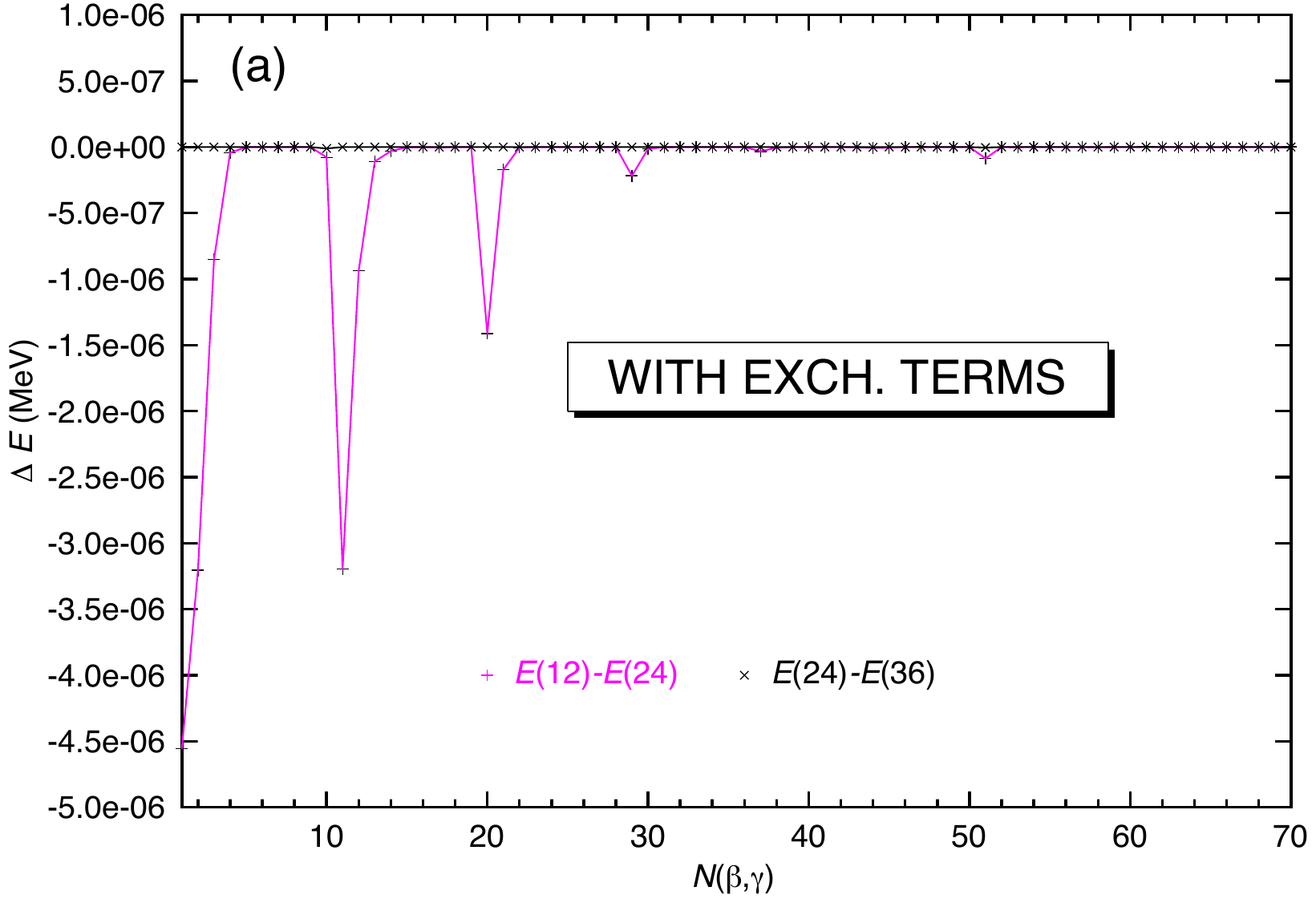}
	\includegraphics[angle=0, width=\columnwidth]{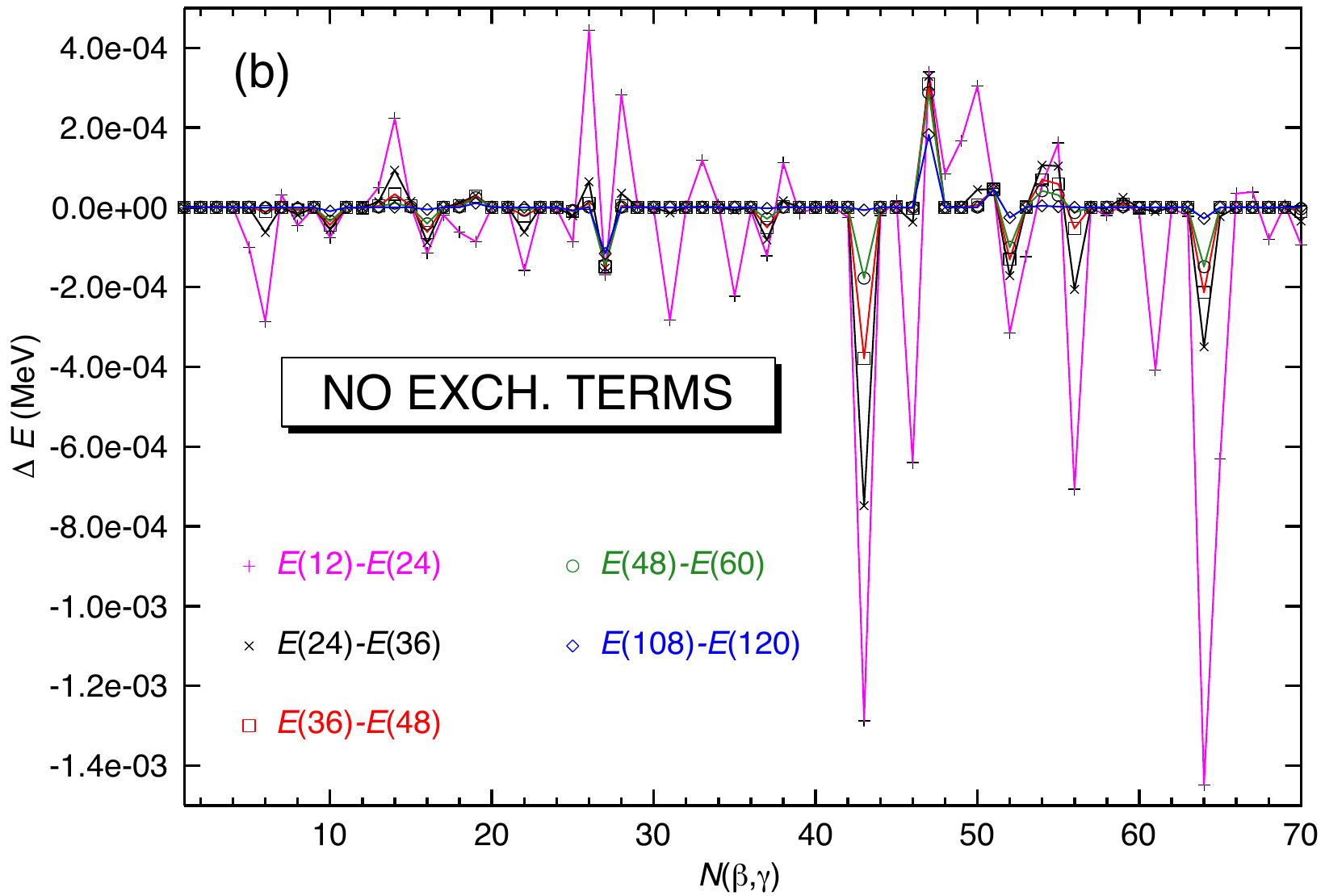}
	\end{center}
	\caption{(Color Online)  Convergence study of the PNP with the number of Fomenko points used in the integracion of Eq.~(\ref{eq:PN}) in the evaluation of the PNP energy. (a) with exchange terms, (b) without exchange terms.}
	\label{Fig:Fom_convg}
\end{figure*}

\subsection{Details on the exchange terms}

Traditionally in calculations with effective forces some exchanges terms have been neglected or calculated in an approximate way.  Sometimes, like in the Skyrme force, this happens because the interaction has two components, one in the particle-hole (p-h) and another in the particle-particle (p-p) channel. In this case the contributions of exchange terms of the p-h (p-p)  part of the interaction to the p-p (p-h) are  neglected.
Other terms which are often neglected are the contribution of the Coulomb force  to the pairing channel  or the exchange terms of the spin-orbit part of the force among others. Lastly other terms, like the Fock term of the Coulomb force, are calculated in an approximate way.

In Ref.~\cite{AER.01P} it was demonstrated that in particle number projected theories the neglect of exchange terms may lead to the presence of poles in the calculations.  It was also shown that the exchange terms {\em  of all components} of the interaction were needed to have a well behaved interaction.  The use of the same interaction in the particle-particle and the particle-hole channels in the Gogny interaction makes this interaction specially attractive for BMFTs. In Ref.~\cite{AER.01E} all the exchange terms of the Gogny force were calculated and finally in Ref.~\cite{AER.01P} PNP calculations (PAV and VAP) were performed.

The presence of poles in PNP theories was shown analytically. The demonstration is easy because the number operator is a scalar and diagonal in the particle basis. In the case of the AMP the situation is different and it is not easy to isolate the potential poles. The existence and impact of potential poles associated to the AMP is still an open question. Since in the case of the Gogny force calculations are lengthy and we want to investigate the potential  poles caused by the neglect of the exchange terms,  any two body force producing reasonable results can be used to perform a detailed study. The pairing plus quadrupole (PPQ) Hamiltonian has been used 
in conjunction with the projected shell model (PSM)  with great success \cite{AXIAL,TRIAX}. We use the Hamiltonian of Ref.~\cite{FE.16_1} and perform two sets of calculations. The first one neglects all exchange terms of the force as it is usual in the PPQ calculations. The second one includes all exchange terms. 
We furthermore perform separately PNP and AMP in both sets of calculations.  According to Ref.~\cite{AER.01P} the PNP calculations without exchange terms  could present poles whereas the PNP with exchange terms cannot have poles. Concerning the AMP calculations, again, with exchange terms there cannot be any poles. Now the big questions is,  are there poles in the AMP calculations without exchange terms ?

\begin{figure*}[tb]
	\begin{center}
		\includegraphics[angle=0, width=\columnwidth]{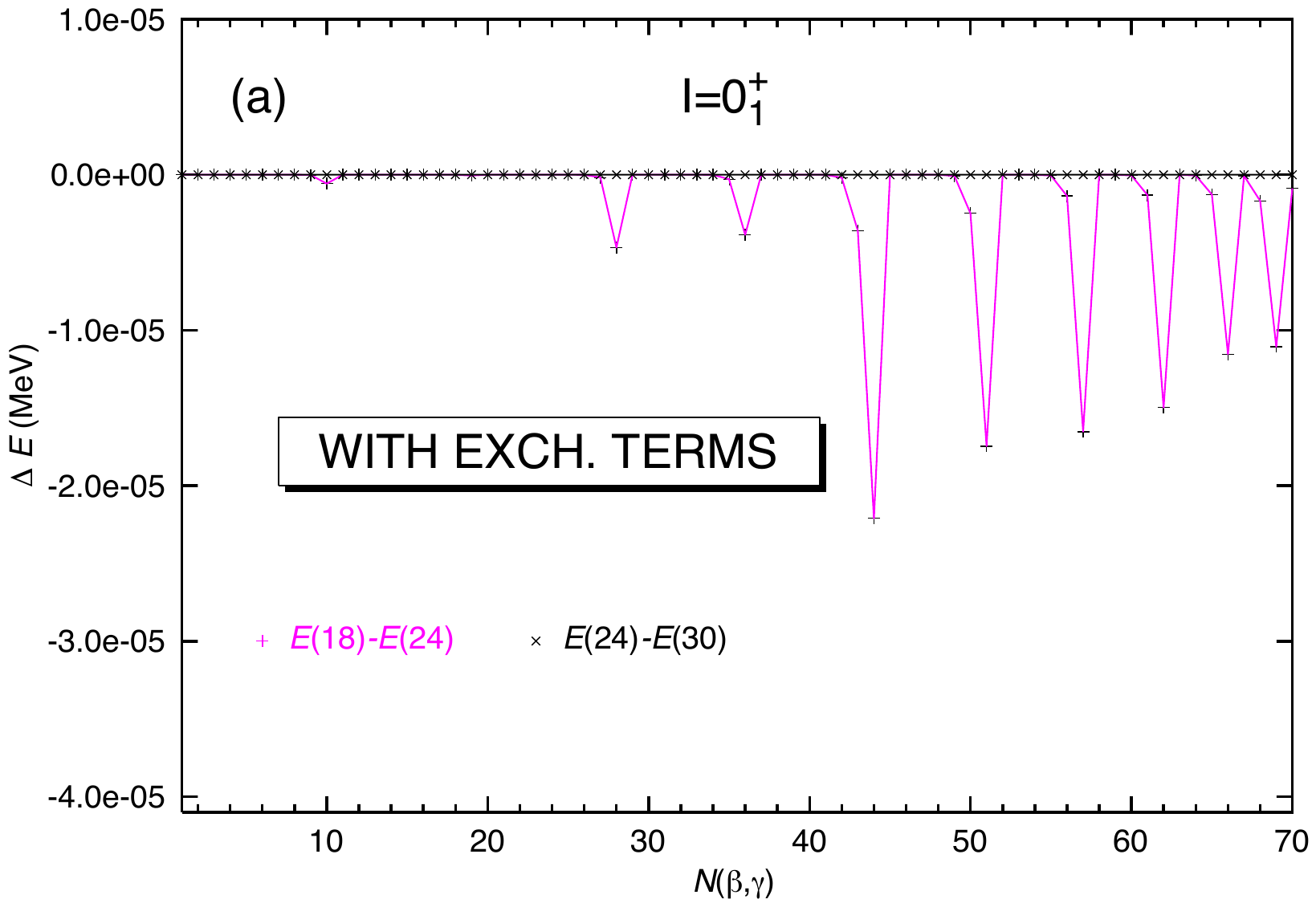}
		\includegraphics[angle=0, width=\columnwidth]{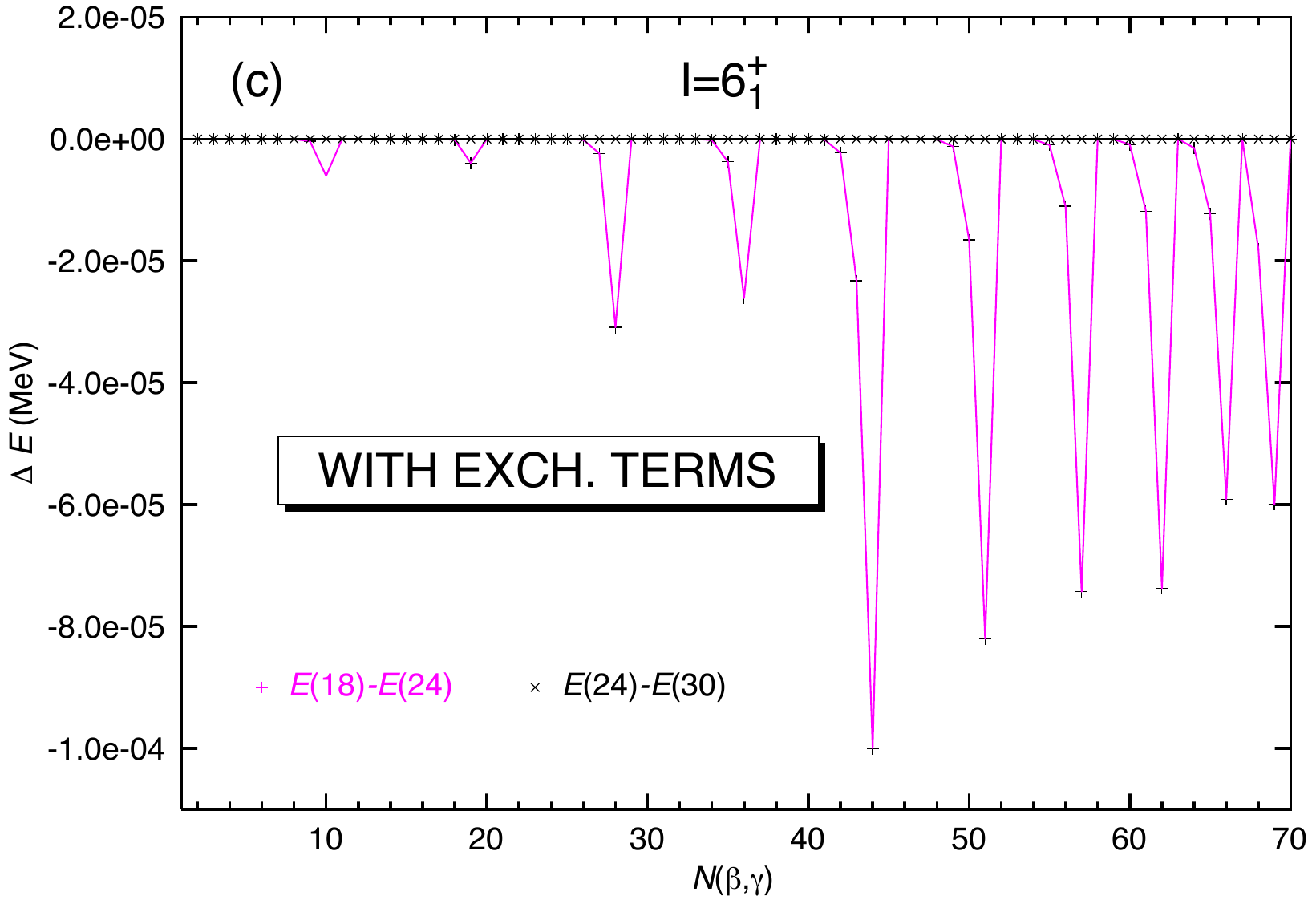}		
		\includegraphics[angle=0, width=\columnwidth]{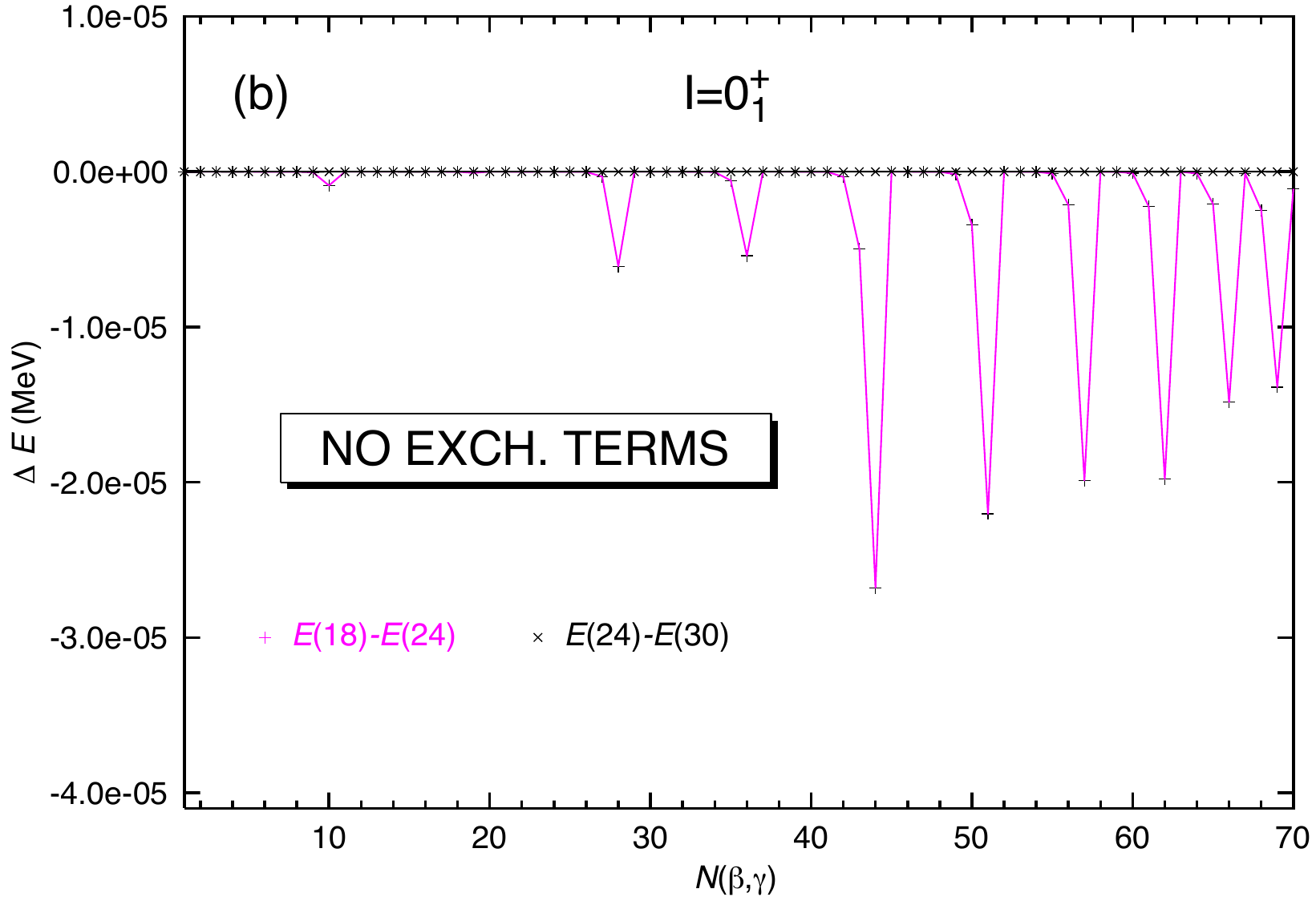}
		\includegraphics[angle=0, width=\columnwidth]{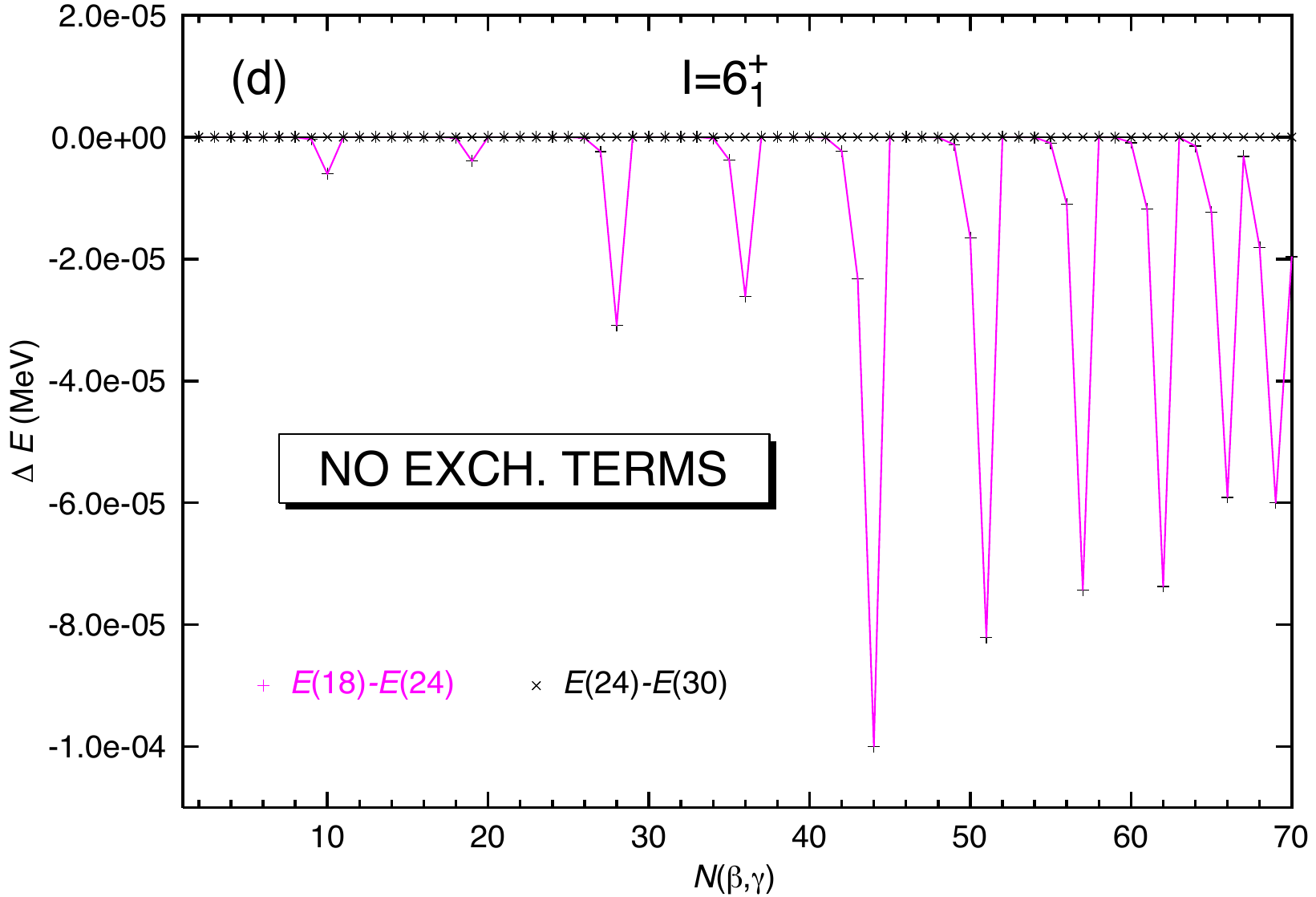}
	\end{center}
	\caption{(Color Online)   Convergence study of the AMP with the number of Euler angles  used in the integration of Eq.~(\ref{AMProj}) in the evaluation of the AMP energy, Eq.~(\ref{E_pro_I}). (a) and (c) with exchange terms, (b) and (d) without exchange terms. For $I=0^+_{1}$, left panels, and for $I=6^+_{1}$, right panels.}
	\label{fig:AMP_conv_K0}
\end{figure*}

Our model space consists of three shells for neutrons  and three shells for protons \cite{FE.16_1}, the interaction strengths have been adjusted to get an overall fit in the Er region.  We have considered a triangular mesh  of 70 points with $\beta_{max.}= 0.7$ distributed in the $(\beta,\gamma)$ plane as shown by the bullets in Fig.~\ref{fig:PNP_mesh}. The calculations  we present here are similar to the ones of Sect.~ \ref{Sect:TCBGC}. We first minimize  the constrained HFB energy
\begin{eqnarray}
{E^{\prime}}(\alpha,\beta) &=&  \langle\phi^{}(\alpha,\beta)|\hat{H}|\phi{}(\alpha,\beta) \rangle \nonumber \\
&- & \langle \phi(\alpha,\beta) |\lambda_{q_{0}}\hat{Q}_{20} + \lambda_{q_{2}}  \hat{Q}_{22} | \phi (\alpha,\beta)\rangle, \label{E_Lagr_bet-gam_hfb}
\end{eqnarray}
with the constraints of Eqs.~(\ref {q0_q2_constr}). Second we calculate the PNP  energies
\begin{eqnarray}
{E^{N}}(\alpha,\beta)= \frac{ \langle\phi^{}(\alpha,\beta)|\hat{H}\hat{P}^{N}|\phi{}(\alpha,\beta) \rangle}{\langle\phi^{}(\alpha,\beta)|\hat{P}^{N}|\phi^{} (\alpha,\beta)\rangle}. \label{E_pro_N}
\end{eqnarray}
As an example of our calculations we have chosen  the nucleus $^{166}$Er, it  PES is shown  in Fig.~\ref{fig:PNP_mesh}. The energy minimum is at $\beta=0.3, \gamma=0$, as expected in this mass region. The energy surface looks reasonably well as compared with other calculations.
 
  In Ref.~\cite{AER.01P} it was shown that for the value $\varphi=\pi/2$ of the canonical angle in the PNP, see Eq.~(\ref{eq:PN}), and for the values of  the occupancies in the canonical basis $v^{2}_k = u^{2}_k=0.5$ one would obtain divergences if the exchange terms were neglected.  Of course, for values of $\varphi$ close to $\pi/2$ and occupancies near $0.5$ one also expects spurious contributions. 
However, in case that the exchange terms were taken into account a compensation will take place and the divergence will disappear.  The integral of Eq.~(\ref{eq:PN}) is calculated using the Fomenko discretization, $\varphi_l=\pi (l+0.5)/L$ with $l=0,...,(L-1)$ and $L$ the total number of points. We choose $L$ even to avoid to have exactly $\pi/2$, i.e. only in the limit $L\rightarrow \infty$ we reach this value.  The usual way to check for divergences is to calculate the mentioned integral for different values of $L$ to see if convergence is found. We performed calculations for $L=12, 24, 36, 48, 60, 72, 84, 96, 108$ and $120$. We denote the  corresponding PNP energies by $E^{N}_{L}(\beta,\gamma)$.  To analyze the convergence we have calculated the quantities $\Delta E_{L}=  E^{N}_{L}(\alpha,\beta)-E^{N}_{L'}(\alpha,\beta)$ at each of the 70 points of the $(\beta,\gamma)$ plane shown in Fig.~\ref{fig:PNP_mesh}.  In Fig.~\ref{Fig:Fom_convg}
$\Delta E_{L}$ is plotted for the different $(\beta,\gamma)$ points numbered $1,2, ...,70$ and denoted by $N(\beta,\gamma)$ in the abscissa. In panel (a) we show the results for the case with exchange terms. The continuous lines are depicted to guide the eye. In general with $L=12$ one founds a good convergence in PNP calculations. This is what is found in panel (a),
where the largest energy difference found for $E^{N}_{12}-E^{N}_{24}$ is of the order of $10^{-6}$, for
 $E^{N}_{24}-E^{N}_{36}$ the largest differences are of the order of $10^{-9}$ or less.  In the case without exchange terms, panel (b), we find that this is not the case,  instead  a tortuous way to convergence is observed for many points.  For some of them, for example, for the point 27 (corresponding to $\beta=0.598, \gamma=5.82^\circ)$ or the 47 $(\beta=0.426, \gamma=34.71^{\circ})$, even  120 Fomenko points aren't sufficient  to find good convergence.
One must mention that in this case the divergence is not a real one, the energy difference between 12 and 120 Fomenko points amounts to at most 3 keV. The maximal  values correspond to the largest peaks in Fig.~\ref{Fig:Fom_convg}. 
 We checked in the HFB wave functions and found that in this $(\beta,\gamma)$ point we never had $v^{2}_k = u^{2}_k$. The closest value that we found was $v^{2}_k -u^{2}_k=0.01$, indicating that we only {\em see} the tail of a pole.
One must realize that the word {\em convergence} in this context is somewhat misleading since for $L \rightarrow \infty$ 
one is sure to have the full contribution of the pole, if any, but for a small $L$ is difficult to know  how much the pole contributes to the energy. We cannot compare either the energy with and without exchange terms because they are different.

We now analyze the potential presence of poles due to the AMP.  Again we calculate the projected energy
\begin{eqnarray}
E^{I}_{MK}(\alpha,\beta)= \frac{ \langle\phi^{}(\alpha,\beta)|\hat{H}\hat{P}^{I}_{MK}|\phi{}(\alpha,\beta) \rangle}{\langle\phi^{}(\alpha,\beta)|\hat{P}^{I}_{MK}|\phi^{} (\alpha,\beta)\rangle}, \label{E_pro_I}
\end{eqnarray}
with and without exchange terms.  We prefer to look for $K$-dependent projected energies rather than for $K$-independent because in the latter case one would have to solve the Hill-Wheeler equation, see Eq.~(\ref{HW_K}), for the weights $g_{\sigma}(\beta,\gamma,K)$ and  loose accuracy in the evaluation of the pole. 

In the case of AMP we only know that if all exchange terms are taken into account there cannot be any pole, but  we do not know if there are poles due to the neglect of exchange terms. To find this out we will apply the same technique as before: to calculate the AMP energy using different number of Euler angles in Eq.~\ref{AMProj}.  We take as standard values for the $(\alpha,\beta,\gamma)$ Euler angles $N_{\alpha}=9$, $N_{\beta}=18$ and  $N_{\gamma}=18$.  The values of $N_\alpha$ and
$N_{\gamma}$ are kept constant at this values and for $N_{\beta}$  we consider the values $18, 24, 30,  36, 54$ and $60$.  In panel (a)  of Fig.~\ref{fig:AMP_conv_K0} we display the results for $I=0^{+}_1$ with  exchange terms. Since the difference between 24 and 30 points is zero, we conclude that with 24 points the convergence is reached and that keeping 18 points our errors are of the order of $10^{-5}$. The results without exchange terms, panel (b), are very similar to the ones of panel (a),  they peak at the same $(\beta,\gamma)$ values and are of the same order of magnitude.  In panel (c) and (d) we show the corresponding results for the  $I=6^{+}_1$ state. Here,  again, the plots with and without exchange terms look very similar. As compared with  $I=0^{+}_1$ the convergence is a bit worse. Interestingly in all four plots the peaks appear at the same $(\beta,\gamma)$ values (with the exception of the point number 19 that does not appear for $I=0^{+}_1$). All these points, without exception, correspond to large deformations $(\beta\ge 0.63)$ and different $\gamma$ values. The  $I=6^{+}_1$  as the  $I=0^{+}_1$ one has only  $K=0$ components.  

\begin{figure}[tb]
	\begin{center}
		\includegraphics[angle=0, width=\columnwidth]{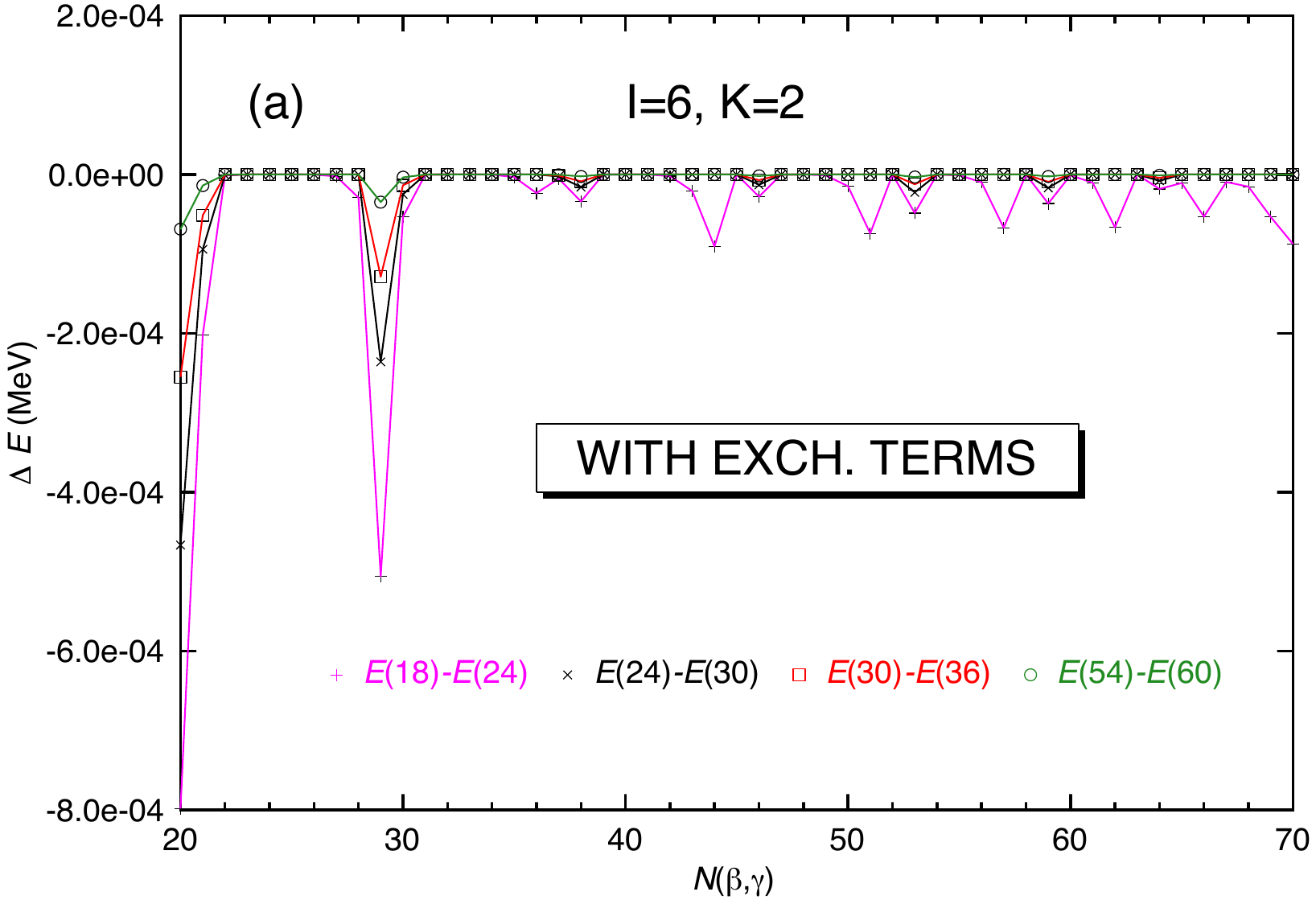}			
		\includegraphics[angle=0,width=\columnwidth] {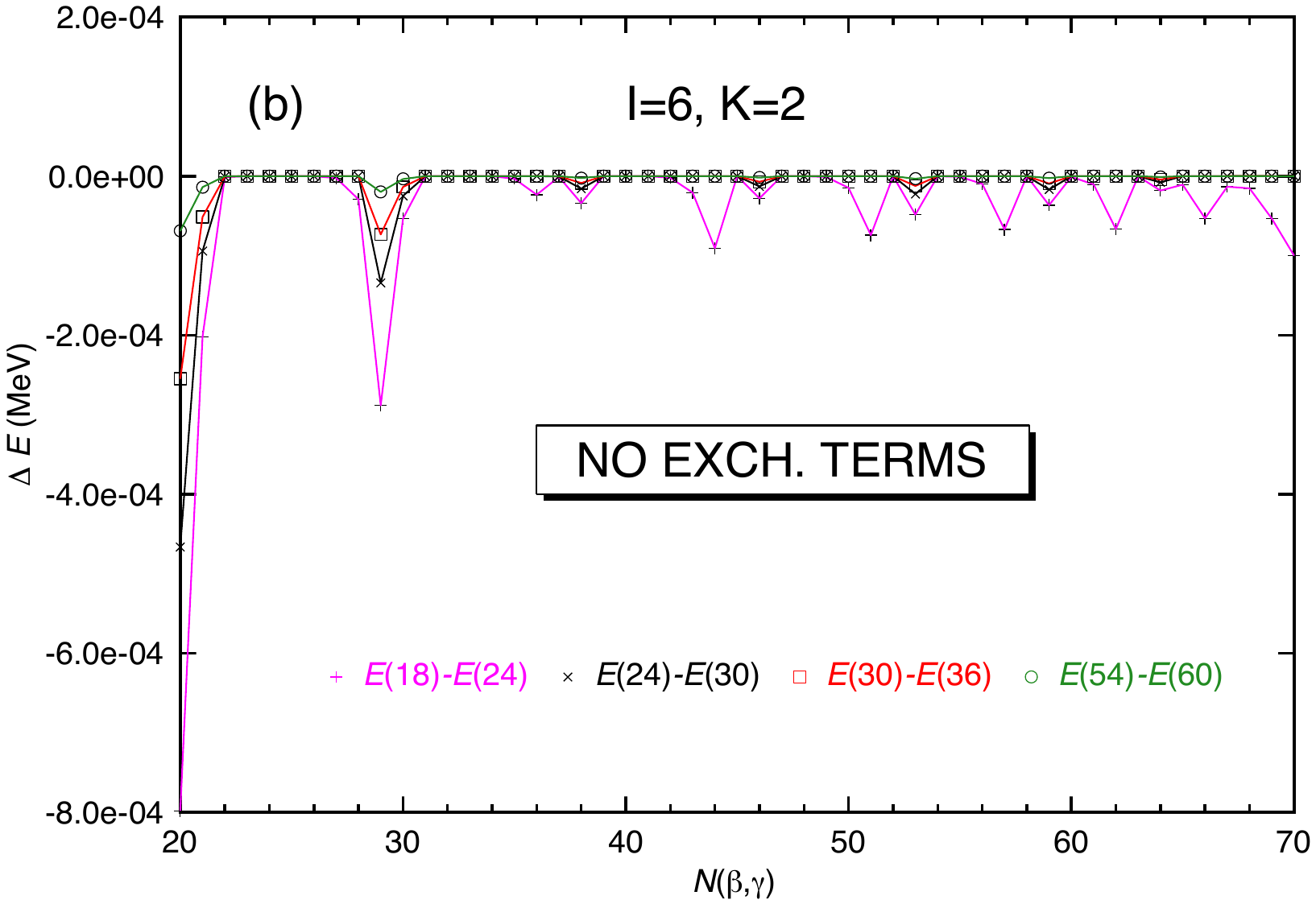}
	\end{center}
	\caption{(Color Online)  Same as Fig.~\ref{fig:AMP_conv_K0} but for $I=6$ and $K=2$.}
	\label{fig:AMP_conv_K2}
\end{figure}

To investigate the $K\ne 0$ states we also display the $I=6^+, K=2$ AMP energy convergence in  Fig.~\ref{fig:AMP_conv_K2}.  The first 19 $(\beta,\gamma)$ points correspond to axially symmetric shapes and have been omitted in the plot. In panel (a) we present the results with exchange terms. With the exception of 2 points (20 and  29) all points are perfectly converged with $N_{\beta}=18$.  These two points have relatively small $\beta$ deformations and large $\gamma$ values. The slower convergence of these two points is due to the oscillations of the Wigner function in Eq.~\ref{AMProj}.
The fact that the point number 20, the point with the worst convergence for $I=6^+, K=2$,  converges perfectly for $I=6^+, K=0$, see right panel of Fig.~\ref{fig:AMP_conv_K0} indicates that the K-independent norm overlap $\langle\phi (a)| \exp^{-i\beta \hat{J}_{y}}|\phi(a^{\prime})\rangle$, behaves properly. In panel (b) we present the results without exchange points. The results are very similar, though a little better, to the previous case. The fact that the results without exchange terms are a little better than with exchange terms corroborates the fact that the slow convergence of these two points has nothing to do  with poles.   

We would like to stress at this point the marked difference between the PNP and the AMP.
In a randomly chosen example ($^{166}$Er), and without an "optimal pole" (remember that the smallest values of $v^{2}_k -u^{2}_k$  is 0.01) we obtain large differences in the PNP calculations with and without exchange terms. At variance in the AMP case, we do not find any  difference. It is also remarkable the  abundance of "poles" in the PNP  case compared with its absence in the AMP case.

The conclusion from the present analysis is that the neglect of exchange terms does not generate divergences in the AMP.  A more detailed and systematic study for many nuclei is underway \cite{FE.16_2}.

\subsection{Details of the density dependent term.} 

We have seen in the previous section that the divergences associated with the PNP due to the neglect of the exchange terms of the interaction can  be straightened out by including {\em all} missing exchange terms of the interaction.  Furthermore we have also seen that in the AMP case there are no divergences even if the exchange terms are neglected. 

The second source of divergences has its origin in the density dependent term of the interaction which we called $V_{DD}$. We discuss this term in this Section.

The density dependent  term was conceived for plain mean field approaches where only expectation values, i.e., diagonal matrix elements, do appear. Consequently in the mean field approach $V_{DD}$ is constructed to depend on the mean field density.
In theories beyond mean field, for example in particle number projection, the contribution to the energy of the 
density dependent term is given by
\begin{eqnarray}
E^P_{DD}   & = & \frac{\langle \Phi^N | {\hat{V}}_{DD}\left [ \overline{\rho} 
	(\vec{r}) \right ] | \Phi^N \rangle} 
{\langle \Phi^N | \Phi^{N} \rangle} \nonumber \\ 
&=&  \frac{  {\displaystyle 
		\int }   d {\varphi}_{} \langle 
	\phi  | {\hat{V}}_{DD}
	\left [ \overline{\rho} (\vec{r}) \right ]  
	e^{i{\varphi}_{} {\hat{N}}_{} } | \phi \rangle  } 
{ {\displaystyle  \int } 
	d {\varphi}_{} \langle 
	\phi | 
	e^{i{\varphi}_{} {\hat{N}}_{} } | \phi \rangle }
\label{eq:epdd}
\end{eqnarray}
where $\left [ \overline{\rho} (\vec{r}) \right ]$ indicates the explicit
dependence of $V_{DD}$ on a density $ \overline{\rho} (\vec{r})$ to be specified. Looking
at these expressions it is not obvious which dependence should be used.
There are two more or less straightforward prescriptions \cite{VER.00}  for 
$ \overline{\rho} (\vec{r})$.

The first prescription is inspired by the following consideration:
In the mean field approximation, the energy is given by
\begin{equation}
E= \frac{\langle \phi | \hat{H} | \phi \rangle }{\langle \phi | \phi \rangle},
\end{equation}
with $\phi$ a HFB wave function, and $V_{DD}$ is assumed to  depend on the density  
\begin{equation}
\rho(\vec{r})=\frac{\langle \phi | \hat{\rho} | \phi \rangle}{\langle \phi | \phi \rangle}.
\end{equation}
On the other hand, if the wave function which describes the nuclear system 
is the projected wave function $| \Phi^N \rangle$,   we 
have to calculate the matrix element
\begin{equation}
E_{DD}= \frac{\langle \Phi^N | \hat{V}_{DD} | \Phi^N \rangle}{\langle \Phi^N | \Phi^N \rangle}.
\end{equation}
It seems reasonable, therefore,  to use in $V_{DD}$ the density 
\begin{equation}
 \overline{\rho} (\vec{r})\equiv \rho^N(\vec{r})= \frac{\langle \Phi^N | \hat{\rho} | \Phi^N \rangle }
 {\langle \Phi^N | \Phi^{N} \rangle}, 
\end{equation}
 i.e. the projected 
density. 
One has to be aware that this prescription  can only  be used in the case of the particle number 
projection where one projects in the gauge space associated to the particle number operator and which has nothing to do
with the spacial coordinates. In the case of symmetries associated with $\vec{r}$ like the angular momentum or parity projection one has to work with the next prescription. 

The second prescription has been guided by the choice usually done in the
Generator Coordinate method with density dependent forces \cite{BD.90}. The
philosophy behind this prescription is the following: to
evaluate Eq.~(\ref{eq:epdd}) we have to calculate matrix elements between different
product wave functions $|\phi \rangle$ and $| \tilde{\phi} \rangle$
($|\tilde{\phi} \rangle =e^{i\varphi\hat{N}}| \phi \rangle$)
(see last term in Eq.~(\ref{eq:epdd})). 
Then, to calculate  matrix elements of the form
\begin{equation}
\frac{\langle \phi | {\hat{V}}_{DD} | \tilde{\phi} \rangle}{\langle \phi |\tilde{\phi} \rangle}
\end{equation}
we choose the mixed density
\begin{equation}
\overline{\rho} (\vec{r})= {\rho}_{\varphi} (\vec{r}) =
 \frac{\langle \phi | \hat{\rho} (\vec{r})| \tilde{\phi} \rangle}{\langle \phi |
	\tilde{\phi} \rangle}
\end{equation}
to be used in $ {\hat{V}}_{DD}$.
This approach is called the mixed density prescription. \\

Both prescriptions have been tested with the Gogny force in the Lipkin
Nogami approach \cite{Valor1997249}, of course pole free,  and practically no difference was found in the numerical applications. One should notice that in the second prescription 
$\overline{\rho} (\vec{r})$ depends on the angle $\varphi$ at variance with the first
prescription.
It has been shown in Ref.~\cite{AER.01P} for the PNP that  the projected prescription is free from divergences while the mixed prescription may present some problems. 
Unfortunately as mentioned above one cannot use the projected density for projectors related with $\vec{r}$, like the AMP or the parity projection, but one can use it for the PNP.

Specifically, in all calculations with the Gogny force presented in this paper we have adopted the following densities:
In the solution of the PN-VAP equations, Eq.~(\ref{eq:E_TOT}), we have used the projected density
\begin{equation}
\rho^{N}(\vec{r})\equiv\frac{\langle\Phi|\hat{\rho}(\vec{r})P^{N}|\Phi\rangle}{\langle\Phi|P^{N}|\Phi\rangle}.
\end{equation}

In the evaluation of the Hamiltonian overlaps of Eq.~(\ref{hamove}) we have used 
the particle number projected spatial density combined with the mixed prescription for the angular momentum projection and GCM part, namely:
\begin{equation}
\rho^{N}(\Omega,\vec{r})\equiv\frac{\langle\Phi|\hat{\rho}(\vec{r})\hat{R}(\Omega)P^{N}|\Phi'\rangle}{\langle\Phi|\hat{R}(\Omega)P^{N}|\Phi'\rangle}.
\end{equation}
This prescription is suitable for dealing with the restoration of broken symmetries  in the coordinate space such as the rotational invariance or the spatial parity

We now collect all information.  We have two sources of problems and two kinds of projections, PNP and AMP. The facts are the following:
 \begin{itemize}
 
 \item  The demonstrations  of Ref.~\cite{AER.01P} apply {\em only} for the PNP {\em but definitively not} for the AMP. We are not aware of any proof for the AMP neither positive or negative. Therefore we admit that it could be that the AMP presents also problems. 
  
   \item  The first source of problems is the neglect of the exchange terms. We have seen in the previous section for the Pairing plus Quadrupole Hamiltonian that if one neglects the exchange terms one has definitively problems with the PNP but not with AMP. Since there is no way to avoid the problems in the PNP, we adopt  the  solution of taking into account  all exchange terms and the problem disappears. 

 \item It has been demonstrated  for the PNP case that the problems 
 arising from the density dependence of the interaction are avoided if one uses the density dependent prescription.  Consequently, we adopt this prescription for the PNP.  For the AMP,  the projected prescription cannot be used and one must use the mixed one. 
\end{itemize}

With this premises and taking into account all exchange terms of the interaction and the above mentioned density prescription the only imaginable source of divergences  could be an eventual pole of the density dependent term in the angular momentum projection.  However, since the origin of the potential poles in this case is the same as with the exchange terms, Ref.~\cite{AER.01P}, namely that the norm overlap vanishes,  our first guess will be that we will not find poles either. To confirm this impression 
we have checked explicitly the convergence of the energy.  We have analyzed many calculations with axial and triaxial angular momentum projection with the Gogny force and we have never found any hint of poles.  In particular in Ref.~\cite{Nuria2} we studied the paradigmatic case of $^{18}$O used by Bender and collaborators \cite{PhysRevC.79.044319}  with different numbers of Fomenko points and Euler angles and we did not find any evidence of poles. After so many negative checks the conclusion one arrives at is that the potential poles of the density dependent part of the interation associated with the AMP behave like the ones of the exchange terms of the interaction discussed in the previous section, i.e., they do not show up.

The conclusion  with respect to the use of the mixed density prescription in the case of the angular momentum projection with the Gogny interaction is that either there are no problems or they appear, contrary to the PNP case, so seldom that the probability of finding them in practical calculations is quite negligible.  

\section{Appendix C: Do we really need Particle Number Projection?}
\label{App:C}
\begin{figure*}[tb]
\begin{center}
\includegraphics[angle=0, scale=0.4]{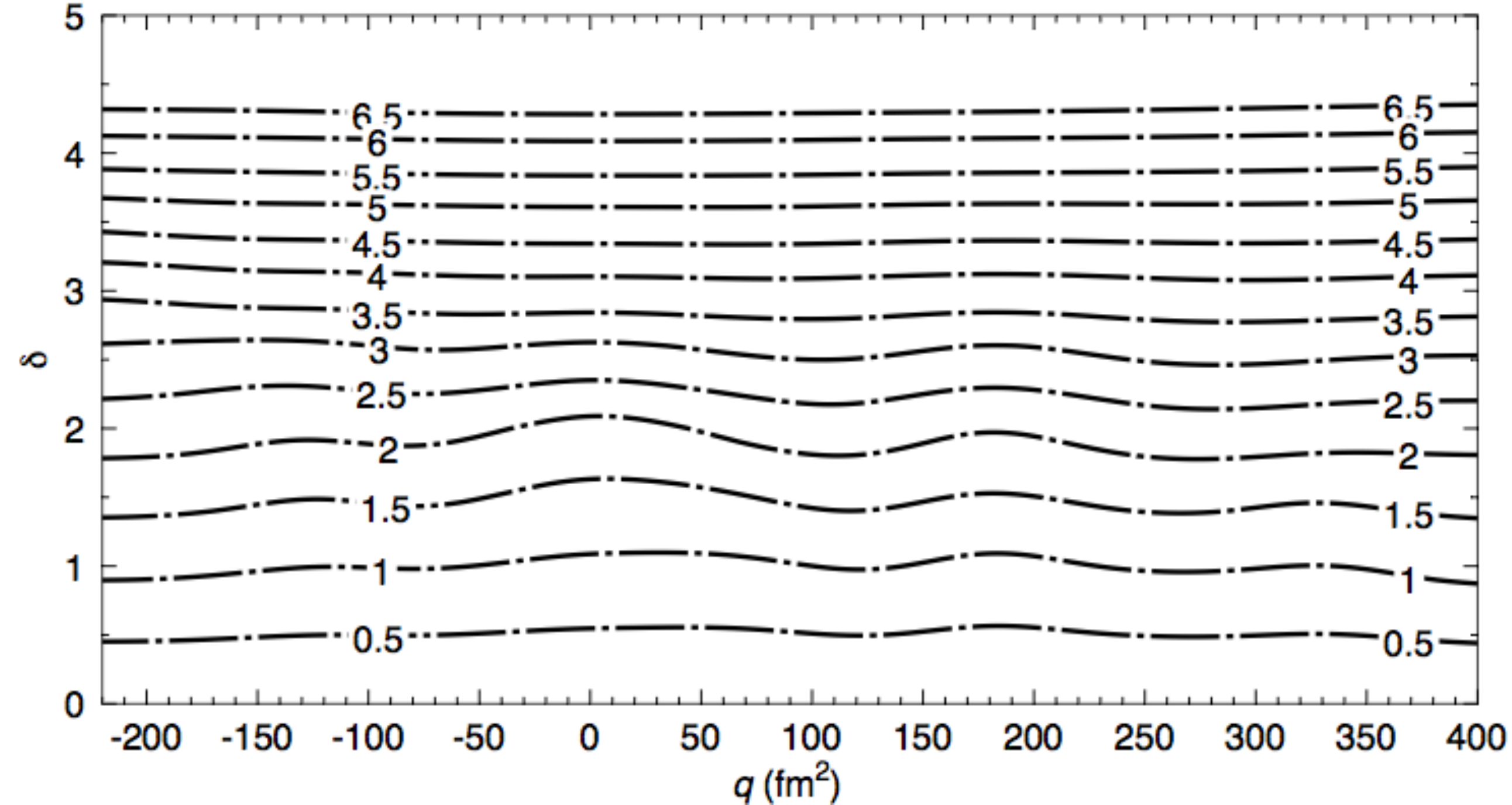}
\end{center}
\caption{(Color Online) Contour plot of the square root of the absolute values of the pairing energies in MeV for $^{52}$Ti in the constrained HFB approach as a function of the
constraining parameters $(\delta,q)$.}
\label{fig:pair_ener_q}
\end{figure*}

It is commonly accepted that  the HFB theory works fine in the strong pairing regime and not that well in the weak pairing.
In this appendix we provide a quantitative  discussion and a detailed analyses  of the validity of the previous sentence
as well as the effects of particle number projection in different physical situations.

   Since the Bogoliubov transformation violates particle number conservation one must do something to obtain trustworthy results. 
  The restoration of the symmetry can be incorporated in the HFB equation  either in a semiclassical  approach or in a full quantum theory.   In the semiclassical way one invokes the Lagrange multipliers theory to ensure particle number conservation on the average.  The minimization of  $E^{\prime}= \langle \phi |\hat{H} |\phi \rangle - \lambda  \langle \phi |\hat{N} |\phi \rangle$, with $\lambda$ determined by the condition    $\langle \phi |\hat{N} |\phi \rangle=N$, provides the wave function $|\phi \rangle$. The energy of the system is given by $E_{HFB} = E^{\prime} + \lambda \langle \phi |\hat{N} |\phi \rangle$.   From the HFB solution one can derive the gap equation and show that a solution different from the trivial one $(\Delta=0)$ can only be found in a strong enough pairing regime where $\langle(\Delta \hat{N})^2 \rangle \gg 1$ \cite{RS.80}. This is a qualitative  justification of the statement above.
     
   It is also well known that the particle number constrained HFB equation can be derived in a quantum theory in the frame of the Kamlah expansion \cite{Ka.68} ( see also Ref.~\cite{RS.80}, p. 466).  In this case the particle number conservation is imposed on the wave function, i.e.,  $|\Phi \rangle = P^N |\phi\rangle$.  Using the Kamlah expansion to determine the intrinsic wave function $|\phi \rangle$,  one obtains the remarkable result that, in a first order approach to an exact {\em variation after projection}, the variational equations are exactly the same as the semiclassical ones. In the quantum case, however, the approximate energy is given by  $E_{\rm HFB+PNP}= \langle \phi |\hat{H} \hat{P}^N|\phi \rangle/ \langle \phi |\hat{P}^N |\phi \rangle$. The first order in the Kamlah expansion is reached only when $\langle \phi |(\Delta \hat{N})^2|\phi\rangle \gg 1$.  Since large $\langle \phi |(\Delta \hat{N})^2|\phi\rangle$ imply large pairing correlations, we obtain in this way an additional justification of the assertion of the mean field practitioners.
   Furthermore, since the Kamlah expansion provides an approximation to a PN-VAP approach, we expect a good agreement
 between the HFB+PNP energy and the PN-VAP one in the limit   $\langle \phi |(\Delta \hat{N})^2|\phi\rangle \gg 1$.  

Thus, we have seen that the condition of strong pairing regime, or equivalently that $\langle \phi |(\Delta \hat{N})^2|\phi\rangle \gg 1$, 
is obtained in the semiclassical  as well as in the quantum approaches. 

There are, however, differences in the interpretation of this requirement and in the calculation of observables in both approaches.

{\em Semiclassical way:}  Expectation values and transition matrix elements are calculated with the wave function $|\phi\rangle$.
If the condition $\langle \phi |(\Delta \hat{N})^2|\phi\rangle \gg 1$ is not satisfied one expects a sharp transition to the non-superfluid
phase and a deterioration of the approach.

{\em Quantum way:}  Expectation values and transition matrix elements are calculated with the projected wave function $P^N|\phi\rangle$.  The right approach is the PN-VAP one, however, if the condition $\langle \phi |(\Delta \hat{N})^2|\phi\rangle \gg 1$ is 
fulfilled, the HFB+PNP approach provides a good approximation to the PN-VAP one.

It therefore seems that the pertinent questions to be answered are: Do we need particle number projection at all and if yes, do we need PN-VAP?
\begin{figure*}[tb]
\begin{center}
\includegraphics[angle=0, scale=0.35]{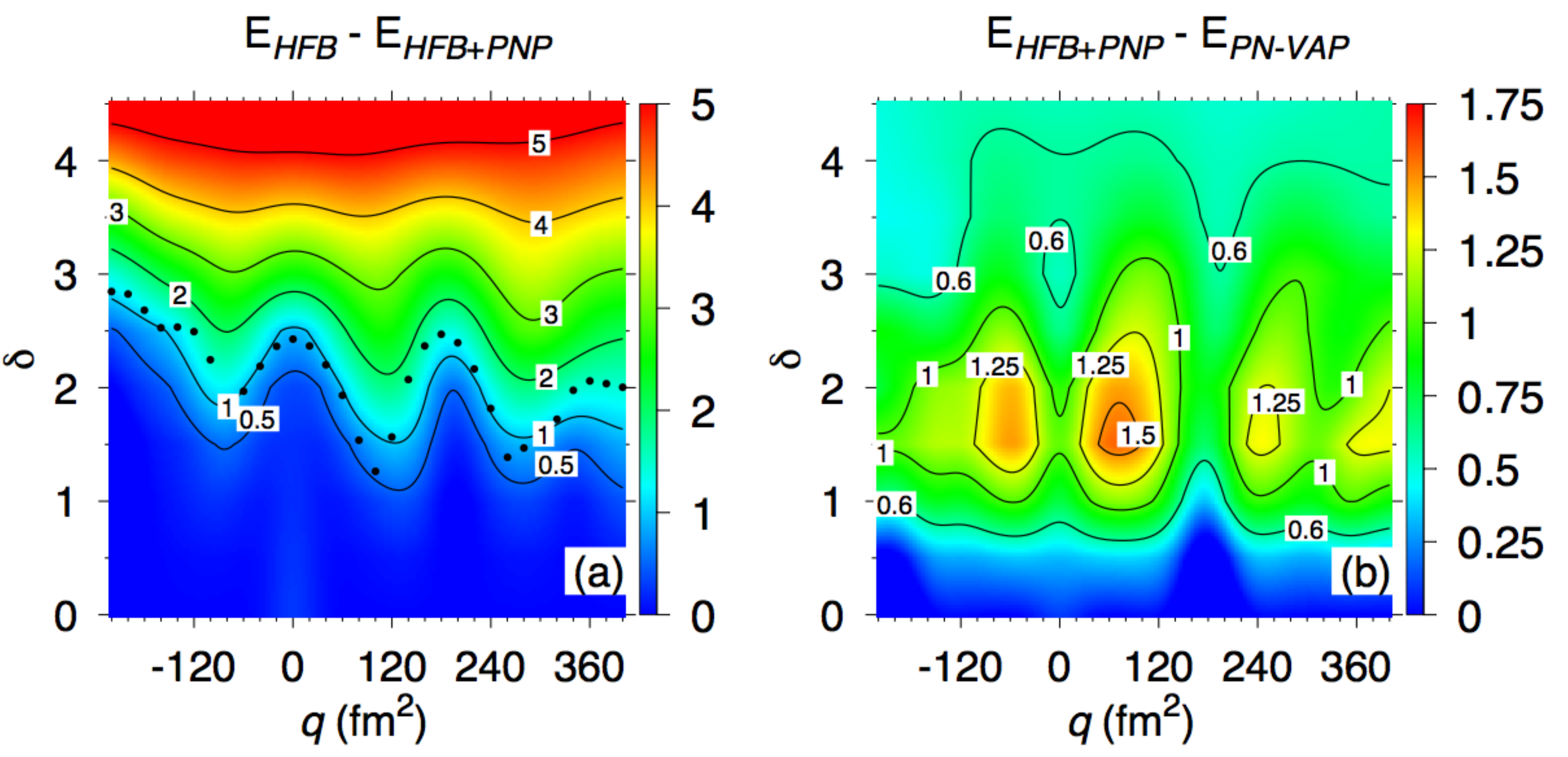}
\end{center}
\caption{(Color Online) (a) Contour plot of the energy differences $E_{\rm HFB}(q,\delta) -E_{\rm HFB+PNP}(q,\delta)$  for $^{52}$Ti in MeV in the 2D constrained HFB approach as a function of the constraining parameters. The dots represent the same energy difference for the 1D calculations
(b) Contour plot of the energy differences $E_{\rm HFB+PNP}(q,\delta) -E_{\rm PN-VAP}(q,\delta)$ for $^{52}$Ti in MeV in the 2D constrained HFB approach as a function of the constraining parameters.}
\label{fig:ener_diff}
\end{figure*}

To provide the right answer one further aspect that must be considered  is the type of calculations  performed. The simplest case takes place when one restricts himself to {\em only one} HFB vacuum  and  its eventual excitations, as two-quasiparticle states or QRPA ones, etc.  In this case since all wave functions are always referred to the same reference (vacuum)  if the condition   $\langle \phi |(\Delta \hat{N})^2|\phi\rangle \gg 1$ is satisfied, issues are simpler and probably alright.      In beyond mean field theories, like in the present work,  we are confronted with many vacua of different character, for example when we perform HFB constrained $(\beta,\gamma)$  calculations where one goes through  different $(\beta,\gamma)$  points of very weak level density or very high one. Or when one considers pairing fluctuations where one has a continuous set of all possible
pairing regimes, like the one displayed by the wave functions $|\phi(q,\delta)\rangle$ in Sect.~(\ref{Sect:ASCMCBPDF}). Another well known example of this situation is found, for instance, in the calculation of the Yrast band of a deformed nucleus, where the vacua depend on the angular momentum. In this case the above condition must be satisfied for each value of the angular momentum. In the past one has found that often this is not the case  and that better approaches  like the  Lipkin-Nogami or the PN-VAP one have to be applied.

  To facilitate the discussion we chose the case of the nucleus $^{52}$Ti with $\beta$ and pairing fluctuations where we can study the change from a weak to a strong pairing regime in a continuous way. This nucleus has been studied in detail in Sects.~(\ref{Sect:ASCMCBDF},
  \ref{Sect:ASCMCBPDF}). To illustrate the properties of our vacua in Fig.~\ref{fig:pair_ener_q} we display contour
  plots of the square root of the  absolute value of the pairing energies  in the $(q,\delta)$ plane  in the HFB approach and for $^{52}$Ti.
Looking at this figure one can conclude that in the HFB approach  $\langle \phi |(\Delta \hat{N})^2|\phi\rangle^{1/2}$ is proportional to the pairing correlation energy. An important remark is that,  due to the constraint on $(\Delta \hat{N})^2$,  in the calculations we have all pairing regimes. Notice also than because of the constraint on $(\Delta \hat{N})^2$ the HFB approach cannot collapse even in the weak pairing regime. The total energy cost to make the constraint, however, is different in each situation and depends on the number of particles and on the deformation (level density).

   Let's now turn back to the question whether particle number projection is needed in beyond mean field calculations.   A first answer  to this question is provided in Fig.~\ref{fig:ener_diff}a where we show contour plots of the energy difference $E_{\rm HFB}(q,\delta) -E_{\rm HFB+PNP}(q,\delta)$. Here we observe that this quantity is very sensitive to $\delta = \langle \phi |(\Delta \hat{N})^2|\phi\rangle^{1/2}$.  In particular we observe that it increases proportionally to $\delta$ causing that the HFB energy surfaces are steeper than the HFB+PNP ones. In particular, for very small   $\delta$ values, i.e., in the absence of pairing correlations, both energies, as it should, do coincide.
   It is important to notice that, for the $\delta$ values where the energy minimum is found, i.e,  $\delta \approx 2$, see Fig.~\ref{fig:pairfluc_Ti52}, one observes also a strong dependence of the energy difference on the deformation. In conclusion we find that in the scope of this approach the PNP strongly affects the energy surface and as a consequence it should be performed. It is interesting to notice that for  very small (or very large) $\delta$ values  the energy difference is independent of the deformation parameter $q$.  In the same plot the $q$-constrained 1D energy differences are represented by dots. In this case the shell structure dictated by the constraining parameter $q$ induces changes in the level density and thereby in the pairing correlations making again a PNP necessary.

   With respect to the second question, whether we need PN-VAP, we can obtain again an impression looking at  Fig.~\ref{fig:ener_diff}(b) where we show contour plots of the energy difference $E_{\rm HFB+PNP}(q,\delta) -E_{\rm PN-VAP}(q,\delta)$. This question is obviously related
  to the convergence of the Kamlah expansion. In this plot we clearly differentiate three regions: for small $\delta$-values, $\delta \leq 0.5$, the PNP and PN-VAP energies differ very little, as expected, since the wave function $|\phi \rangle$ is almost a Slater determinant. For intermediate values, $0.5 \leq \delta \leq 3.0$ there are energy differences up to approximately 2 MeV and for
$\delta \geq 4.0$   the energy differences become again very small.   The first and third region correspond to the limiting situations of 
very small and very large pairing correlations and behave as  expected according to the Kamlah expansion.  The crucial region in this nucleus is around $1.0 \le \delta \le 3.0$.  This region is again the relevant one since the minima (for fixed q) are located in this region and to solve
the Hill-Wheeler equation with the HFB+PNP or PV-VAP energy surface could lead to different results.
See the left and middle spectra of Fig.~\ref{fig:spect_Ti} for a quantitative comparison.

The oscillations in the deformation parameter $q$ have obviously to do with the evolution of the shell structure with this
parameter and the corresponding fluctuations in the level density. With respect to the 1D case the same comments can be made as above.
   
    The fact that the energy differences $E_{\rm HFB+PNP}(q,\delta) -E_{\rm PN-VAP}(q,\delta)$ are, in general, much smaller than the  $E_{\rm HFB}(q,\delta) -E_{\rm HFB+PNP}(q,\delta)$ ones is a clear indication  that the largest source of incorrectness of the HFB approach 
are its wave function components with the wrong number of particles {\em but not} the HFB wave function itself.


\end{document}